\documentclass[a4paper]{article}
 
\usepackage{a4wide}
\usepackage[utf8]{inputenc}
\usepackage{amsfonts}
\usepackage{amsmath, amsthm}
\usepackage{enumitem}
\usepackage{ebezier}
\usepackage{microtype}

\theoremstyle{definition}
\newtheorem{theorem}{Theorem}
\newtheorem{lemma}[theorem]{Lemma}
\newtheorem{proposition}[theorem]{Proposition}
\newtheorem{corollary}[theorem]{Corollary}
\newtheorem{example}[theorem]{Example}
\newtheorem{definition}[theorem]{Definition}

\newcommand{\cA}{\mathcal{A}}
\newcommand{\fA}{\mathfrak{A}}
\newcommand{\cB}{\mathcal{B}}
\newcommand{\fB}{\mathfrak{B}}
\newcommand{\cC}{\mathcal{C}}
\newcommand{\cD}{\mathcal{D}}
\newcommand{\fD}{\mathfrak{D}}
\newcommand{\bD}{\overline{D}}
\newcommand{\cF}{\mathcal{F}}
\newcommand{\cH}{\mathcal{H}}
\newcommand{\cI}{\mathcal{I}}
\newcommand{\cL}{\mathcal{L}}
\newcommand{\cO}{\mathcal{O}}
\newcommand{\cS}{\mathcal{S}}
\newcommand{\bS}{\overline{S}}
\newcommand{\cV}{\mathcal{V}}

\newcommand{\Nat}{\mathbb{N}}
\newcommand{\Int}{\mathbb{Z}}

\newcommand{\<}{\langle}
\renewcommand{\>}{\rangle}

\newcommand{\vars}{\text{\upshape{vars}}}
\newcommand{\consts}{\text{\upshape{consts}}}

\newcommand{\len}{\text{len}}

\newcommand{\hphi}{\widehat{\varphi}}

\newcommand{\ti}{\widetilde{i}}
\newcommand{\tj}{\widetilde{j}}
\newcommand{\tx}{\widetilde{x}}
\newcommand{\ty}{\widetilde{y}}

\newcommand{\subst}[1]{\bigl[#1\bigr]}

\newcommand{\semequiv}{\mathrel{|}\joinrel\Relbar\joinrel\mathrel{|}}

\newcommand{\At}{\text{At}}
\newcommand{\Sk}{\text{Sk}}

\newcommand{\fU}{\mathfrak{U}}
\newcommand{\fa}{\mathfrak{a}}
\newcommand{\fb}{\mathfrak{b}}
\newcommand{\fc}{\mathfrak{c}}

\newcommand{\vu}{\vec{\mathbf{u}}}
\newcommand{\vv}{\vec{\mathbf{v}}}
\newcommand{\vx}{\vec{\mathbf{x}}}
\newcommand{\vy}{\vec{\mathbf{y}}}
\newcommand{\vz}{\vec{\mathbf{z}}}

\newcommand{\fP}{\mathfrak{P}}

\newcommand{\twoup}[2]{{2^{\uparrow #1}(#2)}}

\newcommand{\gnd}{\text{gnd}}

\newcommand{\Mapsto}{{\mathop{\mapsto}}}

\newcommand{\Domino}{\text{DOMINO}}
\newcommand{\Ntime}{\text{NTIME}}
\newcommand{\SatSF}{\text{Sat(SF)}}
\newcommand{\SFdk}{\text{SF}_{\degree \leq k}}
\newcommand{\SFd}[1]{\text{SF}_{\degree \leq #1}}
\newcommand{\SatSFdk}{\text{Sat($\SFdk$)}}

\newcommand{\Succ}{\text{Succ}}
\newcommand{\eq}[1]{\text{eq}^{#1}}
\newcommand{\MinIdx}{\text{MinIdx}}
\newcommand{\MaxIdx}{\text{MaxIdx}}

\newcommand{\lvl}{\text{lvl}}
\newcommand{\degree}{\partial}

\newcommand{\ul}[1]{{\underline{#1}}}
\newcommand{\tfa}{\widetilde{\fa}}
\newcommand{\tfb}{\widetilde{\fb}}

\newcommand{\SF}{\text{SF}}
\newcommand{\poly}{\text{poly}}

\newcommand{\EXPTIME}{\textsc{ExpTime}}
\newcommand{\NEXPTIME}{\textsc{NExpTime}}

\newcommand{\happrox}{\mathrel{\hat{\approx}}}
\newcommand{\AxFin}{{\chi_{\text{fin}}}}

\begin{document}
\title{A Fine-Grained Hierarchy of Hard Problems\\ in the Separated Fragment}

\author{
	\begin{tabular}{l}
		Marco Voigt\\
		\small\textit{Max Planck Institute for Informatics, Saarland Informatics Campus, Saarbr\"ucken, Germany,}\\
		\small\textit{Saarbr\"ucken Graduate School of Computer Science}
	\end{tabular}
}	
\date{}
\maketitle

\begin{abstract}
	Recently, the separated fragment (SF) has been introduced and proved to be decidable.
	Its defining principle is that universally and existentially quantified variables may not occur together in atoms.
	The known upper bound on the time required to decide SF's satisfiability problem is formulated in terms of quantifier alternations:
	Given an SF sentence $\exists \vz\, \forall \vx_1 \exists \vy_1 \ldots \forall \vx_n \exists \vy_n .\, \psi$ in which $\psi$ is quantifier free, satisfiability can be decided in nondeterministic $n$-fold exponential time.
	In the present paper, we conduct a more fine-grained analysis of the complexity of SF-satisfiability.
	We derive an upper and a lower bound in terms of the \emph{degree $\degree$ of interaction of existential variables (short: degree)}---a novel measure of how many separate existential quantifier blocks in a sentence are connected via joint occurrences of variables in atoms.
	Our main result is the $k$-\NEXPTIME-completeness of the satisfiability problem for the set $\SF_{\degree \leq k}$ of all SF sentences that have degree $k$ or smaller.
	Consequently, we show that SF-satisfiability is non-elementary in general, since SF is defined without restrictions on the degree.
	Beyond trivial lower bounds, nothing has been known about the hardness of SF-satisfiability so far.	
\end{abstract}


\section{Introduction}

In \cite{Voigt2016} the separated fragment (SF) of first-order logic with equality is introduced. 
Its defining principle is that universally and existentially quantified variables may not occur together in atoms.
(Topmost existential quantifier blocks are exempt from this rule.)
SF properly generalizes both the Bernays--Sch\"onfinkel--Ramsey (BSR) fragment ($\exists^*\forall^*$-sentences with equality) and the relational monadic fragment without equality (MFO).
Still, the satisfiability problem for SF is decidable.

In computational logic formulas are often classified based on the shape of quantifier prefixes.
There is a wealth of results that separate decidable first-order formulas from undecidable ones in this fashion, see \cite{Borger1997} for references.
The definition of the BSR fragment is only one example.
In the context of computational complexity, hierarchies are defined, such as the polynomial hierarchy, where the hardness of problems is assumed to grow with the number of quantifier alternations that are allowed to occur.

\begin{figure}[hb]
	\begin{center}
		\begin{picture}(190, 190)
			\put(0,0){\framebox(166,185){}}
		
			\put(3,0){\cbezier(0,0)(10,185)(150,185)(160,0)}
			\put(3,0){\cbezier(10,0)(20,133)(140,133)(150,0)}
			\put(3,0){\cbezier(20,0)(30,80)(130,80)(140,0)}
			\put(3,0){\cbezier(30,0)(40,50)(120,50)(130,0)}
			\put(3,0){\cbezier(40,0)(50,20)(110,20)(120,0)}
			
			\put(52,170){\begin{tabular}{c} \textsc{Primitive}\\ \textsc{Recursive}\end{tabular}}
			\put(81,149){$\vdots$}
			\put(54,120){\textsc{Elementary}}
			\put(81,105){$\vdots$}
			\put(55,82){$k$-\textsc{NExpTime}}
			\put(81,66){$\vdots$}
			\put(55,44){\textsc{2-NExpTime}}
			\put(59,23){\textsc{NExpTime}}
			\put(77,3){\textsc{NP}}
			
			\put(83,141.5){\circle*{3}}
			\put(83,141.5){\line(1,0){91}}
			\put(83,98.5){\circle*{3}}
			\put(83,98.5){\line(1,0){91}}
			\put(83,59){\circle*{3}}
			\put(83,59){\line(1,0){91}}
			\put(83,36.5){\circle*{3}}
			\put(83,36.5){\line(1,0){91}}
			\put(83,14){\circle*{3}}
			\put(83,14){\line(1,0){91}}
		
			\put(175,138){\mbox{SF}}
			\put(175,95){\mbox{$\SFd{k}$}}
			\put(175,56){\mbox{$\SFd{2}$}}
			\put(175,33.5){\mbox{$\SFd{1}$}}
			\put(175,11){\mbox{$\SFd{0}$}}
		\end{picture}
		\caption{The subfragments of SF scale over the major nondeterministic complexity classes in \textsc{Elementary}, while SF itself goes even beyond. }
		\label{figure:ComplexityOfSF}
	\end{center}
\end{figure}
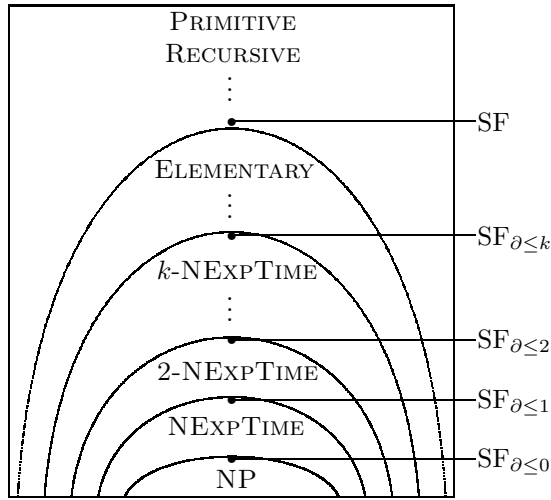

Although the definition of SF breaks with the paradigm of restricting quantifier prefixes, the known upper bound on the complexity of SF-satisfiability is again based on quantifier prefixes:
Deciding whether an SF sentence $\varphi := \exists \vz \,\forall \vx_1 \exists \vy_1 \ldots \forall \vx_n \exists \vy_n. \psi$ with quantifier-free $\psi$ is satisfiable requires a nondeterministic computing time that is at most $n$-fold exponential in the length of $\varphi$ (cf.\ Theorem~17 in \cite{Voigt2016}).
On the one hand, we complement this result with a corresponding lower bound in the present paper.
That is, we show that SF-satisfiability is indeed non-elementary.
On the other hand, we derive a refined upper bound that is based on the \emph{degree $\degree$ of interaction of existential variables}.
An overview of the resulting hierarchy of complete problems is depicted in Figure~\ref{figure:ComplexityOfSF}.
Intuitively, $\varphi$ exhibits a degree $\degree_\varphi = k$, if variables from $k$ distinct existential quantifier blocks interact. 
We say that two variables $x, y$ interact, if they occur together in at least one atom or if there is a third variable $z$ that interacts with both $x$ and $y$ (i.e.\
the property is transitive). 
For instance, in the SF sentence 
	$ \forall x_1 \exists y_1 v_1 \forall x_2 \exists y_2 v_2 \forall x_3 \exists y_3 v_3.\; \bigl(P(x_1, x_2, x_3) \wedge \neg Q(y_1, y_3)\bigr) \vee P(y_2, v_2, v_3) \vee \neg Q(y_3, v_1) $ 
the sets $\{y_1, y_3, v_1\}$ and $\{y_2, v_2, v_3\}$ form the maximal sets of interacting existential variables.
Since each of these sets contains variables from at most two distinct quantifier blocks, the formula exhibits a degree $\degree = 2$.

In Section~\ref{section:DegreeOfInteraction}, and in particular in Theorem~\ref{theorem:SFComplexityDependingOnDepth}, we observe that the satisfiability problem for $\SF_{\degree \leq k}$---the set of all SF sentences $\varphi$ with $\degree_\varphi \leq k$---lies in $k$-\NEXPTIME.
It is worth mentioning that this result adequately accounts for the known complexity of MFO-satisfiability.
For every MFO sentence $\varphi$ we trivially have $\degree_\varphi = 1$, since all occurring predicate symbols have an arity of at most one.
Theorem~\ref{theorem:SFComplexityDependingOnDepth} entails that MFO-satisfiability is in \NEXPTIME, which is well known.
Still, this bound is not reproducible with the analysis of the complexity of SF-satisfiability conducted in \cite{Voigt2016}.
Apparently, non-elementary satisfiability problems are not very widespread among the decidable fragments of classical fist-order logic known today.
We show in Section~\ref{section:ComputationalLowerBounds} that SF falls into this category.
To the present author's knowledge, the only known companion in this respect is the fluted fragment (FL). 
Indeed, Pratt-Hartmann, Szwast, and Tendera show in \cite{PrattHartmann2016} that satisfiability of fluted sentences with at most $2k$ variables is $k$-\NEXPTIME-hard.
Moreover, they argue that satisfiability of fluted sentences with at most $k$ variables lies in $k$-\NEXPTIME.
Although a significant gap between these lower and upper bounds remains to be closed, the fluted fragment seems to comprise a similar hierarchy of hard problems as SF does.

Another characteristic of SF is that it enjoys a \emph{small model property}.
More precisely, given an SF sentence $\varphi$, one can compute a positive integer $n$ that depends on the degree $\degree_\varphi$ and the length of $\varphi$ such that, if there is a model of $\varphi$ at all, then there also is a model whose domain contains at most $n$ elements.
Many first-order fragments are  known to enjoy a small model property. The BSR fragment and MFO are among the classical ones (see \cite{Borger1997} for references). More recently defined fragments include the two-variable fragment (FO$_2$) \cite{Mortimer1975}, \cite{Gradel1997}, the fluted fragment (FL) \cite{Quine1969}, \cite{Quine1976}, \cite{PrattHartmann2016}, the guarded fragment (GF) \cite{Andreka1998}, \cite{Gradel1999}, the guarded negation fragment (GNF) \cite{Barany2015}, and the uniform one-dimensional fragment (UF$_1$) \cite{Kieronski2014}.
While GNF and UF$_1$ are incomparable, GNF extends GF, and UF$_1$ can be considered as a generalization of FO$_2$.
Guarded fragments and the two-variable fragment have received quite some attention due to the fact that modal logics have natural translations into them.
As a continuation of that theme, we shall see in Section~\ref{section:ExpressivenessOfSF} how classes of sentences enjoying a small model property can be effectively translated into subclasses of SF.
During the translation process the length of formulas increases by a factor that is logarithmic in the size of small models of the original.
One benefit of translating non-SF sentences into SF sentences is that in SF one can natively express concepts such as transitivity and basic counting quantifiers (Proposition~\ref{proposition:CountingQuantifiers}). 
This is not always possible in other fragments enjoying a small model property. For example, transitivity cannot be expressed in FO$_2$, GF, and FL. 

Summing up, the main contributions are:\newline
(i) Based on the novel concept of the degree of interaction of existential variables, we substantially refine the existing analysis of the time required to decide SF-satisfiability.
	More concretely, we show that a satisfiable SF sentence $\varphi$ with $\degree_\varphi = k$ has a model whose domain is of a size that is at most $k$-fold exponential in the length of $\varphi$ (Section~\ref{section:TranslationSFintoBSR}, Theorem~\ref{theorem:SFComplexityDependingOnDepth}).
	With this refined approach we can close the complexity gap for the class of \emph{strongly separated} sentences (Corollary~\ref{corollary:ComplexitySSF}) that was left open in~\cite{Voigt2016}. Moreover, the complexity of MFO can be explained in the refined framework. 
	\newline	
(ii) We complement the complexity analysis with corresponding lower bounds in two respects.
	We first derive a lower bound on the length of shortest BSR sentences that are equivalent to a given SF sentence (Section~\ref{section:LowerBoundsSmallestBSRsentences}, Theorem~\ref{theorem:LengthSmallestBSRsentences}).
	In Section~\ref{section:ComputationalLowerBounds}, we prove $k$-\NEXPTIME-hardness of satisfiability for the class of SF sentences $\varphi$ with $\degree_\varphi = k$ (Theorem~\ref{theorem:ComputationalLowerBoundForSF}). 
	Since SF is in general defined without restrictions on the degree $\degree_\varphi$, our result implies that SF-satisfiability is non-elementary.
	\newline
(iii) We devise a simple translation from classes of first-order sentences that enjoy a small model property into SF (Proposition~\ref{proposition:TranslationFMPintoSF}).
	Moreover, we argue that SF can express basic counting quantifiers (Proposition~\ref{proposition:CountingQuantifiers}).

In order to facilitate smooth reading, most proofs are only sketched in the main text and presented in full in the appendix. 
The present paper is the full version of the extended abstract~\cite{Voigt2017}.


\section{Preliminaries}\label{section:Preliminaries}

We mainly reuse the basic notions from \cite{Voigt2016}. We repeat the definition of necessary concepts and notation for the sake of completeness.

We consider first-order logic formulas. The underlying signature shall not be mentioned explicitly, but will become clear from the current context. For the distinguished \emph{equality} predicate we use $\approx$.
We follow the convention that negation binds strongest, that conjunction binds stronger than disjunction, and that all of the aforementioned bind stronger than implication. The scope of quantifiers shall stretch as far to the right as possible.
By $\len(\cdot)$ we denote a reasonable measure of the length of formulas satisfying $\len( \varphi \rightarrow \psi ) = \len( \neg \varphi \vee \psi )$ and $\len( \varphi \leftrightarrow \psi ) = \len( (\neg \varphi \vee \psi) \wedge (\varphi \vee \neg \psi) )$.

We write $\varphi(x_1, \ldots, x_m)$ to denote a formula $\varphi$ whose free variables form a subset of $\{x_1, \ldots, x_m\}$.
In all formulas we tacitly assume that no variable occurs freely and bound at the same time and that no variable is bound by two different occurrences of quantifiers, unless explicitly stated otherwise.
For convenience, we sometimes identify tuples $\vx$ of variables with the set containing all the variables that occur in $\vx$.
We write $\vars(\varphi)$ to address the set of all variable symbols that occur in $\varphi$. Similarly, $\consts(\varphi)$ denotes the set of all constant symbols in $\varphi$.
We denote \emph{substitution} by $\varphi\subst{x/t}$ if every free occurrence of $x$ in $\varphi$ is to be substituted with the term $t$.

A \emph{literal} is an atom or a negated atom, and a \emph{clause} is a disjunction of literals.
We say that a formula is in \emph{conjunctive normal form (CNF)}, if it is a conjunction of clauses, possibly preceded by a quantifier prefix.
A formula in CNF is \emph{Horn} if every clause contains at most one non-negated literal.
It is \emph{Krom} if every clause contains at most two literals at all.

A sentence $\varphi := \forall \vx_1 \exists \vy_1 \ldots \forall \vx_n \exists \vy_n. \psi$ is in \emph{standard form}, if it is in \emph{negation normal form} (i.e.\ every negation symbol occurs directly before an atom) and $\psi$ is quantifier free, contains exclusively the Boolean connectives $\wedge, \vee, \neg$, and does not contain non-constant function symbols.
The tuples $\vx_1$ and $\vy_n$ may be empty, i.e.\ the quantifier prefix does not have to start with a universal quantifier, and it does not have to end with an existential quantifier.
Moreover, we require that every variable occurring in the quantifier prefix does also occur in $\psi$.

As usual, we interpret a formula $\varphi$ with respect to given structures. A \emph{structure} $\cA$ consists of a nonempty \emph{universe} $\fU_\cA$ (also: \emph{domain}) and interpretations $f^\cA$ and $P^\cA$ of all considered function and predicate symbols, respectively, in the usual way. 
Given a formula $\varphi$, a structure $\cA$, and a variable assignment $\beta$, we write $\cA, \beta \models \varphi$ if $\varphi$ evaluates to \emph{true} under $\cA$ and $\beta$.
We write $\cA \models \varphi$ if $\cA, \beta \models \varphi$ holds for every $\beta$. 
The symbol $\semequiv$ denotes \emph{(semantic) equivalence} of formulas, i.e.\ $\varphi \semequiv \psi$ holds whenever for every structure $\cA$ and every variable assignment $\beta$ we have $\cA,\beta \models \varphi$ if and only if $\cA,\beta \models \psi$. 
We call two sentences $\varphi$ and $\psi$ \emph{equisatisfiable} if $\varphi$ has a model if and only if $\psi$ has one.

A structure $\cA$ is a \emph{substructure} of a structure $\cB$ (over the same signature) if (1) $\fU_\cA \subseteq \fU_\cB$, (2) $c^\cA = c^\cB$ for every constant symbol $c$, (3) $P^\cA = P^\cB \cap \fU_\cA^m$ for every $m$-ary predicate symbol $P$, and (4) $f^\cA(\fa_1, \ldots, \fa_m) = f^\cB(\fa_1, \ldots, \fa_m)$ for every $m$-ary function symbol $f$ and every $m$-tuple $\<\fa_1, \ldots, \fa_m\> \in \fU_\cA^m$. 
The following is a standard lemma, see, e.g., \cite{Ebbinghaus1994} for a proof.

\begin{lemma}[Substructure lemma]
	Let $\varphi$ be a first-order sentence in prenex normal form without existential quantifiers and let $\cA$ be a substructure of $\cB$.
	$\cB \models \varphi$ entails $\cA \models \varphi$.
\end{lemma}

\begin{lemma}[Miniscoping]\label{lemma:BasicQuantifierEquivalences}
	Let $\varphi, \psi, \chi$ be arbitrary first-order formulas, and assume that $x$ does not occur freely in $\chi$.\\
	\centerline{$
		\begin{array}{lcl}
			\exists x. (\varphi \vee \psi) 	&	\semequiv	&	(\exists x_1. \varphi) \vee (\exists x_2. \psi) ~,\\
			\exists x. (\varphi \circ \chi) 	&	\semequiv	&	(\exists x. \varphi) \circ \chi ~,\\
			\forall x. (\varphi \wedge \psi) 	&	\semequiv	&	(\forall x_1. \varphi) \wedge (\forall x_2. \psi) ~,\\
			\forall x. (\varphi \circ \chi) 	&	\semequiv	&	(\forall x. \varphi) \circ \chi ~,
		\end{array}
	$}	
	where $\circ \in \{\wedge, \vee\}$.
\end{lemma}

We use the notation $[k]$ to abbreviate the set $\{1, \ldots, k\}$ for any positive integer $k$.
Moreover, $\fP$ shall be used as the power set operator, i.e.\ $\fP S$ denotes the set of all subsets of a given set $S$.
Finally, we need some notation for the \emph{tetration operation}. We define $\twoup{k}{m}$ inductively: $\twoup{0}{m} := m$ and $\twoup{k+1}{m} := 2^{\left(\twoup{k}{m}\right)}$.


\section{The separated fragment}

Let $\varphi$ be a first-order formula. We call two disjoint sets of variables $X$ and $Y$ \emph{separated in $\varphi$} if and only if for every atom $A$ occurring in $\varphi$ we have $\vars(A) \cap X = \emptyset$  or $\vars(A) \cap Y = \emptyset$.
\begin{definition}[Separated fragment (SF), \cite{Voigt2016}]\label{definition:SeparatedFragment}
	The \emph{separated fragment (SF)} of first-order logic consists of all existential closures of prenex formulas without non-constant function symbols in which existentially quantified variables are separated from universally quantified ones. 
	More precisely, SF consists of all first-order sentences with equality but without non-constant function symbols that are of the form $\exists \vz\, \forall\vx_1 \exists\vy_1 \ldots \forall\vx_n \exists\vy_n.\, \psi$, in which $\psi$ is quantifier-free, and in which the sets $\vx_1 \cup \ldots \cup \vx_n$ and $\vy_1 \cup \ldots \cup \vy_n$ are separated.
	
	The tuples $\vz$ and $\vy_n$ may be empty, i.e.\ the quantifier prefix does not have to start with an existential quantifier and it does not have to end with an existential quantifier either.
\end{definition}
Notice that the variables in $\vz$ are not subject to any restriction concerning their occurrences.

In \cite{Voigt2016} the authors show that the satisfiability problem for SF sentences (\emph{SF-satisfiability}) is decidable.
Before we start investigating the complexity issues related to SF-satisfiability, we elaborate on the expressiveness of SF.


\subsection{Expressiveness}\label{section:ExpressivenessOfSF}

Every SF sentence is equivalent to a BSR sentence (\cite{Voigt2016}, Lemma~6). 
We shall outline in Section~\ref{section:TranslationSFintoBSR} how to analyze the blow-up that we have to incur during this translation process and how it depends on the degree of interaction of existential variables.
Since the BSR fragment enjoys a small model property (cf.\ Proposition~\ref{proposition:SmallModelsBSR}), SF inherits the small model property from BSR.
However, regarding the size of minimal models of satisfiable formulas, SF sentences are much more compact.
While satisfiable BSR sentences have models whose domain is linear in the length of the formula, satisfiable SF sentences can enforce domains of a size that cannot be bounded from above by a finite tower of exponentials. 
We provide first evidence for this fact in Theorem~\ref{theorem:LengthSmallestBSRsentences}, where we give a  non-elementary lower bound on the length of equivalent BSR sentences. 
This lower bound even applies to the SF-Horn-Krom subfragment of SF.
Moreover, we exploit the capability of SF sentences $\varphi$ to enforce models of $\degree_\varphi$-fold exponential size in the proof of the $k$-\NEXPTIME-hardness of SF-satisfiability (for every $k \geq 1$).

Apart from compactness of representation, and from the perspective of satisfiability, all first-order fragments that enjoy small model properties share a common ground of expressiveness.
Neglecting efficiency, every sentence $\varphi$ from such a fragment can be effectively translated into a (finite) propositional formula $\phi$ in such a way that from a satisfying variable assignment for $\phi$ one can straightforwardly reconstruct a (Herbrand) model of $\varphi$.
The reason is simply that universal quantification can then be understood as finite conjunction (over a finite domain) and existential quantification can be conceived as finite disjunction.

The following proposition illustrates why SF is to some extent prototypical for first-order fragments that enjoy a small model property. 
\begin{proposition}\label{proposition:TranslationFMPintoSF}
	Consider any nonempty class $\cC$ of first-order formulas without non-constant function symbols for which we know a computable mapping $f : \cC \to \Nat$ such that every satisfiable $\varphi$ in $\cC$ has a model of size at most $f(\varphi)$. Then there exists an effective translation $T$ from $\cC$ into SF such that for every $\varphi \in \cC$
			(a) every model of $T(\varphi)$ is also a model of $\varphi$, 
			(b) every model of $\varphi$ whose size is at most $f(\varphi)$ can be extended to a model of $T(\varphi)$ over the same domain, and
			(c) the length of $T(\varphi)$ lies in $\cO\bigl( \len(\varphi) \cdot \log f(\varphi) \cdot \log \log f(\varphi) \bigr)$.
\end{proposition}	
\begin{proof}
	We outline the translation $T$ for some given input sentence $\varphi$, which we assume to be in negation normal form (without loss of generality). 
	Let $m := \lceil \log_2 f(\varphi) \rceil$ and let $Q_1, \ldots, Q_m$ be unary predicate symbols that do not occur in $\varphi$.
	For all terms $s,t$ we define $s \happrox t$ as abbreviation of $\bigwedge_{i = 1}^m Q_i(s) \leftrightarrow Q_i(t)$.
	In order to restrict the domain to $2^m$ elements, we conjoin the formula
		$\AxFin := \forall x y.\, x \!\happrox\! y \rightarrow x \!\approx\! y$.
	Since in any structure $\cA$ there are at most $2^m$ domain elements that can be distinguished by their membership in the sets $Q_1^\cA, \ldots, Q_m^\cA$, it is clear that $\cA \models \AxFin$ entails $|\fU_\cA| \leq 2^m$.
	Moreover, we observe the following property.
	\begin{itemize}
		\item[($*$)] 	Let $\cA$ be any structure, let $\beta$ be any variable assignment over $\cA$'s domain, and let $s, t$ be two terms. 
			If $\cA \models \AxFin$ holds, then we get $\cA, \beta \models s \happrox t$ if and only if $\cA, \beta \models s \approx t$.
	\end{itemize}	
	This means, if we restrict our attention to domains with at most $2^m$ domain elements, we can now use a separated form of equality.
	
	\begin{itemize}
		\item [($**$)] Let $\psi$ be any first-order formula and let $v$ be some variable that does not occur in $\psi$. Then $\psi$ is equivalent to $\forall v.\, u \approx v$ $\rightarrow\; \psi\subst{u/v}$. 
	\end{itemize}
	We can transform $\varphi$ into an equivalent sentence $\varphi'$ by consecutively replacing each subformula of the form $\exists y. \psi$ in $\varphi$ with $\exists y \forall v.\; y \!\approx\! v \rightarrow \psi\subst{y/v}$, where we assume $v$ to be fresh (one fresh variable for every replaced subformula).
	Consequently, every atom in $\varphi'$ that is not an equation contains exclusively universally quantified variables.
	Moreover, ($**$) implies that $\varphi$ and $\varphi'$ are equivalent.
	
	Let $\varphi''$ be the result of replacing all equations $y \approx v, v \approx y$ in $\varphi'$ in which $y$ is existentially quantified and $v$ universally quantified with the formula $y \happrox v$.
	We then set $\varphi_\SF := \AxFin \wedge \varphi''$.
	By ($*$), any model of $\varphi_\SF$ is also a model of $\varphi$.
	Conversely, any model $\cA$ of $\varphi$ that has at most $2^m$ domain elements can be converted into a model $\cB$ of $\varphi_\SF$ by defining the relations $Q_1^\cB, \ldots, Q_m^\cB$ in an appropriate way.
\end{proof}
	 The $\varphi_\SF$ in the above proof belongs to a subfragment of SF that we call \emph{strongly separated} (cf.\ Definition~\ref{definition:StronglySeparatedFragment}) and whose satisfiability problem is complete for \NEXPTIME\ (cf.\ Corollary~\ref{corollary:ComplexitySSF}).

Unfortunately, the translation methodology of Proposition~\ref{proposition:TranslationFMPintoSF} does not help in the quest for new decidable first-order fragments. 
The reason is simply that we already need arguments leading to a small model property before we can start the translation process, as we need information about the size of the models that have to be considered. 
Nevertheless, such translations can be useful in view of the expressiveness of SF that other first-order fragments, such as FO$_2$, the fluted fragment, and GF, lack.
For instance, SF sentences can naturally express the axioms of equivalence, most prominently, transitivity.
Hence, fundamental and interesting properties of predicates that have to be assumed at the meta-level when dealing with less expressive logics can be formalized directly in SF.
Moreover, basic counting quantifiers can be defined natively in SF and do not have to be introduced via special operators.
More precisely, given any formula $\exists^{\geq n} y.\, \varphi$ with positive $n$ and without non-constant function symbols, its standard translation 
	$\exists y_1 \ldots y_n. \bigwedge_{i=1}^n \varphi\subst{y/y_i} \wedge \bigwedge_{i < j} y_i \!\not\approx\! y_j$ 
is not in conflict with the separateness conditions of SF's definition, if the set $\{y_1, \ldots, y_n\}$ is separated in $\varphi$ from the set of universally quantified variables. 
\begin{proposition}\label{proposition:CountingQuantifiers}
	Counting quantifiers $\exists^{\geq n}$ with positive integer $n$ are expressible in SF.
\end{proposition}


\subsection{Basic complexity considerations}\label{section:ComplexityBasics}
We first recall the well-known small model properties of SF's subfragments BSR and MFO (see \cite{Borger1997} for references).
\begin{proposition}\label{proposition:SmallModelsBSR}
	Let $\varphi := \exists \vz\, \forall \vx. \psi$ be a satisfiable BSR sentence, i.e.\ $\psi$ is quantifier free and does not contain non-constant function symbols.
	There is a model $\cA \models \varphi$ such that $|\fU_\cA| \leq \max \bigl( |\vz| + |\consts(\varphi)|, 1 \bigr)$.
\end{proposition}
We make use of this property when we derive an upper bound on the size of small models for satisfiable SF sentences, as our approach will be based on an effective translation of SF sentences into equivalent BSR sentences.

\begin{proposition}\label{proposition:SmallModelsMFO}
	Let $\varphi := \exists \vz\, \forall\vx_1 \exists\vy_1 \ldots$ $\forall\vx_n \exists\vy_n. \psi$ be a satisfiable monadic sentence without equality and without non-constant function symbols, i.e.\ all predicate symbols in $\varphi$ are of arity $1$.
	Moreover, assume that $\varphi$ contains $k$ distinct predicate symbols.
	There is a model $\cA \models \varphi$ such that $|\fU_\cA| \leq 2^{k}$.
\end{proposition}
Notice that the shape of the quantifier prefix does not contribute to the upper bound.

The following lemma links bounds on the size of models with the computing time that is required to decide satisfiability.
\begin{lemma}[cf.\ \cite{Borger1997}, Proposition~6.0.4]
	\label{lemma:ComplexityWithSmallModelProperty}
	Let $\varphi$ be a first-order sentence in prenex normal form containing $n$ universally quantified variables.
	The question whether $\varphi$ has a model of cardinality $m$ can be decided nondeterministically in time $p\bigl(m^n \cdot \len(\varphi)\bigr)$ for some polynomial $p$.
\end{lemma}
With this lemma at hand, it is enough to prove a small model property for a given class of first-order sentences, in order to bound the worst-case time complexity of the corresponding satisfiability problem from above. This is exactly what the authors of \cite{Voigt2016} have done for SF.

\begin{proposition}[\cite{Voigt2016}, Theorem~17]\label{theorem:SFComplexityOld}
	Let $\varphi := \exists\vz\, \forall \vx_1 \exists \vy_1 \ldots \forall \vx_n \exists \vy_n. \psi$ be an SF sentence for some quantifier-free $\psi$.
	There is some equivalent BSR sentence $\exists \vu\, \forall \vv. \psi'$ in which the number of occurring constant symbols plus the number of existential quantifiers is at most $\len(\varphi) + n \cdot \len(\varphi) \cdot \bigl( \twoup{n}{\len(\varphi)} \bigr)^n$. 
	As a result, satisfiability of $\varphi$ can be decided nondeterministically in time that is at most $n$-fold exponential in $\len(\varphi)$.
\end{proposition}
Clearly, applying this result to an MFO sentence substantially overshoots the actual worst-case time requirements.
To stress it again, the notion of the degree of interaction is a remedy to this sort of inaccuracies, as we shall see in Section~\ref{section:TranslationSFintoBSR}.

A special case that is worth considering, before we investigate the complexity of full SF, is the class of SF sentences that do not contain universal quantifiers.
This species of formulas coincides with the purely existential fragment of first-order logic without non-constant function symbols, and it is a close relative of propositional logic.
Recall that SAT is NP-complete \cite{Cook1971}, Horn-SAT is P-complete \cite{Kasif1986, Plaisted1984}, and 2SAT is NL-complete \cite{Jones1976}.
\begin{proposition}\label{proposition:ComplexityExistentialSF}~
	\begin{enumerate}[label=(\roman{*}), ref=(\roman{*})]
		\item\label{enum:ComplexityExistentialSF:I} Satisfiability for the class of SF sentences without universal quantifiers is NP-complete.
		\item\label{enum:ComplexityExistentialSF:II} Satisfiability for the class of SF-Horn sentences without universal quantifiers is P-complete.
		\item\label{enum:ComplexityExistentialSF:III} Satisfiability for the class of SF-Krom sentences without universal quantifiers and without equality is NL-complete.
	\end{enumerate}
\end{proposition}
\begin{proof}[Proof sketch]
	The proof of \ref{enum:ComplexityExistentialSF:I} -- \ref{enum:ComplexityExistentialSF:III} proceeds by reductions to the corresponding satisfiability problems for propositional logic and back.
	This is straightforwardly done by Skolemization as long as we consider only SF sentences without equality (cf.\ Lemma~\ref{lemma:ComplexitySAT} in the appendix).
	
	In the latter cases, we first Skolemize exhaustively, producing $\varphi_\gnd$, which is ground and contains only Skolem constants and no non-constant function symbols.
	Then we use the standard trick to eliminate the equality predicate $\approx$.
	We introduce a fresh binary relation symbols $E$ and replace
	every equation $c \approx d$ with an atom $E(c,d)$.
	Moreover, we add axioms for reflexivity, symmetry, transitivity, and congruence. 
	Of course, we do not use the universally quantified axioms but rather add their instances with respect to all the constant symbols that occur in $\varphi_\gnd$.
	To avoid an exponential blow-up in the case of the congruence axioms, we only add the instances that affect non-equational atoms which really occur in $\varphi_\gnd$.
	
	Let $\phi$ be the propositional formula that results from $\varphi_\gnd$ by replacing every ground atom $A$ with the propositional variable $p_A$. 
	We observe that $\len(\phi) \in \cO\bigl( \len(\varphi)^3 \bigr )$.
	Moreover, if $\varphi$ is a Horn formula, then $\phi$ is Horn.
	Notice that the outlined elimination of equality does not preserve the Krom property.
\end{proof}


\section{Translation of SF sentences into BSR sentences}\label{section:TranslationSFintoBSR}

In this section, we analyze the transformation process from SF into the BSR fragment from the perspective of the \emph{degree of interaction of existential variables}.
Our aim is to derive upper and lower bounds on the length of the resulting BSR-formulas. 
Roughly speaking, in the first phase of the translation process all quantifiers are moved inwards as far as possible (cf.\ the proof of Lemma~\ref{lemma:TranslationSFintoBSRunderDegree}).
In order to do so, we first transform the sentence in question into a formula in CNF. 
After that, we employ the well-known rules of miniscoping (cf.\ Lemma~\ref{lemma:BasicQuantifierEquivalences}), supplemented by the rule formulated in the following lemma.

\begin{lemma}\label{lemma:AdvancedMiniscoping:refined}
	Let $I$ and $K_i$, $i \in I$, be sets that are finite, nonempty, and pairwise disjoint. 
	The elements of these sets serve as indices.
	Let 
	\[ \varphi := \exists \vy. \bigwedge_{i \in I} \Bigl( \chi_i(\vz) \vee \bigvee_{k \in K_i} \eta_k(\vy, \vz) \Bigr) \]
	be some first-order formula where the $\chi_i$ and the $\eta_k$ denote arbitrary subformulas that we treat as indivisible units in what follows.
	We say that $f : I \to \bigcup_{i \in I} K_i$ is a \emph{selection function} 
	if for every $i \in I$ we have $f(i) \in K_i$.
	We denote the set of all selection functions of this form by $\cF$.
	
	Then $\varphi$ is equivalent to $\varphi' :=$
		\[ \bigwedge_{\text{\scriptsize $\begin{array}{c} S \subseteq I \\ S \neq \emptyset \end{array}$ \normalsize}} \Bigl( \bigvee_{i \in S} \chi_i(\vz) \Bigr) \vee \bigvee_{f \in \cF} \Bigl( \exists \vy. \bigwedge_{i \in S} \eta_{f(i)} (\vy, \vz) \Bigr) ~. \]
\end{lemma}
\begin{proof}[Proof sketch]
The proof of this lemma follows a conceptually simple strategy.
Using the distributivity of $\wedge$ over $\vee$, we first transform $\varphi$ into a disjunction of conjunctions of the indivisible units $\chi_i(\vz)$ and $\eta_k(\vy, \vz)$.
Then, exploiting the equivalences in Lemma~\ref{lemma:BasicQuantifierEquivalences}, we push the existential quantifier block $\exists \vy$ inwards such that it only binds conjunctions of units $\eta_k(\vy,\vz)$. This is possible, because none of the variables in $\vy$ occurs in any of the $\chi_i(\vz)$.
From this point on, we treat the newly emerged subformulas $\exists \vy. \bigwedge_{k'} \eta_{k'}(\vy,\vz)$ as if they were indivisible.
We then transform the formula back into a conjunction of disjunctions of indivisible units, this time using the distributivity of $\vee$ over $\wedge$.
It then remains to show that the result of this transformation exhibits a highly redundant structure and is actually equivalent to $\varphi'$.
\end{proof}

\subsection{Degree of interaction of existential variables and the size of small models}\label{section:DegreeOfInteraction}

Consider the formula $\varphi := \exists \vz\, \forall \vx_1 \exists \vy_1 \ldots$ $\forall \vx_n \exists \vy_n. \psi$ in standard form in which $\psi$ is quantifier free and in which the sets $\vx := \vx_1 \cup \ldots \cup \vx_n$ and $\vy := \vy_1 \cup \ldots \cup \vy_n$ are separated. In addition, we assume that $\vx_1$ and $\vy_1$ are nonempty. The tuple $\vz$, on the other hand, may be empty.

For any $j \in [n]$ and any variable $y \in \vy_j$ we say that \emph{$y$ is a level-$j$ variable}, denoted $\lvl(y) = j$.
For any nonempty set $Y \subseteq \vy$ of existentially quantified variables and any positive integer $k$ we say that \emph{$Y$ has degree $k$ in $\varphi$}, denoted $\degree_{Y,\varphi} = k$, if $k$ is the maximal number of distinct variables $y_1, \ldots, y_k \in Y$ that belong to different levels in $\varphi$, i.e.\ $\lvl(y_1) < \ldots < \lvl(y_k)$.
We say that \emph{$\varphi$'s degree of interaction of existential variables} (short: \emph{degree}) is $k$, denoted $\degree_\varphi = k$, if $k$ is the smallest positive integer such that we can partition $\vy$ into $m > 0$ parts $Y_1, \ldots, Y_m$ that are all pairwise separated in $\varphi$ and for which $k = \max\bigl\{ k_j \bigm| \degree_{Y_j, \varphi} = k_j, 1 \leq j \leq m \bigr\}$.
Sentences $\varphi := \exists \vz\, \forall \vx.\, \psi$ in standard form with quantifier-free $\psi$ are said to have \emph{degree} zero, i.e.\ $\degree_\varphi = 0$, if $\vx$ is empty and we define $\degree_\varphi = 1$ if $\vx$ is nonempty.

\begin{lemma}\label{lemma:TranslationSFintoBSRunderDegree}
	Let $\varphi := \exists \vz\, \forall \vx_1 \exists \vy_1 \ldots \forall \vx_n \exists \vy_n. \psi$ be an SF sentence of positive degree $\degree_\varphi$ in standard form.
	Let $\cL_\varphi (\vy)$ denote the set of all literals in $\varphi$ that contain at least one variable $y \in \vy := \vy_1 \cup \ldots \cup \vy_n$.
	There exists a sentence $\varphi_{\text{BSR}} = \exists \vz\, \exists \vu\, \forall \vv. \psi_{\text{BSR}}$ in standard form with quantifier-free $\psi_{\text{BSR}}$ that is equivalent to $\varphi$ and contains at most $|\vz| + |\vy| \cdot \degree_\varphi \cdot \bigl( \twoup{\degree_\varphi}{|\cL_\varphi (\vy)|} \bigr)^{\degree_\varphi}$ leading existential quantifiers.
\end{lemma}
\begin{proof}[Proof sketch]
	Let $\vx := \vx_1 \cup \ldots \cup \vx_n$.
	We transform $\varphi$ into CNF and then apply Lemma~\ref{lemma:AdvancedMiniscoping:refined} and the rules of miniscoping given in Lemma~\ref{lemma:BasicQuantifierEquivalences} to push all quantifier blocks inwards. 
	Since the sets $\vx$ and $\vy$ are separated in $\varphi$, these operations can be performed in such a way that in the resulting formula $\varphi'$ no universal quantifier lies within the scope of any existential quantifier (other than the ones in $\exists \vz$) and vice versa.
	After removing redundant parts from $\varphi'$, the depth of nestings of existential quantifier blocks (interspersed with conjunctive connectives in $\varphi'$'s syntax tree) can be upper bounded by $\degree_\varphi$.
	As a consequence, $\varphi'$ contains at most $\twoup{\degree_\varphi}{|\cL_\varphi(\vy)|}$ distinct subformulas that are of the form $\exists y. \psi'$ and do not lie within the scope of any quantifier.
	After further transformations, we obtain a formula $\varphi'' := \bigvee_{k} \bigl( \chi_k(\vx) \wedge \bigwedge_{r_k} \eta_{r_k}(\vy) \bigr)$ where the $r_k$ range over at most $\twoup{\degree_\varphi}{|\cL_\varphi(\vy)|}$ indices.
	Moreover, every constituent $\bigwedge_{r_k} \eta_{r_k}$ in $\varphi''$ contains at most $|\vy| \cdot \sum_{k'=1}^{\degree_\varphi} \prod_{d = k'}^{\degree_\varphi} \twoup{d}{|\cL_\varphi(\vy)|}$ occurrences of existential quantifiers.
	Since these existential quantifiers distribute over the topmost disjunction when we move them outwards to the front of the sentence $\varphi''$, and since the universal quantifiers in the $\chi_k$ may also be moved back outwards, one can show that $\varphi$ is equivalent to some BSR sentence with at most $|\vy| \cdot \degree_\varphi \cdot \bigl( \twoup{\degree_\varphi}{|\cL_\varphi(\vy)|} \bigl)^{\degree_\varphi}$ leading existential quantifiers.
\end{proof}

Proposition~\ref{proposition:SmallModelsBSR} now entails that any satisfiable SF-sentence $\varphi$ has a model of size at most 
	\begin{equation}\label{eqn:ModelSizeDependingOnDepth}
		\len(\varphi) + \len(\varphi) \cdot \degree_\varphi \cdot \bigl(\twoup{\degree_\varphi}{\len(\varphi)}\bigr)^{\degree_\varphi} ~.
	\end{equation}	
\begin{theorem}\label{theorem:SFComplexityDependingOnDepth}
	Let $k$ be any positive integer.
	The satisfiability problem for the class of SF sentences $\varphi$ in standard form with degree $\degree_\varphi \!\leq\! k$ can be decided in nondeterministic $k$-fold exponential time.	
\end{theorem}
Together with Proposition~\ref{proposition:ComplexityExistentialSF}\ref{enum:ComplexityExistentialSF:I}, this establishes the upper bounds depicted in Figure~\ref{figure:ComplexityOfSF}.

In cases where $\degree_\varphi = 1$, Expression~(\ref{eqn:ModelSizeDependingOnDepth}) simplifies to $\len(\varphi) + \len(\varphi) \cdot 2^{\len(\varphi)}$.
The syntactic class of sentences satisfying this property is called \emph{strongly separated} in \cite{Voigt2016}.
\begin{definition}[\cite{Voigt2016}]\label{definition:StronglySeparatedFragment}
	Let $\varphi := \forall \vx_1 \exists \vy_1 \ldots$ $\forall \vx_n \exists \vy_n. \psi$ be an SF sentence and assume that $\psi$ is quantifier free.
	We say that $\varphi$ belongs to the \emph{strongly separated fragment (SSF)} if and only if the sets $\vx := \vx_1 \cup \ldots \cup \vx_n$ and $\vy_1, \ldots, \vy_n$ are all pairwise separated in $\varphi$.	
\end{definition}
Since MFO and BSR sentences fall into this syntactic category, and since their decision problem is known to be \NEXPTIME-hard, we obtain the following corollary.

\begin{corollary}\label{corollary:ComplexitySSF}
	The satisfiability problem for SSF is \NEXPTIME-complete.
\end{corollary}

Notice that the presented method can explain the asymptotic complexity of MFO-satisfiability and yields a reasonable upper bound on the size of small models of satisfiable MFO sentences.
This works in spite of the fact that monadic sentences may contain arbitrarily nested alternating quantifiers.
This is a considerable improvement compared to the methods used in~\cite{Voigt2016}.

Let $\varphi$ be any SF sentence with the maximally possible degree $\degree_\varphi = n$, where $n$ is the number of occurring $\forall\exists$-alternations. Then the upper bound shown in Expression~(\ref{eqn:ModelSizeDependingOnDepth}) regarding the number of elements in small models fits the corresponding result entailed by Proposition~\ref{theorem:SFComplexityOld}. 
As one consequence, Theorem~\ref{theorem:SFComplexityDependingOnDepth} in the present paper subsumes Theorem~17 in~\cite{Voigt2016}.
Moreover, Corollary~\ref{corollary:ComplexitySSF} improves the double exponential upper bound on SSF-satisfiability given in Theorem~15 in~\cite{Voigt2016}.
Finally, it is worth noticing that all SF sentences with the quantifier prefix $\exists^* \forall^* \exists^* \forall^*$ belong to the strongly separated fragment. 
Hence, Corollary~\ref{corollary:ComplexitySSF} subsumes Theorem~14 in \cite{Voigt2016}. The latter stipulates \NEXPTIME-completeness of SF sentences with quantifier prefix $\exists^* \forall^* \exists^*$.
Clearly, the refined analysis based on the degree of interaction of existential variables, rather than the number of quantifier alternations, yields significantly tighter results in many cases.


\subsection{Lower bounds on the length of equivalent BSR formulas}\label{section:LowerBoundsSmallestBSRsentences}

Before we derive lower bounds on the time that is required to decide SF-satisfiability in the worst case, we establish lower bounds on the length of the results of the translation from SF into the BSR fragment.
\begin{theorem}\label{theorem:LengthSmallestBSRsentences}
	There is a class of SF sentences that are Horn and Krom such that for every positive integer $n$ the class contains a sentence $\varphi$ of degree $\degree_\varphi = n$ and with a length linear in $n$ for which any equivalent BSR sentence contains at least $\sum_{k=1}^{n}\twoup{k}{n}$ leading existential quantifiers.
\end{theorem}
\begin{proof}[Proof sketch]
	Recall that $[n]$ abbreviates the set $\{1, \ldots, n\}$ and that $\fP S$ denotes the power set of a given set $S$. 
	Let $n \geq 1$ be some positive integer.
	Consider the following first-order sentence in which the sets $\{x_1, \ldots, x_n\}$ and $\{y_1, \ldots, y_n\}$ are separated:
		\[ \varphi := \forall x_n \exists y_n \ldots \forall x_1 \exists y_1. \bigwedge_{i=1}^{4n} \bigl( P_i(x_1, \ldots, x_n) \leftrightarrow Q_i(y_1, \ldots, y_n) \bigr) ~.\]
	Notice that we change the orientation of the indices in the quantifier prefix in this proof.
	
	In order to construct a particular model of $\varphi$, we inductively define the following sets:
		$\cS_1 := \bigl\{ S \subseteq [4n] \bigm| |S| = 2n \bigr\}$, $\cS_{k+1} := \bigl\{ S \in \fP \cS_k \bigm| |S| = \tfrac{1}{2} \cdot |\cS_k| \bigr\}$ for every $k$, $1 < k \leq n$.
	Hence, we observe\\
		$|\cS_1| = {{4n} \choose {2n}} \geq \bigl( \frac{4n}{2n} \bigr)^{2n} = 2^{2n}$,\\
		$|\cS_2| = {{|\cS_1|} \choose {|\cS_1|/2}} \geq \bigl( \frac{|\cS_1|}{|\cS_1|/2} \bigr)^{|\cS_1|/2} \geq 2^{2^{2n-1}}$,\\
		$\strut\qquad\vdots$\\
		$|\cS_n| = {{|\cS_{n-1}|} \choose {|\cS_{n-1}|/2}} \geq 2^{2^{2^{\vdots^{2^{2n-1}-1}}-1}} \geq \twoup{n}{n+1}$.\\
	We now define the structure $\cA$ as follows:	
		\begin{itemize}
			\item $\fU_\cA := \bigcup_{k = 1}^{n} \bigl\{ \fa^{(k)}_{S}, \fb^{(k)}_{S} \bigm| S \in \cS_k \bigr\}$, 
			\item $P_i^\cA := \bigl\{ \<\fa^{(1)}_{S_1}, \ldots, \fa^{(n)}_{S_n}\> \in \fU_\cA^n \bigm| i \in S_1 \in S_2 \in \ldots \in S_n \bigr\}$ for $i = 1, \ldots, 4n$, and
			\item $Q_i^\cA := \bigl\{ \<\fb^{(1)}_{S_1}, \ldots, \fb^{(n)}_{S_n}\> \in \fU_\cA^n \bigm| i \in S_1 \in S_2 \in \ldots \in S_n \bigr\}$ for $i = 1, \ldots, 4n$.
		\end{itemize}		
	One can easily show that $\cA$ is a model of $\varphi$.
	Moreover, employing a game-theoretic argument, one can show the following property:
	\begin{itemize}
		\item[($*$)] the substructure induced by $\cA$'s domain after removing at least one of the $\fb^{(k)}_{S}$ does not satisfy $\varphi$.
	\end{itemize}
	
	We know that $\fU_\cA$ contains at least
		$\sum_{k=1}^{n}\twoup{k}{n}$
	elements of the form $\fb^{(k)}_{S}$.
		
	Using ($*$) and the substructure lemma, one can argue that any BSR sentence $\varphi_*$ that is semantically equivalent to $\varphi$ must contain at least $\sum_{k=1}^{n}\twoup{k}{n}$ leading existential quantifiers.
	
	The key idea is that $\varphi_*$, which is satisfied by $\cA$, must contain one existential quantifier for each and every $\fb^{(k)}_{S}$.
	Otherwise, there would be one $\fb^{(k)}_{S}$, call it $\fb_*$, such that we could remove $\fb_*$ from $\cA$'s domain and any tuple $\< \ldots, \fb_*, \ldots \>$ from the sets $Q_i^\cA$, and the resulting structure would then still be a model of $\varphi_*$.
	But this would contradict ($*$).
		
	Since every atom $Q_i(y_1, \ldots, y_n)$ contains $n$ variables from existential quantifier blocks that are separated by universal ones, the degree $\degree_\varphi$ of $\varphi$ is $n$. 
	Moreover, $\varphi$ can easily be transformed into a CNF that is Horn and Krom at the same time.
	
	Hence, the theorem holds.
\end{proof}

Theorem~\ref{theorem:LengthSmallestBSRsentences} entails that there is no elementary upper bound on the length of the BSR sentences that result from an equivalence-preserving transformation of SF sentences into BSR. On the other hand, by Lemma~\ref{lemma:TranslationSFintoBSRunderDegree}, there is an elementary upper bound, if we only consider SF sentences up to a certain degree.


\section{Lower bounds on the computational complexity of SF-satisfiability}\label{section:ComputationalLowerBounds}

In this section we establish lower bounds on the worst-case time complexity of SF-satisfiability.
Our arguments will be based on a particular form of bounded domino (or tiling) problems developed by Gr\"adel (see \cite{Gradel1990b} and \cite{Borger1997}, Section~6.1.1).
By $\Int_t$ we denote the set of integers $\{0, \ldots, t-1\}$ for any positive $t \geq 1$.
\begin{definition}[\cite{Borger1997}, Definition 6.1.1]
	A \emph{domino system} $\fD := \<\cD, \cH, \cV\>$ is a triple where $\cD$ is a finite set of tiles and $\cH, \cV \subseteq \cD \times \cD$ are binary relations determining the allowed horizontal and vertical neighbors of tiles, respectively. Consider the torus $\Int_t^2 := \Int_t \times \Int_t$ and let $\bD := D_0 \ldots D_{n-1}$ be a word over $\cD$ of length $n \leq t$. The letters of $\bD$ represent tiles.
	We say that $\fD$ \emph{tiles the torus $\Int_t^2$ with initial condition $\bD$} if and only if there exists a mapping $\tau : \Int_t^2 \to \cD$ such that for every $\<x,y\> \in \Int_t^2$ the following conditions hold, where ``$+1$'' denotes increment modulo $t$.
	\begin{enumerate}[label=(\alph{*}), ref=(\alph{*})]
		\item If $\tau(x,y) = D$ and $\tau(x+1, y) = D'$, then $\<D,D'\> \in \cH$.
		\item If $\tau(x,y) = D$ and $\tau(x, y+1) = D'$, then $\<D,D'\> \in \cV$.
		\item $\tau(i,0) = D_{i}$ for $i = 0, \ldots, n-1$.
	\end{enumerate}
\end{definition}

\begin{definition}[\cite{Borger1997}, Definition 6.1.5]
	Let $T : \Nat \to \Nat$ be a function and let $\fD := \<\cD, \cH, \cV\>$ be a domino system.
	The problem $\Domino(\fD, T(n))$ is the set of those words $\bD$ over the alphabet $\cD$ for which $\fD$ tiles $\Int_{T(|\bD|)}^2$ with initial condition $\bD$.
\end{definition}

Domino problems provide a convenient way of deriving lower bounds via reductions. 
Suppose we are given some well-behaved time bound $T(n)$ that grows sufficiently fast.
Further assume there is a reasonable translation from $\Domino(\fD, T(n))$ into some problem $\cL$ where the length of the results is upper bounded by a function $g(n)$.
It follows that the time required to solve the hardest instances of $\cL$ lies in $\Omega \bigl( T(h(n)) \bigr)$, where $h(n)$ may be conceived as an inverse of $g(n)$ from an asymptotic point of view.
The next proposition formalizes this observation.
\begin{proposition}[\cite{Borger1997}, Theorem 6.1.8]\label{theorem:ComputationalLowerBoundViaPolynomialReduction}
	Let $T: \Nat\!\to\!\Nat$ be a time-constructible function with $T(c' n)^2 \in o(T(n))$ for some constant $c' > 0$ and let $\cL$ be a problem such that for every domino system $\fD$ we have $\Domino(\fD,T(n)) \leq_{g(n)} \cL$, i.e.\ $\Domino(\fD,T(n))$ is \emph{polynomially reducible} to $\cL$ via length order g(n) (cf.\ Definition 6.1.7 in \cite{Borger1997}).
	Moreover, let $h: \Nat \to \Nat$ be a function such that $h(d \cdot g(n)) \in \cO(n)$ for any positive $d$.
	There exists a positive constant $c > 0$ such that $\cL \not\in \Ntime(T(c \cdot h(n)))$.
\end{proposition}

Subsections~\ref{section:EnforcingLargeModels} and \ref{section:FormalizingTheTiling} are devoted to the purpose of outlining the following reductions.
\begin{lemma}~
	\begin{enumerate}[label=(\roman{*})]
		\item
			Fix some positive integer $k>0$ and let $\fD$ be an arbitrary domino system. 
			Let $\SatSFdk$ be the set containing all satisfiable SF sentences whose degree $\degree$ is at most $k$.
			We have $\Domino\bigl( \fD, \twoup{k}{n} \bigr) \leq_{n \cdot \log n} \SatSFdk$.
		
		\item 
			Fix some positive integer $m > 1$ and let $\fD$ be an arbitrary domino system. 
			Let $\SatSF$ be the set containing all satisfiable SF sentences.
			We have $\Domino(\fD, \twoup{n}{m}) \leq_{n^2 \cdot \log n} \SatSF$.
	\end{enumerate}	
\end{lemma}
Having these reduction results at hand, Proposition~\ref{theorem:ComputationalLowerBoundViaPolynomialReduction} implies the sought lower bounds on SF-satisfiability for classes of sentences with bounded degree and the class of unbounded SF sentences.
\begin{theorem}\label{theorem:ComputationalLowerBoundForSF}
	There are positive constants $c, d > 0$ for which
		\[ \SatSFdk \not\in \Ntime \bigl( \twoup{k}{c n / \log n} \bigr) \]
	and
		\[ \SatSF \not\in \Ntime \bigl( \twoup{d \cdot \sqrt{n / \log n}}{2} \bigr) ~. \]
\end{theorem}
These lower bounds also hold if we do not allow equality in SF, see Section~\ref{section:ReplaceEqualityInLowerBoundProof}.
The remainder of Section~\ref{section:ComputationalLowerBounds} is devoted to the formalization of sufficiently large tori in SF and to the translation from a given domino system $\fD = \<\cD, \cH, \cV\>$ (for nonempty $\cD, \cH, \cV$) plus an initial condition $\bD$ into an SF sentence $\varphi$ such that $\varphi$ is satisfiable if and only if $\bD \in \Domino(\fD, T_i(|\bD|))$ with $T_1(n) = \twoup{\kappa}{n}$ for any given $\kappa > 0$ and $T_2(n) = \twoup{n}{\mu}$ for any given $\mu>1$.

\subsection{Enforcing a large domain}\label{section:EnforcingLargeModels}

The following description gives a somewhat simplified view. Technical details will follow.

A crucial part in the reduction is that a grid of size $t \times t$ has to be encoded, where $t$ defines the required computing time and we assume $t := \twoup{\kappa}{\mu}$ for positive integers $\kappa$ and $\mu > 1$ that we consider as parameters of the construction.

Every point $p$ on the grid is represented by a pair $p = \<x,y\>$, where each of the coordinates $x$ and $y$ is represented by a bit string of length $\log \bigl(\twoup{\kappa}{\mu}\bigr) = \twoup{\kappa-1}{\mu}$.
Given a bit string $\overline{b}$, we represent the $j$-th bit $b_j$ by the truth value of the atom $J(\ul{\kappa}, \overline{b}, j)$, where $\ul{\kappa}$ is the constant used to address the topmost level of a hierarchy of $\kappa+1$ sets of indices.
The crux of our approach is that we have to enforce the existence of sufficiently many indices $j$, namely $\twoup{\kappa-1}{\mu}$ many, to address the single bits of $\overline{b}$.
Again, we address each of these indices as a bit string, this time of length $\twoup{\kappa-2}{\mu}$.

Thus, we proceed in an inductive fashion, building up a hierarchy of indices with $\kappa+1$ levels.
The lowest level, level zero, is inhabited by $\mu$ indices, which we represent as constants with distinct values.
For every $\ell \geq 1$ any index $j$ on the $\ell$-th level is represented by a bit string consisting of $\twoup{\ell-1}{\mu}$ bits, i.e.\ the $\ell$-th level of the index hierarchy contains $\twoup{\ell}{\mu}$ indices. The $i$-th bit of an $\ell$-th-level index $j$ corresponds to the truth value of the atom $J(\ul{\ell}, j,i)$.

\begin{example}
	Assume $\mu = 2$ and $\kappa = 3$.\\
	\centerline{
		\begin{tabular}{c|c|c}
			index	& 	set of 										& 	number\\
			 level	&	indices									&	of indices	\\
			\hline
			0		&	$\{\fc_1, \fc_2\} \hfill$						&	2			\\
			1		&	$\{00,01,10,11\} \hfill$							&	4			\\
			2		&	$\{0000, 0001, \ldots, 1111\} \hfill$			&	16			\\
			3		&	$\{0, 1\}^{16}$								&	$65536$		
		\end{tabular}
	}	
	On every index level, the bits of one index are indexed by the indices from the previous level. We illustrate this for the word $1010$ on all levels from 2 down to 0. The bits of 1010 on level two are indexed by bit strings from level one, each of them having a length of two. The bits of the indices of level one are themselves indexed by objects of level zero which are some values $\fc_1, \fc_2$ assigned to the constants $c_1, c_2$. To improve readability, we connect the bits of words by dashes.\\[0.5ex]
	
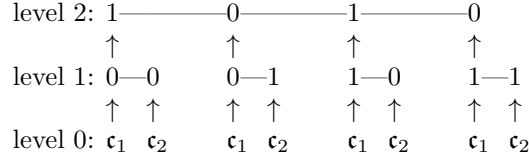
\centerline{
		\begin{tabular}{l@{\hspace{1ex}}l@{\hspace{-0.75ex}}c@{}l@{}c@{}l@{\hspace{-0.75ex}}c@{}l@{}c@{}l@{\hspace{-0.75ex}}c@{}l@{}c@{}l@{\hspace{-0.75ex}}c@{}l}
			level 2:	&	1		&|&	||		&|& 	0 		&|&	||		&|&	1		&|&	||		&|&	0			\\
					&	$\uparrow$	&&			&&	$\uparrow$	&&			&&	$\uparrow$	&&			&&	$\uparrow$		\\
			level 1:	&	0		&|&	0		&&	0		&|&	1		&&	1		&|&	0		&&	1		&|&	1	\\	
					&	$\uparrow$	&&	$\uparrow$	&&	$\uparrow$	&&	$\uparrow$	&&	$\uparrow$	&&	$\uparrow$	&&	$\uparrow$	&&	$\uparrow$	\\
			level 0:	&	$\fc_1$	&&	$\fc_2$	&&	$\fc_1$	&&	$\fc_2$	&&	$\fc_1$	&&	$\fc_2$	&&	$\fc_1$	&&	$\fc_2$	
		\end{tabular}
	}	
\end{example}
For technical reasons the number of indices per level grows slightly slower than described above (cf.\ Lemma~\ref{lemma:IndexSetsProperties}).
The described index hierarchies can be encoded by SF formulas with the quantifier prefix $\exists^*(\forall\exists)^{\kappa}$ that have a length that is polynomial in $\kappa$ and $\mu$.
%
We use the following constant and predicate symbols with the indicated meaning:\\
\centerline{
	\begin{tabular}{ll}
		$\ul{0}, \ul{1}, \ldots, \ul{\kappa}$	&	constants denoting the levels from $0$ to $\kappa$, \\
		$c_1, \ldots, c_\mu$			&	denote the indices at level 0, \\
		$d_1, \ldots, d_\kappa$			&	$d_\ell$ is the min.\ index at level $\ell$, \\
		$e_1, \ldots, e_\kappa$			&	$e_\ell$ is the max.\ index at level $\ell$, \\
		$L(\ul{\ell}, j)$				&	index $j$ belongs to level $\ell$, \\
		$\MinIdx(\ul{\ell}, j)$			& 	$j$ is a min.\ index at level $\ell$, \\
		$\MaxIdx(\ul{\ell}, j)$			& 	$j$ is a max.\ index at level $\ell$, \\
		$J(\ul{\ell}, j, i, b)$				&	the $i$-th bit of the index $j$ at level $\ell$ is $b$, \\
		$J^*(\ul{\ell}, j, i, b)$			&	$b=1$ indicates that all the bits of the index $j$ that \\
								&	are less significant than $j$'s $i$-th bit are 1, \\ 
		$\Succ(\ul{\ell}, j, j')$			&	$j'$ is the successor index of $j$ at level $\ell$.
	\end{tabular}
}
On every level we establish an ordering over the indices of that level.
We use the usual ordering on natural numbers encoded in binary.
Moreover, we formalize the usual successor relation on these numbers by the predicate $\Succ$.

One difficulty that we encounter is that we cannot assert the existence of successors simply by adding $\forall j \exists j'.\, \Succ(\ul{\ell}, j, j')$, as $j$ and $j'$ would not be separated.
Therefore, we fall back on a trick: we start from the equivalent formula $\forall j \exists \tj j'.\, j \!\approx\! \tj \wedge \Succ(\ul{\ell}, \tj, j')$, and replace the atom $j \!\approx\! \tj$ by a subformula $\eq{\ell}_{j,\tj}$ in which $j$ and $\tj$ are separated and which expresses a certain similarity between $j$ and $\tj$ instead of identity.
However, we specify the hierarchy of indices in a sufficiently strong way such that the similarity expressed by $\eq{\ell}_{j,\tj}$ actually coincides with identity.

Next, we formalize the described index hierarchies in $\SF_{\degree \leq \kappa}$.
Every formula is accompanied by a brief description of its purpose.
We shall try to use as few non-Horn sentences as possible.

\begin{align*}
	\psi_1 \!:=
		& \bigwedge_{\ell = 0}^{\kappa} \bigwedge_{\text{\scriptsize $\begin{array}{c} \ell' = 0 \\ \ell' \neq \ell \end{array}$ \normalsize}}^{\kappa} \forall j.\; L(\ul{\ell}, j) \rightarrow  \neg L(\ul{\ell}', j) \\
	&\hspace{-6ex}\text{Every index belongs to at most one level.}\\
	\psi_2 \!:=
		& \bigwedge_{\ell = 0}^{\kappa} \bigl( \forall j.\; \MinIdx(\ul{\ell}, j) \rightarrow L(\ul{\ell}, j) \bigr) \;\;\wedge\;\; \bigl( \forall j j'.\; \MinIdx(\ul{\ell}, j) \rightarrow \neg \Succ(\ul{\ell}, j', j) \bigr) \\
	&\hspace{-6ex}\text{A min.\ index of level $\ell$ belongs to level $\ell$. A min.\ index does not have a predecessor.} \\
	\psi_3 \!:=
		& \bigwedge_{\ell = 0}^{\kappa} \MinIdx(\ul{\ell}, d_\ell) \;\;\wedge\;\; \bigl( \forall j.\; \MinIdx(\ul{\ell}, j) \rightarrow j \approx d_\ell \bigr) \\
	&\hspace{-6ex}\text{There is a unique min.\ index on every level.}\\
	\psi_4 \!:=
		& \bigwedge_{\ell = 0}^{\kappa} \bigl( \forall j.\; \MaxIdx(\ul{\ell}, j) \rightarrow L(\ul{\ell}, j) \bigr) \;\;\wedge\;\; \bigl( \forall j j'.\; \MaxIdx(\ul{\ell}, j) \rightarrow \neg \Succ(\ul{\ell}, j, j') \bigr) \\
	&\hspace{-6ex}\text{A max.\ index of level $\ell$ belongs to level $\ell$. A max.\ index does not have a successor.}
\end{align*}
\begin{align*}		
	\psi_5 \!:=
		& \bigwedge_{\ell = 0}^{\kappa} \MaxIdx(\ul{\ell}, e_\ell) \;\;\wedge\;\; \bigl( \forall j.\; \MaxIdx(\ul{\ell}, j) \rightarrow j \approx e_\ell \bigr) \\
	&\hspace{-6ex}\text{There is a unique max.\ index on every level.} \\
	\psi_6 \!:=
		& \bigwedge_{\ell = 0}^{\kappa} \forall j j'.\; \Succ(\ul{\ell}, j, j') \;\;\rightarrow\;\; L(\ul{\ell}, j) \wedge L(\ul{\ell}, j') \\
	&\hspace{-6ex}\text{If $j'$ is the successor of $j$ at level $\ell$, then both $j$ and $j'$ belong to level $\ell$.}\\
	\psi_7 \!:=
		&\bigwedge_{\ell = 0}^{\kappa} \forall j j' j''.\; \neg \Succ(\ul{\ell}, j, j) \;\;\wedge\;\; \bigl( \Succ(\ul{\ell}, j, j') \wedge \Succ(\ul{\ell}, j, j'') \rightarrow j' \approx j'' \bigr) \\
		&\hspace{26ex}\wedge\;\; \bigl( \Succ(\ul{\ell}, j', j) \wedge \Succ(\ul{\ell}, j'', j) \rightarrow j' \approx j'' \bigr) \\
	&\hspace{-6ex}\text{The successor relation is irreflexive. Every index $j$ has at most one successor and at most}\\[-1ex]
	&\hspace{-6ex}\text{one predecessor.} \\
	\psi_8 \!:=\;
		& \MinIdx(\ul{0}, c_1) \wedge \MaxIdx(\ul{0}, c_\mu) \;\;\wedge\;\; \bigwedge_{i=1}^{\mu-1} \Succ(\ul{0}, c_i, c_{i+1}) \\
	&\hspace{-6ex}\text{At level zero we have the sequence $c_1, \ldots, c_\mu$ of successors, where $c_1$ is min.\ and $c_\mu$ max.} \\
	\psi_9 \!:=\;
		&\bigwedge_{\ell = 1}^{\kappa} \forall j j' i.\; \Succ(\ul{\ell}, j, j') \wedge L(\ul{\ell\!-\!1}, i) \;\;\rightarrow\;\; \Bigl( \bigl( J^*(\ul{\ell}, j, i, 1) \wedge J(\ul{\ell}, j, i, 1) \rightarrow J(\ul{\ell}, j', i, 0) \bigr) \\[-0.5ex]
		&\hspace{41ex}				\wedge \bigl( J^*(\ul{\ell}, j, i, 1) \wedge J(\ul{\ell}, j, i, 0) \rightarrow J(\ul{\ell}, j', i, 1) \bigr) \\
		&\hspace{41ex}				\wedge \bigl( J^*(\ul{\ell}, j, i, 0) \wedge J(\ul{\ell}, j, i, 1) \rightarrow J(\ul{\ell}, j', i, 1) \bigr) \\
		&\hspace{41ex}				\wedge \bigl( J^*(\ul{\ell}, j, i, 0) \wedge J(\ul{\ell}, j, i, 0) \rightarrow J(\ul{\ell}, j', i, 0) \bigr) \Bigr) \\
	&\hspace{-6ex}\text{Define what it means to be a successor at level $\ell$, $\ell > 0$, in terms of the binary increment}\\[-1ex]
	&\hspace{-6ex}\text{operation modulo $\twoup{\ell}{\mu}$.}\\
	\psi_{10} \!:=\;
		&\bigwedge_{\ell = 1}^{\kappa} \forall j i.\; \MinIdx(\ul{\ell}, j) \wedge L(\ul{\ell\!-\!1}, i) \;\;\rightarrow\;\; J(\ul{\ell}, j, i, 0) \\
	&\hspace{-6ex}\text{All bits of a minimal index $j$ are 0.}\\
	\psi_{11} \!:=
		& \bigwedge_{\ell = 1}^{\kappa} \forall j i.\; \MaxIdx(\ul{\ell}, j) \wedge \MaxIdx(\ul{\ell\!-\!1}, i) \;\;\rightarrow\;\; J(\ul{\ell}, j, i, 1) \\
	&\hspace{-6ex}\text{Define what it means to be max.\ (part 1): the most significant bit is 1.} \\
	\psi_{12} \!:=
		& \bigwedge_{\ell = 1}^{\kappa} \forall j i.\;  L(\ul{\ell}, j) \wedge \MaxIdx(\ul{\ell\!-\!1}, i) \wedge J(\ul{\ell}, j, i, 1) \;\;\rightarrow\;\; \MaxIdx(\ul{\ell}, j) \\
	&\hspace{-6ex}\text{Define what it means to be max.\ (part 2): any index with 1 as its most significant bit is max.}\\
	\psi_{13} \!:=
		& \bigwedge_{\ell = 1}^{\kappa} \forall j i.\; L(\ul{\ell}, j) \wedge  L(\ul{\ell\!-\!1}, i) \;\;\rightarrow\;\; \bigl( J(\ul{\ell}, j, i, 0) \rightarrow \neg J(\ul{\ell}, j, i, 1) \bigr) \\[-0.5ex]
		&\hspace{32ex}\wedge \bigl( J^*(\ul{\ell}, j, i, 0) \rightarrow \neg J^*(\ul{\ell}, j, i, 1) \bigr) \\
	&\hspace{-6ex}\text{No bit of an index is $0$ and $1$ at the same time. An analogous requirement is stipulated for $J^*$.} \\
	\psi_{14} \!:=
		& \bigwedge_{\ell = 1}^{\kappa} \forall j i.\; L(\ul{\ell}, j) \wedge \MinIdx(\ul{\ell\!-\!1}, i)  \;\;\rightarrow\;\; J^*(\ul{\ell}, j, i, 1) \\
	&\hspace{-6ex}\text{$J^*(\ul{\ell}, j, d_{\ell-1}, 1)$ holds for every index $j$.}
\end{align*}
\begin{align*}
	\psi_{15} \!:=
		&\bigwedge_{\ell = 1}^{\kappa} \forall j i i'.\; L(\ul{\ell}, j) \wedge \Succ(\ul{\ell\!-\!1}, i, i') \;\;\rightarrow\;\; \bigl( J^*(\ul{\ell}, j, i', 1) \leftrightarrow \bigl( J^*(\ul{\ell}, j, i, 1) \wedge J(\ul{\ell}, j, i, 1) \bigr) \bigr) \\[-0.5ex]
		 &\hspace{40ex}	\wedge \bigl( J(\ul{\ell}, j, i, 0) \rightarrow J^*(\ul{\ell}, j, i', 0) \bigr) \\
		 &\hspace{40ex} 	\wedge \bigl( J^*(\ul{\ell}, j, i, 0) \rightarrow J^*(\ul{\ell}, j, i', 0) \bigr) \\ 
	&\hspace{-6ex}\text{Define the semantics of $J^*$ as indicating that all bits strictly less significant than the $i$-th}\\[-1ex]
	&\hspace{-6ex}\text{bit are 1.} \\
	\eq{1}_{j,\tj} \!:=\;
		& L(\ul{1}, j) \wedge L(\ul{1}, \tj) \;\;\wedge\;\; \bigwedge_{i = 1}^\mu \bigl( J(\ul{1}, j, c_i, 0) \leftrightarrow J(\ul{1}, \tj, c_i, 0) \bigr) \wedge \bigl( J(\ul{1}, j, c_i, 1) \leftrightarrow J(\ul{1}, \tj, c_i, 1) \bigr) \\
	&\hspace{-6ex}\text{Base case of equality of indices.}\\
	\eq{\ell}_{j,\tj} \!:=\;
		&L(\ul{\ell}, j) \wedge L(\ul{\ell}, \tj) \wedge \forall i.\; L(\ul{\ell\!-\!1}, i) \;\;\rightarrow\;\; \exists \ti.\; L(\ul{\ell\!-\!1}, \ti) \wedge \eq{\ell\!-\!1}_{i,\ti}
				 \wedge \bigl( J(\ul{\ell}, j, i, 0) \leftrightarrow J(\ul{\ell}, \tj, \ti, 0) \bigr) \\[-0.5ex]
		&\hspace{57.4ex} \wedge \bigl( J(\ul{\ell}, j, i, 1) \leftrightarrow J(\ul{\ell}, \tj, \ti, 1) \bigr) \\
	&\hspace{-6ex}\text{Inductive case of equality of indices for $\ell > 1$.}\\
	\psi_{16} \!:=\;
		&\bigwedge_{\ell = 1}^{\kappa} \forall j i.\;  L(\ul{\ell}, j) \wedge \MaxIdx(\ul{\ell-1}, i) \wedge J(\ul{\ell}, j, i, 0) \;\;\rightarrow\;\; \exists \tj\, \tj'.\; \eq{\ell}_{j, \tj} \wedge \Succ(\ul{\ell}, \tj, \tj')\\
	&\hspace{-6ex}\text{For every index at level $\ell$ that is not maximal,  i.e. whose most significant bit is 0, there}\\[-1ex]
	&\hspace{-6ex}\text{ exists a successor index.}
\end{align*}

Until now, we have only introduced sentences that can easily be transformed into SF sentences in Horn form, existential variables are separated from universal ones, as all quantifiers occur with positive polarity, and consequents of implications are (conjunctions of) literals.

Regarding the length of the above sentences, we observe the following:
	\begin{itemize}
		\item $\len(\psi_1) \in \cO(\kappa^2 \log \kappa)$
		\item $\len(\psi_2), \ldots, \len(\psi_7), \len(\psi_9), \ldots, \len(\psi_{15}) \in \cO(\kappa \log \kappa)$
		\item $\len(\psi_8) \in \cO\bigl( \mu (\log \kappa + \log \mu) \bigr)$
		\item $\len\bigl( \eq{1}_{j,j'} \bigr) \in \cO\bigl( \mu (\log \kappa + \log \mu) \bigr)$
		\item $\len\bigl( \eq{\ell}_{j,j'} \bigr) \in \cO(\log \kappa) + \len\bigl( \eq{1}_{j,j'} \bigr)$
		\item $\len(\psi_{16}) \in \cO\bigl( \kappa^2 \log \kappa + \kappa \mu (\log \kappa + \log \mu) \bigr)$
	\end{itemize}
In total, this yields $\len(\psi_1 \wedge \ldots$ $\wedge \psi_{16}) \in \cO\bigl( \kappa^2 \log \kappa + \kappa \mu (\log \kappa + \log \mu) \bigr)$.

The following three sentences do not produce Horn formulas when transformed into CNF. 
They serve the purpose of removing spurious elements from the model.
In particular, $\chi_3$ is essential to enforce large models for $\kappa \geq 2$.
\begin{align*}
	\chi_1 \!:=\;
		& \forall j.\; L(\ul{0}, j) \rightarrow \bigvee_{i=1}^\mu j \approx c_i \\
		&\hspace{-6ex}\text{On level 0 there are no indices but $c_1, \ldots, c_\mu$.}\\
	\chi_2 \!:=
		& \bigwedge_{\ell = 1}^{\kappa} \forall j i.\; L(\ul{\ell}, j) \wedge L(\ul{\ell-1}, i) \;\;\rightarrow\;\; J(\ul{\ell}, j, i, 0) \vee J(\ul{\ell}, j, i, 1) \\
		&\hspace{-6ex}\text{We stipulate totality for the predicate $J$.}\\
	\chi_3 \!:=
		&\bigwedge_{\ell = 1}^{\kappa} \forall j j'.\; L(\ul{\ell}, j) \wedge L(\ul{\ell}, j') \rightarrow \exists \tj\, \tj'.\; \eq{\ell}_{j, \tj} \wedge \eq{\ell}_{j', \tj'} \\
		&\hspace{30ex}	\wedge \Bigl( \Bigl( \forall \ti.\; L(\ul{\ell-1}, \ti) \rightarrow \bigl( J(\ul{\ell}, \tj, \ti, 0) \leftrightarrow J(\ul{\ell}, \tj', \ti, 0) \bigr) \Bigr) \rightarrow j \approx j' \Bigr)  \\ 
		&\hspace{-6ex}\text{Two indices at the same level that agree on all of their bits are required to be identical.}	
\end{align*}
Notice that $\chi_3$ is (almost) an SF sentence, since the $\forall \ti$ turns into a $\exists \ti$ as soon as we bring the sentence into prenex normal form.
Regarding the length of $\chi_1, \chi_2, \chi_3$, we observe $\len(\chi_1) \in \cO(\log \kappa + \mu \log \mu)$, $\len(\chi_2) \in \cO(\kappa \log \kappa)$, and $\len(\chi_3) \in \cO\bigl( \kappa^2 \log \kappa + \kappa \mu (\log \kappa + \log \mu) \bigr)$.
Hence, we overall  have $\len(\chi_1 \wedge \chi_2 \wedge \chi_3) \in  \cO\bigl( \kappa^2 \log \kappa + \kappa \mu (\log \kappa + \log \mu) \bigr)$.

Consider any model $\cA$ of $\psi_{1} \wedge \ldots$ $\wedge \psi_{16}\wedge \chi_1 \wedge \chi_2 \wedge \chi_3$.
\begin{definition}\label{definition:IndexSetsAndRelatedNotation}
	We define the following sets and relations:
		$\cI_{\ell} := \bigl\{ \fa \in \fU_\cA \mid \cA, [j \Mapsto \fa] \models L(\ul{\ell}, j) \bigr\}$ for every $\ell = 0, \ldots, \kappa$;
		${\prec_\ell} \subseteq \cI_{\ell} \times \cI_{\ell}$ for every $\ell = 0, \ldots, \kappa$ such that $\fa \prec_\ell \fa'$ holds if and only if $\cA, [j \Mapsto \fa, j' \Mapsto \fa'] \models \Succ(\ul{\ell}, j, j')$.	
\end{definition}

\begin{lemma}\label{lemma:IndexSetsProperties}
	For every $\ell = 1, \ldots, \kappa$ we have $|\cI_{\ell}| = p$ where $p := 2^{|\cI_{\ell-1}|-1}+1 = \twoup{\ell}{\mu-1}+1$.
	Moreover, there is a unique chain $\fa_1 \prec_\ell \ldots \prec_\ell \fa_p$ comprising all elements in $\cI_\ell$.
\end{lemma}

Leaving out the non-Horn parts $\chi_1, \chi_2, \chi_3$ renders the lemma invalid for $\ell > 1$. 
On the other hand, for $\kappa = 1$ the sentence $\psi_1 \wedge \ldots \wedge \psi_{16}$---which can be transformed into an equivalent Horn sentence---has only models $\cA$ for which $\cI_1$ contains at least $2^{\mu-1}+1$ elements.
Notice that this could be used to derive \EXPTIME-hardness of satisfiability for the class of Horn SF sentences of degree $1$.
But such lower bounds are already known for the Horn subfragments of MFO and of the BSR fragment, which are proper subsets of SF's Horn subfragment.

\subsection{Formalizing a tiling of the torus}\label{section:FormalizingTheTiling}

In order to formalize a given domino problem $\fD = \<\cD, \cH, \cV\>$ and an initial condition $\bD$, we introduce the following constant and predicate symbols:\\
\centerline{
	\begin{tabular}{ll}
		$H(x, y, x', y')$			& 	$\<x',y'\>$ is the horiz.\ neighbor of $\<x,y\>$, \\
							&	i.e.\ $x' = x + 1$ (mod $\twoup{\kappa}{\mu-1}+1$) and $y' = y$, \\
		$V(x, y, x', y')$			& 	$\<x',y'\>$ is the vert.\ neighbor of $\<x,y\>$, \\
		$\ul{D}(x,y)$			&	$\<x,y\>$ is tiled with $D \in \cD$, \\
		$f_1, \ldots, f_{|\bD|}$		&	constants addressing points $\<0,0\>, \ldots, \<|\bD|-1,0\>$.
	\end{tabular}
}
With the ideas we have seen when formalizing the index hierarchy, it is now fairly simple to formalize the torus.
For instance, the following sentence makes sure that every point that is not on the ``edge'' of the torus has a horizontal neighbor.
\begin{align*}	
	\eta_{3} \!:=\;	
		& \forall x y i.\; L(\ul{\kappa}, x)\wedge L(\ul{\kappa}, y) \wedge \MaxIdx(\ul{\kappa-1}, i) \wedge J(\ul{\kappa}, x, i, 0) \\
		&\hspace{10ex} \rightarrow\;\; \exists \tx\, \ty\, \tx'.\; \eq{\kappa}_{x, \tx} \wedge \eq{\kappa}_{y, \ty} \wedge \Bigl( \bigwedge_{D \in \cD} \ul{D}(x,y) \leftrightarrow \ul{D}(\tx, \ty) \Bigr) \wedge H(\tx, \ty, \tx', \ty)
\end{align*}
The next sentence, on the other hand, makes sure that the rules of the domino system $\fD$ are obeyed.
\begin{align*}
		\eta_{15} \!:=\;
		& \forall x x' y.\; H(x, y, x', y) \;\;\rightarrow\;\; \bigvee_{\<D, D'\> \in \cH} \ul{D}(x,y) \wedge \ul{D}'(x',y)
\end{align*}
Proceeding this way, the formalization $\eta$ of a domino system in SF requires a length in $\cO \bigl( \widehat{n} \log \widehat{n} \bigr)$, where $\widehat{n} := \max \{ \kappa, \mu, |\bD|, |\cD|^2 \}$.

\begin{lemma}
	Assume that $\cD$, $\cH$, and $\cV$ are nonempty and
	let $\cA$ be a model of the sentence $\psi_{1} \wedge \ldots \wedge \psi_{16}\wedge \chi_1 \wedge \chi_2 \wedge \chi_3 \wedge \eta$.
	$\cA$ induces a tiling $\tau$ of $\Int_t^2$ with initial condition $\bD := D_1, \ldots, D_n$, where $t := \twoup{\kappa}{\mu-1}+1$.
\end{lemma}

\subsection{Replacing the equality predicate}\label{section:ReplaceEqualityInLowerBoundProof}

Since SF can express reflexivity, symmetry, transitivity, and congruence properties, it is easy to formulate an SF sentence without equality that is equisatisfiable to $\psi_1 \wedge \ldots \wedge \psi_{16} \wedge \chi_1 \wedge \chi_2 \wedge \chi_3 \wedge \eta$ and uses atoms $E(s,t)$ instead of $s \approx t$.
In addition to replacing equational atoms as indicated, we add the usual axioms concerning the fresh predicate symbol $E$.
Overall, the additional formulas have a length that lies in $\cO \bigl( \kappa \log \kappa + |\cD| \log |\cD| \bigr)$.

Consequently, the hardness result that we have obtained for SF with equality can be directly transferred to SF without equality.
Moreover, notice that all the above formulas can be transformed into Horn form. 
Hence, one could also replace $\approx$ by $E$ in hardness proofs for the Horn subfragment of SF.


\section{Conclusion}

We stress in this paper that an analysis of the computational complexity of satisfiability problems can greatly benefit from an analysis of how variables occur together in atoms instead of exclusively considering the number of occurring quantifier alternations.
What we have not yet taken into account is the Boolean structure of sentences.
This may widen the scope of our methods considerably and may moreover help understand where the hardness of satisfiability problems stems from.

Consider a quantified Boolean formula $\varphi := \forall \vx_1 \exists \vy_1  \ldots \forall \vx_n \exists \vy_n . \psi$ with quantifier-free $\psi$.
All satisfiable formulas of this shape together form a hard problem residing on the $n$-th level of the polynomial hierarchy.
But what if, for instance, $\psi$ has the form $\bigl( \bigwedge_i K_i \bigr) \wedge \bigl( \bigvee_j L_j \bigr)$, where the $K_i$ and the $L_j$ are literals and none of the existential variables in $\bigwedge_i K_i$ occurs in $\bigvee_j L_j$?
Since Boolean variables cannot jointly occur in atoms, $\varphi$ can be transformed into the equivalent formula $\exists \vy_1 \ldots \vy_n  \forall \vx_1 \ldots \vx_n . \psi$ by application of the rules of miniscoping (cf.\ Lemma~\ref{lemma:BasicQuantifierEquivalences}). Apparently, $\varphi$ belongs to a class of QBF sentences that resides on the first level of the polynomial hierarchy rather than on the \mbox{$n$-th}.

Perhaps it is time to reconsider some of the definitions that are based on the shape of quantifier prefixes alone.


\section*{Acknowledgments}
The author would like to thank the anonymous reviewers for valuable hints and suggestions.



\newpage
\appendix
\section*{Appendix}

\newcounter{auxTheorem}

\setcounter{auxTheorem}{\value{theorem}}
\setcounter{theorem}{9}
\setcounter{theorem}{\value{auxTheorem}}


\section{Details regarding Section~\ref{section:ComplexityBasics}}

\begin{lemma}\label{lemma:ComplexitySAT}~
	\begin{enumerate}[label=(\roman{*}), ref=(\roman{*})]
		\item Satisfiability for the class of SF sentences $\varphi$ without universal quantifiers and without equality can be decided nondeterministically in $\poly\bigl(\len(\varphi)\bigr)$ time.
		\item Satisfiability for the class of SF-Horn sentences $\varphi$ without universal quantifiers and without equality can be decided deterministically in $\poly\bigl(\len(\varphi)\bigr)$ time.
		\item Satisfiability for the class of SF-Krom sentences $\varphi$ without universal quantifiers and without equality can be decided nondeterministically using space that is logarithmic in $\len(\varphi)$.
	\end{enumerate}
\end{lemma}
\begin{proof}
	We show that we can reduce the above SF-satisfiability problems to the respective satisfiability problems for propositional logic and vice versa.

	Let $\varphi$ be a first-order sentence without non-constant function symbols that does not contain any universal quantifiers.
	Skolemization of all its existential quantifiers leads to the equisatisfiable ground sentence $\varphi_\gnd$ in which every atom has the shape $P(c_1, \ldots, c_m)$.
	Let $A_1, \ldots, A_k$ be a  complete enumeration of all the atoms---without duplicates---that occur in $\varphi_\gnd$.
	Let $q_1, \ldots, q_k$ be a list of pairwise distinct propositional variables.
	We construct the propositional formula $\phi$ from $\varphi_\gnd$ by replacing every atom $A_i$ with $q_i$.
	Clearly, any model of $\cA$ of $\varphi_\gnd$ induces a model $\cB$ of $\phi$: $\cB \models q_i$ if and only if $\cA \models A_i$.
	Conversely, any model $\cB'$ of $\phi$ induces a Herbrand model $\cA'$ of $\varphi$: $\cA' \models A_i$ if and only if $\cB' \models q_i$.
	Consequently, deciding satisfiability of $\varphi$ can be reduced to deciding satisfiability of $\phi$.
	Moreover, we observe the following properties.
	\begin{enumerate}[label=(\alph{*}), ref=(\alph{*})]
		\item $\len(\phi) \leq \len(\varphi)$.
		\item If $\varphi$ is a Horn formula, then $\phi$ is Horn.
		\item If $\varphi$ is a Krom formula, then $\phi$ is Krom.
			\qedhere
	\end{enumerate}
\end{proof}

\begin{lemma}\label{lemma:ComplexitySATwithEquality}
	There is an effective translation $T$ from the class of SF sentences with equality but without universal quantifiers into the class of ground first-order sentences without equality such that for every $\varphi$ from the former class we have that
	\begin{enumerate}[label=(\alph{*}), ref=(\alph{*})]
		\item every model of $T(\varphi)$ is a model of $\varphi$,
		\item every model of $\varphi$ can be extended to a model of $T(\varphi)$ over the same domain,
		\item $\len(T(\varphi)) \in \cO\bigl( \len(\varphi)^3 \bigr)$,
		\item If $\varphi$ is Horn, then $T(\varphi)$ is Horn as well.
	\end{enumerate}	
\end{lemma}
\begin{proof}
	We describe the translation $T$ informally.
	Let $\varphi$ be some SF sentence that contains equality but no universal quantifiers.
	Let $\varphi_\Sk$ be the result of Skolemizing the leading existential quantifier block in $\varphi$.
	Notice that $\varphi_\Sk$ neither contains quantifiers nor non-constant function symbols.
	Let $E$ be a binary predicate symbol that does not occur in $\varphi_\Sk$.
	We construct the following ground formulas:
		\begin{align*}
			\psi_\text{refl} 	&:= \bigwedge_{c \,\in\, \consts(\varphi_\Sk)} 		E(c,c) ~, \\
			\psi_\text{symm} 	&:= \bigwedge_{c,d \,\in\, \consts(\varphi_\Sk)} 		\bigl( E(c,d) \rightarrow E(d,c) \bigr) ~, \\
			\psi_\text{trans} 	&:= \bigwedge_{c,d,e \,\in\, \consts(\varphi_\Sk)} 	\bigl( E(c,d) \wedge E(d,e) \rightarrow E(c,e) \bigr) ~.
		\end{align*}
	Let $\psi_\text{cong}$ be the conjunction of all ground formulas of the form $E(c_1, d_1) \wedge \ldots \wedge E(c_m, d_m) \wedge P(c_1, \ldots, c_m) \;\rightarrow\; P(d_1, \ldots, d_m)$ where $c_1, d_1, \ldots, c_m, d_m \in \consts(\varphi_\Sk)$ and $P$ is an $m$-ary predicate symbol in $\varphi_\Sk$.
	We write $\psi'_\text{cong}$ to denote the restriction of $\psi_\text{cong}$ to formulas $E(c_1, d_1) \wedge \ldots \wedge E(c_m, d_m) \wedge P(c_1, \ldots, c_m) \;\rightarrow\; P(d_1, \ldots, d_m)$ whose constituents $P(c_1, \ldots, c_m)$ and $P(d_1, \ldots, d_m)$ actually occur in $\varphi_\Sk$.
	
	Let $\varphi'$ be the result of replacing every equation $c \approx d$ in $\varphi_\Sk$  with the atom $E(c,d)$.
	\begin{description}
		\item \underline{Claim I:}
			$\varphi_\Sk$ is satisfiable if and only if $\varphi' \wedge \varphi_\text{refl} \wedge \varphi_\text{symm} \wedge \varphi_\text{trans} \wedge \varphi_\text{cong}$ is satisfiable.
			
		\item \underline{Proof:}
			Let $\cA$ be a model of $\varphi_\Sk$. 
			By the Substructure Lemma, we may assume that $\cA$'s universe is $\bigl\{ c^\cA \bigm| c \in \consts(\varphi_\Sk) \bigr\}$.
			We now construct a model $\cB \models \varphi' \wedge \varphi_\text{refl} \wedge \varphi_\text{symm} \wedge \varphi_\text{trans} \wedge \varphi_\text{cong}$ from $\cA$.
			We take over $\cA$'s domain and its interpretation of the predicate symbols and constant symbols in $\varphi_\Sk$.
			We define $E$'s interpretation under $\cB$ such that $E^\cB := \bigl\{ \<\fa,\fa\> \bigm| \fa \in \fU_\cA \bigr\}$.
			Hence, for all $c,d \in \consts(\varphi_\Sk)$ we observe $\cB \models E(c,d)$ if and only if $\cB \models c \approx d$.
			Consequently, $\cB$ must be a model of $\varphi_\Sk$ and $\varphi' \wedge \varphi_\text{refl} \wedge \varphi_\text{symm} \wedge \varphi_\text{trans} \wedge \varphi_\text{cong}$.
			
			\smallskip
			Let $\cB'$ be a model of $\varphi' \wedge \varphi_\text{refl} \wedge \varphi_\text{symm} \wedge \varphi_\text{trans} \wedge \varphi_\text{cong}$. 
			By the Substructure Lemma, we may assume that $\cB'$'s universe is $\fU_{\cB'} = \bigl\{ c^{\cB'} \bigm| c \in \consts(\varphi') \bigr\}$.
			We now construct a model $\cA' \models \varphi_\Sk$ from $\cB'$.\
			Because of $\cB' \models \varphi_\text{refl} \wedge \varphi_\text{symm} \wedge \varphi_\text{trans} \wedge \varphi_\text{cong}$, we know that $E^{\cB'}$ is a congruence relation over $\fU_{\cB'}$. 
			We define the universe of $\cA'$ to be the quotient set $\fU_{\cA'} := \fU_{\cB'} /_{E^{\cB'}}$.
			Moreover, we define $c^{\cA'} := [c^{\cB'}]_{E^{\cB'}}$ for every $c \in \consts(\varphi_\Sk)$.
			For every congruence class $[\fa]_{E^{\cB'}}$ we know that two domain elements $d^{\cB'}, e^{\cB'} \in [\fa]_{E^{\cB'}}$ are indistinguishable by the relations $P^{\cB'}$ for which $P$ occurs in $\varphi'$.
			Therefore, we can use the following definition for every $m$-ary predicate symbol $P$ in $\varphi'$ (including $E$): $P^{\cA'} := \bigl\{ \bigl\<[\fa_1]_{E^{\cB'}}, \ldots, [\fa_m]_{E^{\cB'}} \bigr\> \bigm| \<\fa_1, \ldots, \fa_m\> \in P^{\cB'} \bigr\}$.
			This yields $\cA' \models \varphi'$ and for all $c,d \in \consts(\varphi')$ we observe $\cA' \models E(c,d)$ if and only if $\cA' \models c \approx d$.
			Hence, $\cA' \models \varphi_\Sk$.
				\hfill$\Diamond$
	\end{description}		

	It now remains to show equisatisfiability of $\varphi' \wedge \varphi_\text{refl} \wedge \varphi_\text{symm} \wedge \varphi_\text{trans} \wedge \varphi_\text{cong}$ and $\varphi' \wedge \varphi_\text{refl} \wedge \varphi_\text{symm} \wedge \varphi_\text{trans} \wedge \varphi'_\text{cong}$.
	The direction from left to right is obvious.
	\begin{description}
		\item \underline{Claim II:}
			Any model $\cA \models \varphi' \wedge \varphi_\text{refl} \wedge \varphi_\text{symm} \wedge \varphi_\text{trans} \wedge \varphi'_\text{cong}$ gives rise to a model $\cB \models \varphi' \wedge \varphi_\text{refl} \wedge \varphi_\text{symm} \wedge \varphi_\text{trans} \wedge \varphi_\text{cong}$.
			
		\item \underline{Proof:}
			Let $P(c_1, \ldots, c_m)$ and $P(d_1, \ldots, d_m)$ be two atoms that occur in $\varphi'$.
			We observe the following property:
			\begin{itemize}
				\item[($*$)] If $E^\cA$ contains the pairs $\<c_1, d_1\>, \ldots, \<c_m, d_m\>$, then $\cA \models \psi'_\text{cong}$ entails that $\cA \models P(c_1, \ldots, c_m)$ holds if and only if $\cA \models P(d_1, \ldots, d_m)$ does.
			\end{itemize}
			
			We define $\cB$ such that $\fU_\cB := \fU_\cA$, $E^\cB := E^\cA$, and $c^\cB := c^\cA$ for every $c \in \consts(\varphi')$.
			Moreover, for every $m$-ary predicate symbol $P$ occurring in $\varphi'$ and every tuple $\<\fa_1, \ldots, \fa_m\> \in \fU_\cB$ we set $\<\fa_1, \ldots, \fa_m\> \in P^\cB$ if and only if there is some atom $P(c_1, \ldots, c_m)$ in $\varphi'$ for which we have $\<c_1^\cB, \fa_1\>, \ldots, \<c_m^\cB, \fa_m\> \in E^\cB$.
			Due to ($*$), we know that $\cB \models \varphi'$ still holds.
			By construction of $\cB$, we moreover observe $\cB \models \varphi_\text{refl} \wedge \varphi_\text{symm} \wedge \varphi_\text{trans} \wedge \varphi_\text{cong}$.			
			\hfill$\Diamond$
	\end{description}		
	
	We set $T(\varphi) := \varphi' \wedge \varphi_\text{refl} \wedge \varphi_\text{symm} \wedge \varphi_\text{trans} \wedge \varphi'_\text{cong}$.
	It is easy to see that if $\varphi$ is Horn then $T(\varphi)$ is Horn as well.
	Moreover, the length of $T(\varphi)$ is upper bounded by $k \cdot \bigl(\len(\varphi') + |\consts(\varphi_\Sk)|^3 + \len(\varphi') \cdot |\At(\varphi_\Sk)|^2 \bigr)$ for some positive integer $k$, where $\At(\varphi_\Sk)$ denotes the set of all non-equational atoms that occur in $\varphi_\Sk$.
\end{proof}

\setcounter{auxTheorem}{\value{theorem}}
\setcounter{theorem}{9}
\begin{proposition}~
	\begin{enumerate}[label=(\roman{*}), ref=(\roman{*})]
		\item Satisfiability for the class of SF sentences without universal quantifiers is NP-complete.
		\item Satisfiability for the class of SF-Horn sentences without universal quantifiers is P-complete.
		\item Satisfiability for the class of SF-Krom sentences without universal quantifiers and without equality is NL-complete.
	\end{enumerate}
\end{proposition}
\begin{proof}
	The membership in the respective complexity classes is settled in Lemmas~\ref{lemma:ComplexitySAT} and~\ref{lemma:ComplexitySATwithEquality}. It remains to reduce the respective SAT problems to the corresponding satisfiability problems for SF.

	Let $\phi'$ be some propositional sentence and let $q_1, \ldots, q_k$ be a complete list of all propositional variables occurring in $\phi'$ (without duplicates).
	Let $c_1, \ldots, c_k$ be pairwise distinct constant symbols.
	We construct the first-order sentence $\varphi'$ from $\phi'$ by replacing every $q_i$ by the atom $P(c_i)$.
	By similar arguments as we have used in the proof of Lemma~\ref{lemma:ComplexitySAT}, we can show that $\phi'$ is satisfiable if and only if $\varphi'$ is satisfiable.
	Moreover, we observe the following properties.
	\begin{enumerate}[label=(\alph{*}), ref=(\alph{*})]
		\item $\len(\varphi) \in \cO\bigl( \len(\phi) \bigr )$.
		\item If $\phi$ is a Horn formula, then $\varphi'$ is Horn.
		\item If $\phi$ is a Krom formula, then $\varphi$ is Krom.
			\qedhere
	\end{enumerate}	
\end{proof}
\setcounter{theorem}{\value{auxTheorem}}

\section{Details regarding Section~\ref{section:TranslationSFintoBSR}}
\setcounter{auxTheorem}{\value{theorem}}
\setcounter{theorem}{10}

\begin{lemma}
	Let $I$ and $K_i$, $i \in I$, be sets that are finite, nonempty, and pairwise disjoint. 
	The elements of these sets serve as indices.
	Let 
	\[ \varphi := \exists \vy. \bigwedge_{i \in I} \Bigl( \chi_i(\vz) \vee \bigvee_{k \in K_i} \eta_k(\vy, \vz) \Bigr) \]
	be some first-order formula where the $\chi_i$ and the $\eta_k$ denote arbitrary subformulas that we treat as indivisible units in what follows.
	We say that $f : I \to \bigcup_{i \in I} K_i$ is a \emph{selection function} 
	if for every $i \in I$ we have $f(i) \in K_i$.
	We denote the set of all selection functions of this form by $\cF$.
	
	$\varphi$ is equivalent to 
		\[ \varphi' := \bigwedge_{\text{\scriptsize $\begin{array}{c} S \subseteq I \\ S \neq \emptyset \end{array}$ \normalsize}} \Bigl( \bigvee_{i \in S} \chi_i(\vz) \Bigr) \vee \bigvee_{f \in \cF} \Bigl( \exists \vy. \bigwedge_{i \in S} \eta_{f(i)} (\vy, \vz) \Bigr) ~. \]
\end{lemma}
\begin{proof}
	For the sake of readability we sometimes reuse variables in different occurrences of quantifiers in this proof.
	Using distributivity of $\wedge$ over $\vee$, we transform $\varphi$ into an equivalent disjunction of conjunctions:
		\[ \exists \vy.\!\!\!\! \bigvee_{\text{\scriptsize $\begin{array}{c} \<T, f\> \in \quad\strut\\ \strut\quad (\fP I) \times \cF \end{array}$ \normalsize}} \!\!\! 
			\Bigl( \bigwedge_{i \in T} \chi_i(\vz) \Bigr) 	\wedge 	\Bigl( \bigwedge_{i \in I \setminus T} \eta_{f(i)}(\vy, \vz) \Bigr) ~. \]
	Since the existential quantifier block distributes over the topmost disjunction, we can move this block inwards and obtain the equivalent formula
		\begin{equation}\label{eqn:bigDisjunction}
			\hspace{-2ex}
			\bigvee_{\text{\scriptsize $\begin{array}{c} \<T, f\> \in \quad\strut\\ \strut\quad (\fP I) \times \cF \end{array}$ \normalsize}} \!\!\!
			\Bigl( \bigwedge_{i \in T} \chi_i(\vz) \Bigr) 	\wedge 	\Bigl( \exists \vy. \bigwedge_{i \in I \setminus T} \eta_{f(i)}(\vy, \vz) \Bigr) ~.
		\end{equation}	
	At this point, we employ distributivity of $\vee$ over $\wedge$ to transform this result into an equivalent conjunction of disjunctions $\varphi'' := \bigwedge_j \bigvee_\ell \psi_{j,\ell}$ in which for every index $j$ and every pair $\<T, f\> \in (\fP I) \times \cF$ there is exactly one $\ell$ such that
	\begin{itemize}
		\item either $\psi_{j, \ell} = \chi_i$  for some $i \in T$
		\item or $\psi_{j, \ell} = \exists \vy. \bigwedge_{i \in I\setminus T} \eta_{f(i)} (\vy, \vz)$.
	\end{itemize}

	In order to show that $\varphi''$ is semantically equivalent to $\varphi'$,	we prove the following claims.
	
	\begin{description}
		\item \underline{Claim I:}
			Every disjunction $\bigvee_\ell \psi_{j,\ell}$ in $\varphi''$ is subsumed by a disjunction
				\[ \psi'_S := \Bigl( \bigvee_{i \in S} \chi_i(\vz) \Bigr) \vee \bigvee_{f \in \cF} \Bigl( \exists \vy. \bigwedge_{i \in S} \eta_{f(i)} (\vy, \vz) \Bigr) \]
			for some nonempty $S \subseteq I$.
		
		\item \underline{Proof:}
			Fix some index $j$ and consider $\bigvee_\ell \psi_{j,\ell}$.
			We set $S := \bigl\{ i \in I \bigm| \text{$\psi_{j,\ell} = \chi_i$ for some $\ell$} \bigr\}$.
			Consider the set $\bS := I \setminus S$.
			By definition of $S$, we know that none of the $\chi_i$ with $i \in \bS$ is a constituent of $\bigvee_\ell \psi_{j,\ell}$.
			For every selection function $f \in \cF$ there is some disjunct
				\[ \Bigl( \bigwedge_{i \in \bS} \chi_i (\vz) \Bigr) \wedge \Bigl( \exists \vy. \bigwedge_{i \in I  \setminus \bS} \eta_{f(i)} (\vy, \vz) \Bigr) \]
			in (\ref{eqn:bigDisjunction}) of which we know that none of the $\chi_i$ in it has been picked as constituent of $\bigvee_\ell \psi_{j,\ell}$ when constructing $\varphi''$.
			Hence, due to the definition of $\varphi''$, there must be some $\ell_*$ such that $\psi_{j,\ell_*} = \exists \vy. \bigwedge_{i \in I  \setminus \bS} \eta_{f(i)} (\vy, \vz)$, where $I  \setminus \bS = S$.
			
			Consequently, $\bigvee_\ell \psi_{j,\ell}$ is subsumed by $\psi'_S$.
			\strut\hfill$\Diamond$

		\item \underline{Claim II:} 
			Each of the subsuming disjunctions $\psi'_S$ in Claim~I is indeed equivalent to some disjunction $\bigvee_\ell \psi_{j,\ell}$ in $\varphi''$.
		
		\item \underline{Proof:}
				Fix some nonempty $S_* \subseteq I$ and consider $\psi'_{S_*}$.
				We obtain the equivalent disjunction $\psi_*$ from the disjuncts in (\ref{eqn:bigDisjunction}) as follows. For every $T \subseteq I$ with nonempty $T \cap S_*$ we pick one of the $\chi_i$ with $i \in T \cap S_*$ as constituent of $\psi_*$. 
				For every $T \subseteq I$ for which $T \cap S_*$ is empty and any $f \in \cF$ we pick $\exists \vy. \bigwedge_{i \in I\setminus T} \eta_{f(i)} (\vy,\vz)$ as constituent of $\psi_*$.
				Since $S_*$ is nonempty, $T$ must be a proper subset of $I$ and thus $I \setminus T$ is also nonempty.
				
				For every constituent of the form $\exists \vy. \bigwedge_{i \in T'} \eta_{f(i)} (\vy,\vz)$ that belongs to the disjunction $\psi_*$ we know that $S_* \subseteq T'$.
				Hence, $\psi_*$ is of the form 
					\begin{align*}
						\Bigl( \bigvee_{i \in S_*} \chi_i (\vz) \Bigr) \vee \bigvee_{j} \bigvee_{f \in \cF} &\exists \vy. \Bigl( \bigwedge_{i \in S_*} \eta_{f(i)}(\vy,\vz) \Bigr) \wedge \Bigl( \bigwedge_{i \in S'_j} \eta_{f(i)}(\vy,\vz) \Bigr)
					\end{align*}	
				for certain sets $S'_j \subseteq I\setminus S_*$. Among the $S'_j$ is, in particular, the empty set, originating from $T = I \setminus S_*$. In this case, we have $S'_j = \bigl( I \setminus T \bigr) \setminus S_* = \bigl( I \setminus (I \setminus S_*) \bigr) \setminus S_* = \emptyset$.
				Hence, we can equivalently transform $\psi_*$ into 
					\begin{align*}
						 &\Bigl( \bigvee_{i \in S_*} \chi_i (\vz) \Bigr) \vee \bigvee_{f \in \cF} \bigvee_{j} \exists \vy. \Bigl( \bigwedge_{i \in S_*} \eta_{f(i)}(\vy,\vz) \Bigr) \wedge \Bigl( \bigwedge_{i \in S'_j} \eta_{f(i)}(\vy,\vz) \Bigr) \\
						 &\semequiv\; \Bigl( \bigvee_{i \in S_*} \chi_i (\vz) \Bigr) \vee \bigvee_{f \in \cF} \exists \vy. \bigvee_{j} \biggl( \Bigl( \bigwedge_{i \in S_*} \eta_{f(i)}(\vy,\vz) \Bigr) \wedge \Bigr( \bigwedge_{i \in S'_j} \eta_{f(i)}(\vy,\vz) \Bigr)\biggr) \\
						 &\semequiv\; \Bigl( \bigvee_{i \in S_*} \chi_i (\vz) \Bigr) \vee \bigvee_{f \in \cF} \exists \vy. \Bigl( \bigwedge_{i \in S_*} \eta_{f(i)}(\vy,\vz) \Bigr) \vee \bigvee_{j} \Bigl( \bigwedge_{i \in S_*} \eta_{f(i)}(\vy,\vz) \wedge\!\!\! \bigwedge_{\text{\scriptsize $\begin{array}{c} i \in S'_j \\ S'_j \neq \emptyset \end{array}$ \normalsize}}\!\!\! \eta_{f(i)}(\vy,\vz) \Bigr) \\
						 &\semequiv\; \Bigl( \bigvee_{i \in S_*} \chi_i (\vz) \Bigr) \vee \bigvee_{f \in \cF} \exists \vy. \Bigl( \bigwedge_{i \in S_*} \eta_{f(i)}(\vy,\vz) \Bigr) \vee \biggl(\Bigl( \bigwedge_{i \in S_*} \eta_{f(i)}(\vy,\vz) \Bigr) \wedge \bigvee_{j} \!\!\!\bigwedge_{\text{\scriptsize $\begin{array}{c} i \in S'_j \\ S'_j \neq \emptyset \end{array}$ \normalsize}}\!\!\! \eta_{f(i)}(\vy,\vz) \biggr) ~.
					\end{align*}	 
				By the absorption axiom of Boolean algebra, we finally obtain the equivalent disjunction
					\[ \Bigl( \bigvee_{i \in S_*} \chi_i (\vz) \Bigr) \vee \bigvee_{f \in \cF} \Bigl( \exists \vy. \bigwedge_{i \in S_*} \eta_{f(i)}(\vy,\vz) \Bigr) ~. \]
			 	Thus, the claimed equivalence holds.
				
				We have not yet explicitly explained why the first subformula $\bigvee_{i \in S_*} \chi_i (\vz)$ of $\psi_*$ covers $S_*$ completely. But this is easy to see, when one takes the singleton sets $T = \{i\}$ for every $i \in S_*$ into account, for which we pick the $\chi_i$ as a constituents of $\psi_*$.
				\strut\hfill$\Diamond$
	\end{description}	
	This completes the proof of the lemma.
\end{proof}
\setcounter{theorem}{\value{auxTheorem}}

\setcounter{auxTheorem}{\value{theorem}}
\setcounter{theorem}{11}
\begin{lemma}
	Let $\varphi := \exists \vz\, \forall \vx_1 \exists \vy_1 \ldots \forall \vx_n \exists \vy_n. \psi$ be an SF sentence of positive degree $\degree_\varphi$ in standard form.
	Let $\cL_\varphi (\vy)$ denote the set of all literals in $\varphi$ that contain at least one variable $y \in \vy := \vy_1 \cup \ldots \cup \vy_n$.
	There exists a sentence $\varphi_{\text{BSR}} = \exists \vz\, \exists \vu\, \forall \vv. \psi_{\text{BSR}}$ in standard form with quantifier-free $\psi_{\text{BSR}}$ that is equivalent to $\varphi$ and contains at most $|\vz| + |\vy| \cdot \degree_\varphi \cdot \bigl( \twoup{\degree_\varphi}{|\cL_\varphi (\vy)|} \bigr)^{\degree_\varphi}$ leading existential quantifiers.
\end{lemma}
\begin{proof}
	For convenience, we pretend that $\vz$ is empty. The argument works for nonempty $\vz$ as well.
	Let $\vx := \vx_1 \cup \ldots \cup \vx_n$.
	We transform $\varphi$ into an equivalent CNF formula of the form 
		\[ \forall \vx_1 \exists \vy_1 \ldots \forall \vx_n \exists \vy_n.\; \bigwedge_{i \in I} \Bigl( \chi_i(\vx) \vee \bigvee_{k \in K_i} L_{k}(\vy) \Bigr) \]
	where $I$ and the $K_i$ are finite, pairwise disjoint sets of indices, the subformulas $\chi_i$ are disjunctions of literals, and the $L_k$ are literals.
	By Lemma~\ref{lemma:AdvancedMiniscoping:refined}, we can construct an equivalent formula of the form 
		\begin{align*}
			\varphi' := \forall \vx_1 \exists \vy_1 \ldots \forall \vx_n. \bigwedge_{S \in \fP I\setminus \emptyset} &\Bigl( \bigvee_{i \in S} \chi_i(\vx) \Bigr) \vee \bigvee_{f \in \cF} \Bigl( \exists \vy_n. \bigwedge_{i \in S} \eta_{f(i)} (\vy) \Bigr)
		\end{align*}	
	where $\cF$	is the set of all selection functions over the index sets $K_i$, $i \in I$.
	For the sake of readability we sometimes reuse variables in different occurrences of quantifiers in this proof.
	Applying ordinary miniscoping, we may move inward the universal quantifier block $\forall \vx_n$ and thus obtain 
		\begin{align*}
			\varphi'' :=
			\forall \vx_1 \exists \vy_1 \ldots \exists \vy_{n-1}. &\bigwedge_{S \in \fP I\setminus \emptyset} \Bigl( \forall \vx_n. \bigvee_{i \in S} \chi_i(\vx) \Bigr) \vee \bigvee_{f \in \cF} \Bigl( \exists \vy_n. \bigwedge_{i \in S} \eta_{f(i)} (\vy) \Bigr) ~.
		\end{align*}
	We can now iterate these two steps in an alternating fashion until all quantifier blocks have been moved inwards in the described way.
	The constituents of the result $\varphi^{(3)} := \bigwedge_q \Bigl( \chi^{(3)}_q \vee \bigvee_p \eta^{(3)}_{qp} \Bigr)$ of this process have the form 
		\begin{align*}
			\chi^{(3)}_q =
			\forall \vx_1. &\bigvee_{\ell_1} \forall \vx_2. \bigvee_{\ell_2} \Bigl( \ldots \Bigl( \bigvee_{\ell_{n-1}} \forall \vx_n. \bigvee_{i \in S_{\ell_1, \ldots, \ell_{n-1}}} \chi_i (\vx) \Bigr) \ldots \Bigr)
		\end{align*}	
		where the $S_{\ell_1, \ldots, \ell_n-1}$ are certain subsets of $I$,
	and 
		\begin{align*}
			\eta^{(3)}_{qp} =
			\exists \vy_1. &\bigwedge_{\ell'_1} \exists \vy_2. \bigwedge_{\ell'_2} \Bigl( \ldots \Bigl( \bigwedge_{\ell'_{n-1}} \exists \vy_n. \bigwedge_{k \in K_{\ell'_1, \ldots, \ell'_{n-1}}} L_{k} (\vy) \Bigr) \ldots \Bigr)
		\end{align*}	
	where	the $K_{\ell'_1, \ldots, \ell'_{n-1}}$ are certain subsets of $\bigcup_{i \in I} K_i$.
	
	By definition of $\degree_\varphi$, we may assume that there is some positive integer $m$ and a partition of the set $\vy$ into $m$ nonempty subsets $Y_1, \ldots, Y_m$ such that $Y_1, \ldots, Y_m$ are all pairwise separated in $\varphi$ and for every $j$, $1\leq j\leq m$, we have $\degree_{Y_j} \leq \degree_\varphi$.
	Since the sets $Y_1, \ldots, Y_m$ are pairwise separated in $\varphi$, we may partition the set $\cL_\varphi(\vy)$ into subsets $\cL_\varphi(Y_1), \ldots, \cL_\varphi(Y_m)$ such that each $\cL_\varphi(Y_j)$ contains exactly the literals in $\varphi$ that contain at least one variable from $Y_j$.
	This means, we can rewrite every $\eta^{(3)}_{qp}$ into the form
		\begin{align*}
			\eta^{(4)}_{qp} =\; &\exists \vy_1. \bigwedge_{\ell'_1} \exists \vy_2. \bigwedge_{\ell'_2} \Bigl( \ldots \Bigl( \bigwedge_{\ell'_{n-1}} \exists \vy_n. \bigwedge_{j \in [m]} \bigwedge_{k \in K^j_{\ell'_1, \ldots, \ell'_{n-1}}} L_{k} (Y_j) \Bigr) \ldots \Bigr)
		\end{align*}		
	where the sets $K^1_{\ell'_1, \ldots, \ell'_{n-1}}, \ldots, K^m_{\ell'_1, \ldots, \ell'_{n-1}}$ constitute a partition of $K_{\ell'_1, \ldots, \ell'_{n-1}}$.
	(Some of these parts may be empty.)
	We then observe the following equivalences.	
	\begin{align*}
		&\exists \vy_1. \bigwedge_{\ell'_1} \exists \vy_2. \bigwedge_{\ell'_2} \Bigl( \ldots \Bigl( \bigwedge_{\ell'_{n-1}} \exists \vy_n. \bigwedge_{j \in [m]}\; \bigwedge_{k \in K^j_{\ell'_1 \ldots \ell'_{n-1}}} L_{k} (Y_j) \Bigr) \ldots \Bigr) \\
		&\semequiv\;
			\exists \vy_1. \bigwedge_{\ell'_1} \exists \vy_2. \bigwedge_{\ell'_2} \Bigl( \ldots \Bigl( \bigwedge_{\ell'_{n-1}} \bigwedge_{j \in [m]} \exists \bigl(\vy_n \cap Y_j \bigr). \bigwedge_{k \in K^j_{\ell'_1 \ldots \ell'_{n-1}}} L_{k} (Y_j) \Bigr) \ldots \Bigr) \\
		&\semequiv\;
			\exists \vy_1. \bigwedge_{\ell'_1} \exists \vy_2. \bigwedge_{\ell'_2} \Bigl( \ldots \Bigl( \bigwedge_{j \in [m]}\; \bigwedge_{\ell''_{n-1}} \exists \bigl(\vy_n \cap Y_j \bigr). \bigwedge_{k \in K^j_{\ell'_1 \ldots \ell'_{n-2} \ell''_{n-1}}} L_{k} (Y_j) \Bigr) \ldots \Bigr) \\
		&\qquad\vdots \\
		&\semequiv\;
			\bigwedge_{j \in [m]} \exists \bigl(\vy_1 \cap Y_j \bigr). \bigwedge_{\ell''_1} \exists \bigl(\vy_2 \cap Y_j \bigr). \bigwedge_{\ell''_2} \Bigl( \ldots \Bigl( \bigwedge_{\ell''_{n-1}} \exists \bigl(\vy_n \cap Y_j \bigr). \bigwedge_{k \in K^j_{\ell''_1 \ldots \ell''_{n-1}}} L_{k} (Y_j) \Bigr) \ldots \Bigr) 
	\end{align*}
	For every $\eta^{(4)}_{qp}$, we call the result of the above transformation $\eta^{(5)}_{qp}$.
	In cases where the set $\vy_i \cap Y_j$ is empty, the existential quantifier block vanishes.
	For every $j \in [m]$ there are at most $\degree_{Y_j}$ nonempty sets $\vy_i \cap Y_j$.
	Hence, every $\eta^{(5)}_{qp}$ contains at most $\degree_\varphi$ nested existential quantifier blocks that are separated by in-between conjunctive connectives in the syntax tree.
	
	We obtain $\varphi^{(5)}$ from $\varphi^{(3)}$ by replacing every constituent $\eta^{(3)}_{qp}$ with the corresponding $\eta^{(5)}_{qp}$ after applying the idempotence axioms of Boolean Algebra exhaustively to remove redundant conjuncts.
	
	Let $\kappa := \max \bigl\{ |\cL_\varphi(Y_j)| \bigm| j \in [m] \bigr\}$. Due to the idempotence axioms, the following upper bounds can be shown inductively for any positive integer $d$, starting from $d = 1$:
	Modulo idempotence, there are at most $\twoup{d}{\kappa}$ formulas of the form
		\begin{align*} 
			&\exists \bigl(\vy_{i_1} \cap Y_j \bigr). \bigwedge_{\ell_1} \exists \bigl(\vy_{i_2} \cap Y_j \bigr). \bigwedge_{\ell_2} \Bigl( \ldots \Bigl( \bigwedge_{\ell_{d-1}} \exists \bigl(\vy_{i_d} \cap Y_j \bigr). \bigwedge_{k \in K'^j_{\ell_1, \ldots, \ell_{d-1}}} L_{k} (Y_j) \Bigr) \ldots \Bigr) ~.
		\end{align*}	
	For the sentence $\varphi^{(5)} = \bigwedge_q \Bigl( \chi^{(3)}_q \vee \bigvee_p \eta^{(5)}_{qp} \Bigr)$ this means that it contains at most $\twoup{\degree_\varphi}{\kappa}$ distinct subformulas (not occurrences thereof!) that are of the form $\exists y. \psi'$ and do not lie within the scope of any quantifier. 
	We treat every such subformula $\exists y. \psi'$ and every subformula $\chi^{(3)}_q$ as indivisible unit and, employing distributivity of $\wedge$ over $\vee$, transform $\varphi^{(5)}$ into a disjunction of conjunctions $\varphi^{(6)} := \bigvee_{s} \Bigl( \bigwedge_{r_1}\chi^{(6)}_{r_1} \wedge \bigwedge_{r_2} \eta^{(6)}_{r_2} \Bigr)$ where the $\chi^{(6)}_{r_1}$ have the same shape as the $\chi^{(3)}_q$, and the $\eta^{(6)}_{r_2}$ are of the form
	\begin{align*}
		&\exists \bigl(\vy_{i_1} \cap Y_j \bigr). \bigwedge_{\ell_1} \exists \bigl(\vy_{i_2} \cap Y_j \bigr). \bigwedge_{\ell_2} \Bigl( \ldots \Bigl( \bigwedge_{\ell_{\degree_{Y_j}-1}} \exists \bigl(\vy_{i_{\degree_{Y_j}}} \cap Y_j \bigr). \bigwedge_{k \in K'^j_{\ell_1, \ldots, \ell_{\degree_{Y_j}-1}}} L_{k} (Y_j) \Bigr) \ldots \Bigr)
	\end{align*}
	for a certain $j$ and certain indices $i_1, \ldots, i_{\degree_{Y_j}}$ with $1 \leq i_1 < \ldots < i_{\degree_{Y_j}} \leq n$, all depending on $r_2$.
	
	Due to previous observations, we know that, modulo idempotence, $r_2$ ranges over at most $\twoup{\degree_\varphi}{\kappa}$ indices.
	Moreover, any $\ell_k$ in any $\eta^{(6)}_{r_2}$ ranges over at most $\twoup{\degree_\varphi-k}{\kappa}$ indices.
	Consequently, every constituent $\bigwedge_{r_2} \eta^{(6)}_{r_2}$ in $\varphi^{(6)}$ contains at most $\max_{i,j} |\vy_i \cap Y_j| \cdot \sum_{k'=1}^{\degree_\varphi} \prod_{d = k'}^{\degree_\varphi} \twoup{d}{\kappa}$ occurrences of existential quantifiers.
	
	Since these existential quantifiers distribute over the topmost disjunction when we move them outwards to the front of the sentence $\varphi^{(6)}$, and since the universal quantifiers in the $\chi^{(6)}_q$ may also be moved back outwards, we have shown that $\varphi$ is equivalent to some BSR sentence with at most $|\vy| \cdot \degree_\varphi \cdot \bigl( \twoup{\degree_\varphi}{\kappa} \bigl)^{\degree_\varphi}$ leading existential quantifiers.
\end{proof}
\setcounter{theorem}{\value{auxTheorem}}

\section{Details regarding Section~\ref{section:LowerBoundsSmallestBSRsentences}}
\setcounter{auxTheorem}{\value{theorem}}
\setcounter{theorem}{15}

\begin{theorem}
	There is a class of SF sentences that are Horn and Krom such that for every positive integer $n$ the class contains a sentence $\varphi$ of degree $\degree_\varphi = n$ and with a length linear in $n$ for which any equivalent BSR sentence contains at least $\sum_{k=1}^{n}\twoup{k}{n}$ leading existential quantifiers.
\end{theorem}
\begin{proof}
	Let $n \geq 1$ be some positive integer.
	Consider the following first-order sentence in which the sets $\{x_1, \ldots, x_n\}$ and $\{y_1, \ldots, y_n\}$ are separated:
	\begin{align*}
		\varphi := \forall &x_n \exists y_n \ldots \forall x_1 \exists y_1. \bigwedge_{i=1}^{4n} \bigl( P_i(x_1, \ldots, x_n) \leftrightarrow Q_i(y_1, \ldots, y_n) \bigr) ~.
	\end{align*}	
	Notice that we change the orientation of the indices in the quantifier prefix in this proof.
	
	In order to construct a particular model of $\varphi$, we inductively define the following sets:
		$\cS_1 := \bigl\{ S \subseteq [4n] \bigm| |S| = 2n \bigr\}$, $\cS_{k+1} := \bigl\{ S \in \fP \cS_k \bigm| |S| = \tfrac{1}{2} \cdot |\cS_k| \bigr\}$ for every $k > 1$.
	Hence, we observe that
		\begin{description}
			\item $|\cS_1| \;=\; {{4n} \choose {2n}} \;\geq\; \bigl( \frac{4n}{2n} \bigr)^{2n} \;=\; 2^{2n}$,
			\item $|\cS_2| \;=\; {{|\cS_1|} \choose {|\cS_1|/2}} \;\geq\; \bigl( \frac{|\cS_1|}{|\cS_1|/2} \bigr)^{|\cS_1|/2} \;=\; 2^{|\cS_1|/2} \;\geq\; 2^{2^{2n}/2} \;=\; 2^{2^{2n-1}}$,
			\item[] $\strut\qquad\vdots$
			\item $|\cS_n| \;=\; {{|\cS_{n-1}|} \choose {|\cS_{n-1}|/2}} \;\geq\; 2^{|\cS_{n-1}|/2} \;\geq\; 2^{2^{2^{\vdots^{2^{2n-1}-1}}-1}} \;\geq\; \twoup{n}{2n-(n-1)} \;=\; \twoup{n}{n+1}$,
		\end{description}	
	where the inequality ${n \choose k} \geq (n/k)^{k}$ can be found in 
	T.~Cormen et al. \emph{Introduction to Algorithms} (McGraw-Hill Higher Education, 2nd edition, 2001.), page 1097, for example.
	
	\medskip
	Having the sets $\cS_k$, we now define the structure $\cA$ as follows:	
		\begin{itemize}
			\item $\fU_\cA := \bigcup_{k = 1}^{n} \bigl\{ \fa^{(k)}_{S}, \fb^{(k)}_{S} \bigm| S \in \cS_k \bigr\}$, 
			\item $P_i^\cA := \bigl\{ \<\fa^{(1)}_{S_1}, \ldots, \fa^{(n)}_{S_n}\> \in \fU_\cA^n \bigm| i \in S_1 \in S_2 \in \ldots \in S_n \bigr\}$ for $i = 1, \ldots, 4n$, and
			\item $Q_i^\cA := \bigl\{ \<\fb^{(1)}_{S_1}, \ldots, \fb^{(n)}_{S_n}\> \in \fU_\cA^n \bigm| i \in S_1 \in S_2 \in \ldots \in S_n \bigr\}$ for $i = 1, \ldots, 4n$.
		\end{itemize}		
	Clearly, for any choice of $S_1, \ldots, S_n$  and every $i$, $1\leq i\leq 4n$, we have 
		\begin{align*}
			\cA, \bigl[ &x_1 \Mapsto \fa^{(1)}_{S_1}, \ldots, x_n \Mapsto \fa^{(n)}_{S_n}, y_1 \Mapsto \fb^{(1)}_{S_1}, \ldots, y_n \Mapsto \fb^{(n)}_{S_n} \bigr]  \models P_i(x_1, \ldots, x_n) \leftrightarrow Q_i(y_1, \ldots, y_n) ~.
		\end{align*}	
	For any other choice of tuples $\< \fc_1, \ldots, \fc_n\>$, i.e.\ there do not exist sets $S_1 \in \cS_1, \ldots, S_n \in \cS_n$ such that $\< \fc_1, \ldots, \fc_n \>$ equals $\< \fa^{(1)}_{S_1}, \ldots, \fa^{(n)}_{S_n} \>$ or $\< \fb^{(1)}_{S_1}, \ldots, \fb^{(n)}_{S_n} \>$, we observe $\cA, [x_1 \Mapsto \fc_1, \ldots, x_n \Mapsto \fc_n] \not\models P_i(x_1, \ldots, x_n)$ and $\cA, [y_1 \Mapsto \fc_1, \ldots, y_n \Mapsto \fc_n] \not\models Q_i(y_1, \ldots, y_n)$ for every $i$.
	Hence, 
		\begin{align*}
			\cA, [&x_1 \Mapsto \fc_1, \ldots, x_n \Mapsto \fc_n, y_1 \Mapsto \fc_1, \ldots, y_n \Mapsto \fc_n] \models \bigwedge_{i = 1}^{4n} P_i(x_1, \ldots, x_n) \leftrightarrow Q_i(y_1, \ldots, y_n) ~.
		\end{align*}
	Consequently, $\cA$ is a model of $\varphi$.
		
	Consider the following simple two-player game with Players $\fA$ and $\fB$ where both players have complete and instantaneous knowledge about all moves that are made by either player.
	In the first round $\fA$ moves first by picking some domain element $\fa^{(n)}_{S_{\fA, n}}$ for some set $S_{\fA,n} \in \cS_n$. $\fB$ knows about $\fA$'s choice and answers by picking a domain element $\fb^{(n)}_{S_{\fB, n}}$ for some set $S_{\fB,n} \in \cS_n$. 
	The game continues for $n-1$ more rounds, where in every round Player $\fA$ picks a domain element $\fa^{(j)}_{S_{\fA, j}}$ with $S_{\fA,j} \in S_{\fA, j+1}$ and $\fB$ answers by picking some $\fb^{(j)}_{S_{\fB,j}} \in S_{\fB,j+1}$.	
	Hence, in the last round the chosen domain elements $\fa^{(1)}_{S_{\fA, 1}}$ and $\fb^{(1)}_{S_{\fB, 1}}$ are such that $S_{\fA, 1}$ and $S_{\fB, 1}$ are both nonempty subsets of $[4n]$.
	Player $\fA$ wins if and only if 
		\begin{align*}
			\cA, \bigl[ &x_1 \Mapsto \fa^{(1)}_{S_{\fA, 1}}, \ldots, x_n \Mapsto \fa^{(n)}_{S_{\fA, n}}, y_1 \Mapsto \fb^{(1)}_{S_{\fB, 1}}, \ldots, y_n \Mapsto \fb^{(n)}_{S_{\fB, n}} \bigr] \not\models P_i(x_1, \ldots, x_n) \leftrightarrow Q_i(y_1, \ldots, y_n)
		\end{align*}	
	for some $i \in [4n]$,
	and Player $\fB$ wins if and only if 
		\begin{align*}
			\cA, \bigl[ &x_1 \Mapsto \fa^{(1)}_{S_{\fA, 1}}, \ldots, x_n \Mapsto \fa^{(n)}_{S_{\fA, n}}, y_1 \Mapsto \fb^{(1)}_{S_{\fB, 1}}, \ldots, y_n \Mapsto \fb^{(n)}_{S_{\fB, n}} \bigr] \models P_i(x_1, \ldots, x_n) \leftrightarrow Q_i(y_1, \ldots, y_n)
		\end{align*}	
	for every $i \in [4n]$.
	Since $\cA$ is a model of $\varphi$, there must exist a winning strategy for $\fB$.
	
	\begin{description}
		\item \underline{Claim:}
			There is exactly one winning strategy for $\fB$, namely, for every $j = n, \ldots, 1$ Player $\fB$ picks the element $\fb^{(j)}_{S_{\fA,j}}$ in round $n-j+1$, i.e.\ for every $j$ we have $S_{\fB,j} = S_{\fA,j}$.
				
		\item \underline{Proof:}
			It is easy to see that the described strategy is a winning strategy for $\fB$.
			
			Assume $\fB$ deviates from this strategy.
			This means there exists some $j_*$, $1 \leq j_* \leq n$, such that $\fB$ did not adhere to the described strategy in the $(n-j_*+1)$st round, i.e.\ $S_{\fB, j_*} \neq S_{\fA, j_*}$.
			
			We show by induction on $j_*$ that $\fA$ has a winning strategy from this deviation point on.
			
			For the base case $j = 1$ we consider two distinct nonempty sets $S_{\fA,1}, S_{\fB,1} \subseteq [4n]$.
			There must be some index $i_*$ that belongs to one of the two sets but not to the other, i.e.\ $i_* \in (S_{\fA,1} \cup S_{\fB,1}) \setminus (S_{\fA,1} \cap S_{\fB,1})$.
			
			Suppose that $i_* \in S_{\fA,1} \setminus S_{\fB,1}$. Hence, we can construct the chain $i_* \in S_{\fA,1} \in \ldots \in S_{\fA,n}$, by definition of the allowed moves.
			This entails $\cA, [x_1 \Mapsto \fa^{(1)}_{S_{\fA,1}}, \ldots, x_n \Mapsto \fa^{(n)}_{S_{\fA,n}}] \models P_{i_*}(x_1, \ldots, x_n)$.
			On the other hand, we know $\cA, [y_1 \Mapsto \fb^{(1)}_{S_{\fB,1}}, \ldots, y_n \Mapsto \fb^{(n)}_{S_{\fB,n}}] \not\models Q_{i_*}(y_1, \ldots, y_n)$, because of $i_* \not\in S_{\fB,1}$.
			Hence, $\fA$ wins and the chosen strategy cannot be a winning strategy for $\fB$.

			The case where $i_* \in S_{\fB,1} \setminus S_{\fA,1}$ is symmetric and $\fA$ also wins.
			
			For the inductive case we fix some $j_* > 1$.
			Since $S_{\fA,j_*}$ and $S_{\fB,j_*}$ are distinct but have the same number of elements, there is some set $S' \in S_{\fA,j_*} \setminus S_{\fB,j_*}$.
			If $\fA$ picks $\fa^{(j_*-1)}_{S_{\fA,j_*-1}} := \fa^{(j_*-1)}_{S'}$ in the following round, we have $S_{\fB,j_*-1} \neq S_{\fA,j_*-1}$ for any choice $\fb^{(j_*-1)}_{S_{\fB,j_*-1}}$ that $\fB$ could possibly make.
			By induction, $\fA$ has a winning strategy starting from the next round of the game. Hence, there is a winning strategy starting from the current round.			
			\strut\hfill$\Diamond$		
	\end{description}	
	The just proved claim would still hold true if we allowed $\fB$ to freely pick any element of the domain $\fU_\cA$ at every round. 
	The reason is that for any choice of elements $\fa^{(n)}_{S_{\fA,n}}, \ldots, \fa^{(1)}_{S_{\fA,1}}$ made by $\fA$ with $S_{\fA,1} \in \ldots \in S_{\fA,n} \in \cS_n$ we know that $S_{\fA,1}$ is nonempty.
	Hence, we can always find some $i_* \in S_{\fA,1}$ such that $\< \fa^{(n)}_{S_{\fA,n}}, \ldots, \fa^{(1)}_{S_{\fA,1}} \> \in P_{i_*}^\cA$.
	On the other hand, for any sequence $\fc_n, \ldots, \fc_1$ picked by $\fB$ that does not comply with the rules of the described game, we have $\< \fc_n, \ldots, \fc_1 \> \not\in Q_{i_*}^\cA$.
	
	This result proves the following observation for any of the $\fb^{(k)}_{S}$: 
	\begin{itemize}
		\item[($*$)] removing $\fb^{(k)}_{S}$ from $\cA$'s domain and restricting the sets $Q_i^\cA$ to subsets of $\bigl(\fU_\cA \setminus \{\fb^{(k)}_{S}\} \bigr)^n$ yields a structure that does not satisfy $\varphi$.
	\end{itemize}
	The reason is simply that in this case player $\fA$ can always prevent $\fB$ from reaching a state of the game where $\fB$ can apply the described winning strategy.
	
	We have already analyzed the size of the sets $\cS_k$. Due to the observed lower bounds, we know that $\fU_\cA$ contains at least
		$\sum_{k=1}^{n}\twoup{k}{n}$
	elements of the form $\fb^{(k)}_{S}$.

	\medskip
	Next, we argue that any sentence $\varphi_*$ (in prenex normal form) that is semantically equivalent to $\varphi$ and starts with a quantifier prefix of the form $\exists^*\forall^*$ contains at least $\sum_{k=1}^{n}\twoup{k}{n}$ leading existential quantifiers.
		
	 Let $\varphi_* := \exists y_1 \ldots y_m \forall x_1 \ldots x_\ell. \chi_*$ (with $\chi_*$ being quanti\-fier-free) be a sentence with minimal $m$ that is semantically equivalent to $\varphi$. Since $\cA$ is also a model of $\varphi_*$, we know that there is a sequence of elements $\fc_1, \ldots, \fc_m$ taken from the domain $\fU_\cA$ such that $\cA, [y_1 \Mapsto \fc_1, \ldots, y_m \Mapsto \fc_m] \models \forall x_1 \ldots x_\ell. \chi_* $.
	Consequently, we can extend $\cA$ to a model $\cA_*$ (over the same domain) of the Skolemized formula $\varphi_{\text{Sk}} := \forall x_1 \ldots x_\ell. \chi_*\subst{y_1/c_1, \ldots, y_m/c_m}$ by adding $c_j^{\cA_*} := \fc_j$ for $j = 1, \ldots, m$. On the other hand, every model of the Skolemized formula $\varphi_{\text{Sk}}$ immediately yields a model of $\varphi_*$.
	
	The signature underlying $\varphi_{\text{Sk}}$ comprises exactly the constant symbols $c_1, \ldots, c_m$ and does not contain any other function symbols. 
	Suppose $m < \sum_{k=1}^{n}\twoup{k}{n}$. Hence, there is some $\fb^{(k)}_{S}$ with $S \in \cS_k$ such that for every $j$ we have $c_j^{\cA_*} \neq \fb^{(k)}_{S}$. By the Substructure Lemma, the following substructure $\cB$ of $\cA_*$ constitutes a model of $\varphi_{Sk}$: $\fU_\cB := \fU_{\cA_*} \setminus \{\fb^{(k)}_{S}\}$, $P_i^\cB := P_i^{\cA_*} \cap \fU_\cB^n = P_i^{\cA_*}$ and $Q_i^\cB := Q_i^{\cA_*} \cap \fU_\cB^n$ for every $i$, and $c_j^\cB := c_j^{\cA_*}$ for every $j$.
	
	But then $\cB$ must also be a model of both $\varphi_*$ and $\varphi$, since every model of $\varphi_{\text{Sk}}$ is a model of $\varphi_*$, and because we assumed $\varphi_*$ and $\varphi$ to be semantically equivalent.
	This contradicts Observation~($*$), and thus we must have $m \geq \sum_{k=1}^{n}\twoup{k}{n}$.
	
	Since every atom $Q_i(y_1, \ldots, y_n)$ contains $n$ variables from existential quantifier blocks that are interspersed with universal quantifier blocks, the degree $\degree_\varphi$ of $\varphi$ is $n$. 
	Moreover, $\varphi$ can easily be transformed into a CNF that is Horn and Krom at the same time.
	Hence, the theorem holds.
\end{proof}
\setcounter{theorem}{\value{auxTheorem}}

\section{Details regarding Section~\ref{section:ComputationalLowerBounds}}

The following theorems illustrate why the domino problems defined in Section~\ref{section:ComputationalLowerBounds} readily lend themselves to deriving lower bounds on worst-case computing time.
\begin{theorem}[\cite{Borger1997}, Theorem 6.1.2]
	Let $M$ be a \emph{simple} nondeterministic one-tape Turing machine with input alphabet $\Gamma$. Then there is a domino system $\fD = \<\cD, \cH, \cV\>$ and a linear-time reduction which takes any input $w \in \Gamma^*$ to a word $\bD \in \cD^*$ with $|w| = |\bD|$ such that
	\begin{itemize}
		\item If $M$ accepts $w$ in time $t_0$ with space $s_0$, then $\fD$ tiles $\Int_{s} \times \Int_{t}$ with initial condition $\bD$ for all $s \geq s_0 + 2$ and $t \geq t_0 + 2$;
		\item If $M$ does not accept $w$, then $\fD$ does not tile $\Int_s \times \Int_t$ with initial condition $\bD$ for any $s,t \geq 2$.
	\end{itemize}
\end{theorem}
By a \emph{simple} Turing machine the authors of \cite{Borger1997} mean a nondeterministic one-tape Turing machine $M$ over the input alphabet $\Gamma$ that meets the following conditions:
	\begin{quotation}
		``The alphabet of $M$ contains $\Gamma$ and at least one other symbol $\Box$ (blank). $M$ works on a semi-infinite tape and never tries to move left from the left-most tape cell. At every stage of the computation there is some $s$ such the tape cells $0, \ldots, s$ contain only non-blank symbols, all other tape cells contain $\Box$; in particular, to the right of a blank only other blanks may appear. Furthermore, we assume that $M$ has a unique accepting configuration: the machine is in the unique accepting state $q_a$, the tape contains only blanks and the head is in position $0$.
		
		These conditions do not restrict computational power. Every language accepted in time $T(n)$ and space $S(n)$ by some one-tape nondeterministic Turing machine is accepted within the same time and space bounds by a simple Turing machine, as long as $S(n), T(n) \geq 2n$.''
		\hfill\cite{Borger1997}, page 243
	\end{quotation}

\begin{theorem}[\cite{Borger1997}, Theorem 6.1.6]
	We call a function $T: \Nat \to \Nat$ \emph{time constructible} if there exists a deterministic Turing machine making precisely $T(n)$ steps on inputs of length $n$.

	Let $T: \Nat \to \Nat$ be a time-constructible function with $T(c' n)^2 \in o(T(n))$ for some constant $c' > 0$.
	There exists a domino system $\fD$ and a constant $c > 0$ such that $\Domino(\fD, T(n)) \not\in \Ntime(T(c n))$.
\end{theorem}


\newcommand{\Twoup}[1]{{2^{\uparrow #1}}}

Recall our notation for the \emph{tetration operation} $\twoup{k}{m}$, which we defined inductively: $\twoup{0}{m} := m$ and $\twoup{k+1}{m} := 2^{\left(\twoup{k}{m}\right)}$. In addition, we shall use the short-hand $\Twoup{k}$ to abbreviate $\twoup{k}{2}$.

Since we intend to plug in the function $\Twoup{n}$ for $T(n)$, we need to verify the condition $\bigl( \Twoup{c'n} \bigr)^2 \in o(\Twoup{n})$ for some positive constant $c'$. Setting $c' := \tfrac{1}{2}$ entails the following. 
\begin{lemma}
	For every positive constant $c > 0$ there exists some positive integer $n_0 > 0$ such that for every $n \geq n_0$ we have
		\[ \Bigl( \Twoup{\lceil n/2  \rceil} \Bigr)^2 \leq c\cdot \Twoup{n} ~. \]
\end{lemma}
\begin{proof}
	We distinguish two cases:
	\begin{description}
		\item Suppose $c \geq 1$.
			We observe 
			\begin{enumerate}[label=(\arabic{*}), ref=(\arabic{*})]
				\item $ \Bigl( \Twoup{\lceil n/2  \rceil} \Bigr)^2 		\;=\;	2^{2\cdot 2^{\Twoup{\lceil n/2  \rceil-2}}} \;=\; 2^{2^{1+\Twoup{\lceil n/2  \rceil-2}}}$, and
				\item $c\cdot \Twoup{n} 	\;\;\geq\;\;	 \Twoup{n} 	\;\;=\;\;	2^{2^{\Twoup{n-2}}}$.
			\end{enumerate}
			Hence, it suffices to show there is some $n_0 \geq 2$ such that $1+\Twoup{\lceil n/2  \rceil-2} \;\leq\; \Twoup{n-2}$ holds for every $n \geq n_0$.
			One possible choice is $n_0 = 4$
				
		\item Suppose $0 < c < 1$.
			We set $d := \tfrac{1}{c}$. Due to $d > 1$, $\log_2 d$ is defined.
			Moreover, we observe
			\begin{align*}
				d \cdot \Bigl( \Twoup{\lceil n/2  \rceil} \Bigr)^2 	
				\;&=\; 					2^{\log_2 d} \cdot \Bigl( \Twoup{\lceil n/2  \rceil} \Bigr)^2	\\
				&=\; 					2^{\log_2 d + 2\cdot \Twoup{\lceil n/2 \rceil -1}}	\\
				&\leq\;				 	2^{\log_2 d\cdot\Twoup{\lceil n/2 \rceil -1} + 2\cdot \Twoup{\lceil n/2 \rceil -1}}	\\
				&=\; 					2^{(\log_2 d + 2)\cdot \Twoup{\lceil n/2 \rceil -1}}	\\
				&=\; 					2^{2^{\log_2 (\log_2 d + 2)} \cdot 2^{\Twoup{\lceil n/2 \rceil -2}}}	\\
				&=\; 					2^{2^{\log_2 (\log_2 d + 2) + \Twoup{\lceil n/2 \rceil -2}}}	~.
			\end{align*}
			Hence, in order to prove that there is some $n_0$ such that $d \cdot \Bigl( \Twoup{\lceil n/2  \rceil} \Bigr)^2 \leq \Twoup{n}$ holds for every $n \geq n_0$, it suffices to show that there is some $n_0$ such that 
				\[ d' + \Twoup{\lceil n/2 \rceil -2} \;\;\leq\;\; \Twoup{n-2} \]			
			where $d' := \log_2 (\log_2 d + 2)$.
			The proof thus boils down to asking whether the difference $\Twoup{n-2} - \Twoup{\lceil n/2 \rceil - 2}$ exceeds any constant value $d'$ for sufficiently large $n$. This is certainly the case. 
			\qedhere
	\end{description}
\end{proof}

\begin{lemma}
	Let $k \geq 1$ be some positive integer.
	For every positive constant $c > 0$ there exists some positive integer $n_0 \geq 1$ such that for every $n \geq n_0$ we have
		\[ \Bigl( \twoup{k}{\lceil n/4  \rceil} \Bigr)^2 \leq c\cdot \twoup{k}{n} ~. \]
\end{lemma}
\begin{proof}
	\begin{description}
		\item Let $k = 1$.
			We observe 
				\[ \Bigl( \twoup{1}{n/4} \Bigr)^2 = 2^{2\cdot n/4} = 2^{n/2} ~.\]
			If $c \geq 1$, then $2^{n/2} \leq c\cdot 2^n$ is obvious.
			
			Assume $0 < c < 1$ and set $d := \tfrac{1}{c}$. Hence, $d > 1$.
			If remains to show $d \cdot 2^{n/2} \leq 2^n$ for every sufficiently large $n$.
			Due to $2^n = 2^{n/2 + n/2} = 2^{n/2} \cdot 2^{n/2}$, we observe $d \cdot 2^{n/2} \leq 2^{n/2} \cdot 2^{n/2}$ if and only if $d \leq 2^{n/2}$.
			But the latter certainly holds for sufficiently large $n$.
			
		\item Let $k = 2$.
			If $c \geq 1$, then $\twoup{2}{n} \leq c\cdot \twoup{2}{n}$.
			It thus suffices to show $2^{2^{n/4}} \leq 2^{2^n}$ or, equivalently, $\tfrac{1}{4} n + 1 \leq n$ for sufficiently large $n$.
			But this is obviously true.
			
			Assume $0 < c < 1$ and define $d := \tfrac{1}{c}$.
			It follows that $d > 1$.
			It remains to show $d \cdot 2^{2^{n/4+1}} \leq 2^{2^n}$ for sufficiently large $n$.
			Due to $2^{2^{n}} = 2^{2^{n/2} \cdot 2^{n/2}} \geq 2^{2^{n/2} + 2^{n/2}} = 2^{2^{n/2}} \cdot 2^{2^{n/2}}$
			with $d \leq 2^{2^{n/2}}$ and $2^{2^{n/4 + 1}} \leq 2^{2^{n/2}}$ for sufficiently large $n$, we also observe $d \cdot 2^{2^{n/4+1}} \leq 2^{2^{n/2}} \cdot 2^{2^{n/2}} \leq 2^{2^n}$ for sufficiently large $n$.
	\end{description}
	Analogous arguments holds for every $k \geq 2$.
\end{proof}

\section{Details regarding Section~\ref{section:EnforcingLargeModels}}

We investigate the properties of models of the SF sentences given in Section~\ref{section:EnforcingLargeModels}.
For the remainder of this subsection we therefore consider a model $\cA$ of the sentence $\psi_{1} \wedge \ldots \wedge \psi_{16}\wedge \chi_1 \wedge \chi_2 \wedge \chi_3$.

\begin{definition}
	We define the following sets and relations:
		\begin{itemize}
			\item $\cI_{\ell} := \bigl\{ \fa \in \fU_\cA \mid \cA, [j \Mapsto \fa] \models L(\ul{\ell}, j) \bigr\}$ for every $\ell = 0, \ldots, \kappa$.
			
			\item ${\prec_\ell} \subseteq \cI_{\ell} \times \cI_{\ell}$ for every $\ell = 0, \ldots, \kappa$ such that $\fa \prec_\ell \fa'$ holds if and only if $\cA, [j \Mapsto \fa, j' \Mapsto \fa'] \models \Succ(\ul{\ell}, j, j')$.
			
			\item $\cF_{\ell} := \{f : \cI_\ell \to \{0,1\} \mid \text{$f$ is total and for every $\fa$ for which $\cA, [j \Mapsto \fa] \models \MaxIdx(\ul{\ell}, j)$ we}$ \linebreak
				\strut\hspace{27.5ex} $\text{have that $f(\fa) = 1$ entails $f(\fb) = 0$ for every $\fb \neq \fa$} \}$ \\
				for every $\ell = 0, \ldots, \kappa-1$.
				
			\item For every $\ell = 0, \ldots, \kappa-1$ and two total functions $f, g : \cI_{\ell} \to \{0,1\}$ we write $f \sqsubset_\ell g$ if and only if 
				\begin{itemize}
					\item there is some integer $p$ and a unique chain $\fa_1 \prec_\ell \ldots \prec_\ell \fa_p$ comprising all elements in $\cI_\ell$, and
					\item incrementing the bit string $f(\fa_1) f(\fa_2) \ldots f(\fa_p)$ (interpreted as number encoded in binary where the leftmost bit is the least significant one) by one yields $g(\fa_1) g(\fa_2) \ldots g(\fa_p)$.
				\end{itemize}	
				
			\item $\cS_{\ell, f} := \bigl\{ \fa \in \cI_\ell \mid \text{$\cA, [j \Mapsto \fa, i \Mapsto \fb] \models$}$ $\text{$J(\ul{\ell}, j, i, f(\fb))$ for every $\fb \in \cI_{\ell-1}$} \bigr\}$\\
				 for every $\ell = 1, \ldots, \kappa$ and every $f \in \cF_{\ell-1}$.
			
			\item For every $\ell = 1, \ldots, \kappa$ and two elements $\fa, \fa' \in \cI_\ell$ we write $\fa \sim_\ell \fa'$ if and only if for every $\fb \in \cI_{\ell-1}$ we observe
				\begin{itemize}
					\item $\cA, [j \Mapsto \fa, i \Mapsto \fb] \models J(\ul{\ell}, j, i, 0)$ if and only if $\cA, [j' \Mapsto \fa', i \Mapsto \fb] \models J(\ul{\ell}, j', i, 0)$ and
					\item $\cA, [j \Mapsto \fa, i \Mapsto \fb] \models J(\ul{\ell}, j, i, 1)$ if and only if $\cA, [j' \Mapsto \fa', i \Mapsto \fb] \models J(\ul{\ell}, j', i, 1)$.
				\end{itemize}	
		\end{itemize}	
\end{definition}

\begin{proposition}
	For every $\ell$, $1 \leq \ell \leq \kappa$, $\sim_\ell$ is an equivalence relation.
\end{proposition}

\begin{proposition}
	For all distinct $\ell, \ell'$, $0 \leq \ell, \ell' \leq \kappa$, $\cI_\ell$ and $\cI_{\ell'}$ are disjoint.
\end{proposition}

\begin{lemma}\label{lemma:SuccEntailsProgressInSlice}
	Let $\fa, \fa' \in \cI_\ell$ for some $\ell$, $1\leq \ell \leq \kappa$.
	Let $f, g : \cI_{\ell-1} \to \{0,1\}$ be two total functions such that for every $\fb \in \cI_{\ell-1}$ we have $\cA, [j \Mapsto \fa, i \Mapsto \fb] \models J(\ul{\ell}, j, i, f(\fb))$ and $\cA, [j' \Mapsto \fa', i \Mapsto \fb] \models J(\ul{\ell}, j', i, g(\fb))$.
	Moreover, assume there is a unique chain $\fc_1 \prec_{\ell-1} \ldots \prec_{\ell-1} \fc_p$ comprising all elements in $\cI_{\ell-1}$.
	
	Then $\fa \prec_\ell \fa'$ implies $f \sqsubset_{\ell-1} g$.
\end{lemma}	
\begin{proof}
	Since $\fc_1$ is the only element in $\cI_{\ell-1}$ for which there is no element $\fc' \in \cI_{\ell-1}$ with $\fc' \prec_{\ell-1} \fc_1$, $\cA \models \psi_2 \wedge \psi_3$ implies $\cA, [i \mapsto \fc_1] \models \MinIdx(\ul{\ell-1}, i)$.	
	
	Let $f^* : \cI_{\ell-1} \to \{0,1\}$ be defined such that 
		\begin{itemize}
			\item $f^*(\fb) = 0$ if and only if $\cA, [j \Mapsto \fa, i \Mapsto \fb] \models J^*(\ul{\ell}, j, i, 0)$ and
			\item $f^*(\fb) = 1$ if and only if $\cA, [j \Mapsto \fa, i \Mapsto \fb] \models J^*(\ul{\ell}, j, i, 1)$.
		\end{itemize}	
	This function is well-defined because of $\cA \models \psi_{13}$.
	Due to $\cA \models \chi_2 \wedge \psi_{14} \wedge \psi_{15}$ it is also total.
	
	$\cA \models \psi_{14}$ enforces $\cA, [j \Mapsto \fa, i \Mapsto \fc_1] \models J^*(\ul{\ell}, j, i, 1)$, i.e.\ $f^*(\fc_1) = 1$.
	Moreover, for any $k$ with $1 < k \leq p$, we have $f^*(\fc_k) = 1$ if and only if $f(\fc_1) = \ldots = f(\fc_{k-1}) = 1$, because of $\cA \models \psi_{15}$.
	
	$\cA \models \psi_{9}$ together with our assumption $\fa \prec_\ell \fa'$ translates to the following property, which we phrase in terms of operations on bits:
	\begin{itemize}
		\item[($**$)] for every $k$, $1\leq k\leq p$, we observe $g(\fc_k) = f(\fc_k) \oplus f^*(\fc_k)$
	\end{itemize}
	where $\oplus$ denotes exclusive or. But this corresponds to an increase of the bit string $f(\fc_1) \ldots f(\fc_p)$ by one (where $f(\fc_1)$ is the least significant bit).
	
	Hence, $f \sqsubset_{\ell-1} g$.
\end{proof}

\begin{lemma}\label{lemma:SameSliceEntailsEquivalence}
	Let $\fa, \fa' \in \cI_\ell$ for some $\ell$, $1\leq \ell \leq \kappa$.
	
	If $\fa$ and $\fa'$ belong to the same $\cS_{\ell, f}$ for some function $f \in \cF_{\ell-1}$, then $\fa \sim_\ell \fa'$.	
\end{lemma}
\begin{proof}
	By totality of $f$, it follows that for every $\fb \in \cI_{\ell-1}$ we have $\cA, [j \Mapsto \fa, i \Mapsto \fb] \models J(\ul{\ell}, j, i, f(\fb))$ and $\cA, [j' \Mapsto \fa', i \Mapsto \fb] \models J(\ul{\ell}, j', i, f(\fb))$.
	Moreover, $\cA \models \psi_{13}$ entails $\cA, [j \Mapsto \fa, i \Mapsto \fb] \not\models J(\ul{\ell}, j, i, \overline{f(\fb)})$ and $\cA, [j' \Mapsto \fa', i \Mapsto \fb] \not\models J(\ul{\ell}, j', i, \overline{f(\fb)})$ where $\overline{f(\fb)} := 1 - f(\fb)$ is the complement of $f(\fb)$.
	This results in $\fa \sim_\ell \fa'$.
\end{proof}

\begin{lemma}\label{lemma:EquivalenceEntailsSameSlice}
	Let $\fa, \fa' \in \cI_\ell$ for some $\ell$, $1\leq \ell \leq \kappa$.
	Moreover, let $f$ be some function in $\cF_{\ell-1}$.
		
	If $\fa$ belongs to $\cS_{\ell, f}$ and we have $\fa \sim_\ell \fa'$, then $\fa' \in \cS_{\ell, f}$.
\end{lemma}
\begin{proof}
	By totality of $f$, it follows that for every $\fb \in \cI_{\ell-1}$ we have $\cA, [j \Mapsto \fa, i \Mapsto \fb] \models J(\ul{\ell}, j, i, f(\fb))$.
	Moreover, $\cA \models \psi_{13}$ entails $\cA, [j \Mapsto \fa, i \Mapsto \fb] \not\models J(\ul{\ell}, j, i, \overline{f(\fb)})$ where $\overline{f(\fb)} := 1 - f(\fb)$ is the complement of $f(\fb)$.
	
	Since we assume $\fa \sim_\ell \fa'$, we know that these properties transfer to $\fa'$. 
	Hence,  for every $\fb \in \cI_{\ell-1}$ we observe $\cA, [j' \Mapsto \fa', i \Mapsto \fb] \models J(\ul{\ell}, j', i, f(\fb))$ and $\cA, [j' \Mapsto \fa', i \Mapsto \fb] \not\models J(\ul{\ell}, j', i, \overline{f(\fb)})$.
	Consequently, $\fa' \in \cS_{\ell,f}$.
\end{proof}

\begin{lemma}\label{lemma:ArrangeFunctions}
	Consider some $\ell$, $0 \leq \ell \leq \kappa-1$.
	If there is a unique chain $\fa_1 \prec_\ell \ldots \prec_\ell \fa_p$ comprising all elements in $\cI_\ell$, then we can uniquely arrange all functions in $\cF_\ell$ into a chain
	$f_1 \sqsubset_\ell \ldots \sqsubset_\ell f_{p'}$ where $p' = |\cF_\ell| = 2^{p-1}+1$.
\end{lemma}
\begin{proof}
	Let $\{0,1\}^p$ be the set of all bit strings of length $p$.
	If we interpret each of them as a number encoded in binary (where we assume the leftmost bit to be the least significant one), we can uniquely arrange the $2^{p-1}+1$ smallest bit strings in $\{0,1\}^p$ into a chain
		\[ \overline{b}_0 < \overline{b}_1 < \ldots < \overline{b}_{2^{p-1}-1} < \overline{b}_{2^{p-1}} \]
	where the indices reflect the represented numerical value and $<$ is intended to be the usual ordering based on this value.
	Since we assume the rightmost bit to be the most significant one, it is $0$ in $\overline{b}_0, \ldots, \overline{b}_{2^{p-1}-1}$.
	Accordingly, $\overline{b}_{2^{p-1}}$ is the bit string with all zeros except for the most significant bit, i.e.\ $\overline{b}_{2^{p-1}} = 0 \ldots 0 1$.
	
	Obviously, the following mapping $\mu$ induces a one-to-one correspondence between bit strings and all the mappings in $\cF_{\ell}$:
	$\mu(f) := f(\fa_1) f(\fa_2) \ldots f(\fa_p)$.			
	By definition of $\sqsubset_\ell$, we have $f \sqsubset_\ell g$ if and only if $\mu(f) + 1 = \mu(g)$.
	
	Consequently, we obtain the chain $\mu^{-1}(\overline{b}_0) \sqsubset_\ell \mu^{-1}(\overline{b}_1) \sqsubset_\ell \ldots \sqsubset_\ell \mu^{-1}(\overline{b}_{2^{p-1}-1}) \sqsubset_\ell \mu^{-1}(\overline{b}_{2^{p-1}})$.		
\end{proof}
		
\begin{lemma}\label{lemma:IndexSetsProperties:One}
	We observe that $\cI_0 = \{c_1^\cA, \ldots, c_\mu^\cA\}$ with $|\cI_0| = \mu$.
	Moreover, there is a unique chain $c_1^\cA \prec_0 \ldots \prec_0 c_\mu^\cA$.
\end{lemma}
\begin{proof}
	$\cA \models \chi_1$ entails that $\cI_0 \subseteq \{c_1^\cA, \ldots, c_\mu^\cA\}$.
	Due to $\cA \models \psi_8$ we have $c_1^\cA \prec_0 \ldots \prec_0 c_\mu^\cA$.
	By $\cA \models \psi_6$, $\fa \prec_0 \fb$ entails $\fa, \fb \in \cI_0$.
	Hence, $\{c_1^\cA, \ldots, c_\mu^\cA\} \subseteq \cI_0$
	
	We next show that all $c_1^\cA, \ldots, c_\mu^\cA$ are pairwise distinct.
	\begin{description}
		\item\underline{Claim:} For every index $j \geq 2$ the first $j$ elements $c_1^\cA, \ldots, c_j^\cA$ are distinct.
		\item\underline{Proof:}
			We proceed by induction on $j$.
	
			For $j=2$, $c_1^\cA \neq c_2^\cA$ must hold, for otherwise $c_1^\cA \prec_0 c_2^\cA$ contradicts $\cA \models \psi_7$ which entails $\cA \models \forall j.\; \neg \Succ(\ul{0}, j, j)$.
			
			Let $j \geq 3$ and assume, by induction, that the elements $c_1^\cA, \ldots, c_{j-1}^\cA$ are all pairwise distinct.
			Suppose there is some index $i_*$, $1\leq i_* < j$, such that $c_{i_*}^\cA = c_j^\cA$.
			We distinguish two cases:
			In case $i_* = 1$, we have $c_{j-1}^\cA \prec_0 c_1^\cA$. 
			But this contradicts $\cA \models \psi_2$.
			
			In case of $i_* > 1$, $\cA \models \psi_7$ entails $c_{i_*-1}^\cA = c_{j-1}^\cA$, since we have $c_{i_*-1}^\cA \prec_0 c_{i_*}^\cA$ and $c_{j-1}^\cA \prec_0 c_{j}^\cA$, and since we assumed $c_{i_*}^\cA = c_j^\cA$.
			But this contradicts our inductive hypothesis, because $i_*-1$ and $j-1$ are distinct indices and thus the inductive hypothesis implies that $c_{i_*-1}^\cA$ and $c_{j-1}^\cA$ are distinct.			
			\strut\hfill$\Diamond$
	\end{description}
	The above claim entails $|\cI_0| = \mu$.
	
	$\cA \models \psi_8$ entails $c_1^\cA \prec_0 \ldots \prec_0 c_\mu^\cA$.
	By the above arguments, we know that this chain comprises all elements in $\cI_0$.
	Moreover, due to $\cA \models \psi_7$, this chain is the only chain satisfying the desired properties.	
\end{proof}

\begin{lemma}\label{lemma:IndexSetsProperties:Two}
	For every $\ell = 1, \ldots, \kappa$ the following properties are satisfied:
	\begin{enumerate}[label=(\roman{*}), ref=(\roman{*})]
		\item\label{enum:IndexSetsProperties:Two:Ia} For all $\fa, \tfa \in \cI_{\ell}$ for which $\cA, [j \Mapsto \fa, \tj \Mapsto \tfa] \models \eq{\ell}_{j, \tj}$ it follows $\fa \sim_\ell \tfa$.

		\item\label{enum:IndexSetsProperties:Two:Ib} All $\fa, \tfa \in \cI_{\ell}$ with  $\fa \sim_\ell \tfa$ satisfy $\cA, [j \Mapsto \fa, \tj \Mapsto \tfa] \models \eq{\ell}_{j, \tj}$.
						
		\item\label{enum:IndexSetsProperties:Two:II} For every $f \in \cF_{\ell-1}$ the set $\cS_{\ell,f}$ is nonempty.
		
		\item\label{enum:IndexSetsProperties:Two:III} $\cI_\ell = \bigcup_{f \in \cF_{\ell-1}} \cS_{\ell, f}$.
		
		\item\label{enum:IndexSetsProperties:Two:IV} For every $f \in \cF_{\ell-1}$ the set $\cS_{\ell, f}$ contains exactly one element.
		
		\item\label{enum:IndexSetsProperties:Two:V}	There is a unique chain $\fa_1 \prec_\ell \ldots \prec_\ell \fa_p$ comprising all elements in $\cI_\ell$, and $\cA, [j \Mapsto \fa_1] \models \MinIdx(\ul{\ell}, j)$ and $\cA, [j' \Mapsto \fa_p] \models \MaxIdx(\ul{\ell}, j')$ hold.
	\end{enumerate}	
\end{lemma}
\begin{proof}
	We proceed by induction on $\ell$.	
	\begin{description}
		\item Base case: $\ell = 1$.
			\begin{description}
				\item \textit{Ad \ref{enum:IndexSetsProperties:Two:Ia}.}
					Due to the assumption $\cA, [j \Mapsto \fa, \tj \Mapsto \tfa] \models \eq{1}_{j,\tj}$, the construction of $\eq{1}_{j,\tj}$ entails
						\begin{align*}
							\cA, &[j \Mapsto \fa, \tj \Mapsto \tfa] \models \bigwedge_{i = 1}^\mu \Bigl( \bigl( J(\ul{1}, j, c_i, 0) \leftrightarrow J(\ul{1}, \tj, c_i, 0) \bigr) \wedge \bigl( J(\ul{1}, j, c_i, 1) \leftrightarrow J(\ul{1}, \tj, c_i, 1) \bigr) \Bigr) ~.
						\end{align*}	
					By Lemma~\ref{lemma:IndexSetsProperties:One}, we know that $\fb \in \cI_0$ entails $\fb = c_i^\cA$ for some $i$, $1 \leq i \leq \mu$.
					Consequently, for every $\fb \in \cI_0$, $\cA, [j \Mapsto \fa, i \Mapsto \fb] \models J(\ul{1}, j, i, 0)$ holds if and only if $\cA, [\tj \Mapsto \tfa, i \Mapsto \fb] \models J(\ul{1}, \tj, i, 0)$ does.
					This entails $\fa \sim_\ell \tfa$, because of $\cA \models \psi_{13} \wedge \chi_2$.

				\item \textit{Ad \ref{enum:IndexSetsProperties:Two:Ib}.}
					By definition of $\sim_1$, $\fa \sim_1 \tfa$ entails that for every $\fb \in \cI_{0}$ we have
					\begin{itemize}
						\item $\cA, [j \Mapsto \fa, i \Mapsto \fb] \models J(\ul{1}, j, i, 0)$ if and only if $\cA, [\tj \Mapsto \tfa, i \Mapsto \fb] \models J(\ul{1}, \tj, i, 0)$ and
						\item $\cA, [j \Mapsto \fa, i \Mapsto \fb] \models J(\ul{1}, j, i, 1)$ if and only if $\cA, [\tj \Mapsto \tfa, i \Mapsto \fb] \models J(\ul{1}, \tj, i, 1)$.
					\end{itemize}
					Moreover, Lemma~\ref{lemma:IndexSetsProperties:One} states that $\cI_0 = \{c_1^\cA, \ldots, c_\mu^\cA\}$.
					Since $\fa, \tfa$ belong to $\cI_1$, we conclude $\cA, [j \Mapsto \fa, \tj \Mapsto \tfa] \models \eq{1}_{j,\tj}$.

				\item \textit{Ad \ref{enum:IndexSetsProperties:Two:II}.}					
					$\cA \models \psi_8$ entails $c_1^\cA \prec_0 \ldots \prec_0 c_\mu^\cA$.
					By Lemma~\ref{lemma:IndexSetsProperties:One}, we know that this chain comprises all elements in $\cI_0$.
					Moreover, due to $\cA \models \psi_7$, this chain is the only chain satisfying this property.					
					 Hence, by Lemma~\ref{lemma:ArrangeFunctions}, we can arrange all functions in $\cF_0$ into a sequence $f_1 \sqsubset_0 \ldots \sqsubset_0 f_p$ for $p = 2^{\mu-1}+1$. Clearly, $f_1$ maps every element $\fb\in \cI_0$ to $f_1(\fb) = 0$, and $f_p$ maps every element $\fb\in \cI_0 \setminus \{c_\mu^\cA\}$ to $f_p(\fb) = 0$ and $c_\mu^\cA$ to $f_p(c_\mu^\cA) = 1$.
					By $\cA \models \psi_{3} \wedge \psi_{10}$, we know that $\cA, [j \Mapsto d_{1}^\cA, i \Mapsto \fb] \models J(\ul{1}, j, i, 0)$ for every $\fb \in \cI_0$. Hence, $d_{1}^\cA \in S_{1, f_1}$.
					
					We next show that for every $k$, $1 \leq k < p$, if $S_{1, f_k}$ is nonempty, then $S_{1, f_{k+1}}$ is nonempty.
					Let $\fa$ be an element of $S_{1, f_k}$.
					Because of $k < p$, we know that $\cA, [j \Mapsto \fa, i \Mapsto c_\mu^\cA] \models J(\ul{1}, j, i, 0)$.
					By virtue of \ref{enum:IndexSetsProperties:Two:Ia} and due to $\cA \models \psi_{16}$ we conclude that there are elements $\tfa, \tfa' \in \cI_1$ such that $\fa \sim_1 \tfa$ and $\tfa \prec_1 \tfa'$. 
					By Lemma~\ref{lemma:SuccEntailsProgressInSlice}, this results in $\tfa' \in \cS_{\ell, f_{k+1}}$.

				\item \textit{Ad \ref{enum:IndexSetsProperties:Two:III}.}
					As a consequence of $\cA \models \psi_{13}$
					there is a unique partial function $g_\fa : \cI_0 \to \{0,1\}$ for every $\fa \in \cI_1$ such that for every $\fb \in \cI_0$ we have $\cA, [j \Mapsto \fa, i \Mapsto \fb] \models J(\ul{1}, j, i, g_\fa(\fb))$ if and only if $g_\fa$ is defined for $\fb$. Because of $\cA \models \chi_2$, we know that $g_\fa$ must be total.
					
					For every $\fa \in \cI_1$ where $g_\fa(c_\mu^\cA) = 0$ we have $g_\fa \in \cF_0$ and thus also $\fa \in \cS_{1, g_\fa}$.
					
					Because of $\cA \models \psi_{5} \wedge \psi_{11} \wedge \psi_{12}$ we know that $\cA \models \MaxIdx(\ul{1}, e_{1})$ and that $e_{1}^\cA$ is the only element in $\cI_1$ for which $\cA, [j \Mapsto e_{1}^\cA, i \Mapsto c_\mu^\cA] \models J(\ul{1}, j, i, 1)$.
					
					It remains to show that for every $\fb \in \cI_0$ with $\fb \neq c_\mu^\cA$ we have $\cA, [j \Mapsto e_{1}^\cA, i \Mapsto \fb] \models J(\ul{1}, j, i, 0)$.
					But this is a consequence of \ref{enum:IndexSetsProperties:Two:II} and the fact that the total function $f_*$ mapping all elements $\fb$ in $\cI_0$ but $c_\mu^\cA$ to $f_*(\fb) = 0$ belongs to $\cF_0$, and thus $\cS_{1, f_*}$ is nonempty. In particular, $\cS_{1, f_*} = \{e_{1}^\cA\}$.
					
					Consequently, $\cI_1$ cannot contain any elements that do not lie in $\bigcup_{f \in \cF_{0}} \cS_{1, f}$.
					
				\item \textit{Ad \ref{enum:IndexSetsProperties:Two:IV}.}
					Consider any set $\cS_{1,f}$. By virtue of \ref{enum:IndexSetsProperties:Two:II}, $\cS_{1,f}$ contains at least one element.
					Suppose we are given two elements $\fa, \fa'$ in $\cS_{1,f}$.
					By virtue of Lemma~\ref{lemma:SameSliceEntailsEquivalence}, this means $\fa \sim_1 \fa'$.
					
					Because of $\cA \models \chi_3$, there are two elements $\tfa, \tfa' \in \cI_1$ for which \ref{enum:IndexSetsProperties:Two:Ia} entails $\fa \sim_1 \tfa$ and $\fa' \sim_1 \tfa'$.
					By symmetry and transitivity of $\sim_1$, we have $\tfa \sim_1 \tfa'$.
					Hence, we observe 
						\begin{align*}
							\cA, &[\tj \Mapsto \tfa, \tj' \Mapsto \tfa'] \models \forall \ti.\; L(\ul{\ell-1}, \ti) \rightarrow \bigl( J(\ul{\ell}, \tj, \ti, 0) \leftrightarrow J(\ul{\ell}, \tj', \ti, 0) \bigr) ~. 
						\end{align*}		
					Consequently, $\cA \models \chi_3$ leads to $\fa = \fa'$.					
				
				\item \textit{Ad \ref{enum:IndexSetsProperties:Two:V}.}
					As we have seen in the proof of Lemma~\ref{lemma:ArrangeFunctions}, we can define a bijective mapping $\mu$ that maps the functions in $\cF_{0}$ to bit strings of length $\mu$ that either have a $0$ as most significant bit or correspond to $0 \ldots 0 1$ (where the most significant bit is the rightmost one).
					
					Let $p := 2^{\mu-1}+1$.
					By virtue of Lemma~\ref{lemma:ArrangeFunctions}, we can uniquely construct a chain
						\[ f_1 \sqsubset_0 f_2 \sqsubset_0 \ldots \sqsubset_0 f_p \]
					comprising all functions in $\cF_0$. 
					
					Properties \ref{enum:IndexSetsProperties:Two:III} and \ref{enum:IndexSetsProperties:Two:IV} together yield that $\cI_1 = \{ \fa_1, \ldots, \fa_p \}$ where $\fa_k \in \cS_{1,f_k}$ for every $k = 1, \ldots, p$.
					Lemma~\ref{lemma:SuccEntailsProgressInSlice} says that for any $\fa_k, \fa_{k'}$ with $\fa_k \prec_1 \fa_{k'}$ we also observe $f_k \sqsubset_0 f_{k'}$.
					By definition of $\sqsubset_0$ and the fact that all the $f_k, f_{k'}$ are distinct, it clearly holds that $f_k \sqsubset_0 f_{k'}$ implies $k' = k+1$ and $1\leq k < k' \leq p$.
					Hence, $\fa_k \prec_1 \fa_{k'}$ can only hold if $k' = k+1$ and $1\leq k < k' \leq p$.

					Consider any element $\fa \in \cI_1$ for which $\cA, [j \Mapsto \fa, i \Mapsto c_\mu^\cA] \models J(\ul{1}, j, i, 0)$.
					By $\cA \models \psi_{16}$, we know that there are elements $\tfa, \tfa' \in \cI_1$ such that $\tfa \prec_1 \tfa'$ and $\cA, [j \Mapsto \fa, \tj \Mapsto \tfa] \models \eq{1}_{j,\tj}$. 
					By \ref{enum:IndexSetsProperties:Two:Ia}, the latter translates to $\fa \sim_1 \tfa$. 
					
					Let $g, \widetilde{g}$ be functions such that $\fa \in \cS_{1, g}$ and $\tfa \in \cS_{1, \widetilde{g}}$.
					Such functions exist by virtue of \ref{enum:IndexSetsProperties:Two:III}.
					However, $\fa \sim_1 \tfa$ entails $g = \widetilde{g}$, by Lemma~\ref{lemma:EquivalenceEntailsSameSlice}.
					But then \ref{enum:IndexSetsProperties:Two:IV} leads to $\fa = \tfa$.
					Consequently, we have $\fa \prec_1 \tfa'$.					
					By \ref{enum:IndexSetsProperties:Two:IV}, this means that all but one element in $\cI_1$ must have a successor in $\cI_1$ and all but one elements in $\cI_1$ are successor in $\cI_1$.
					Hence, we obtain the chain $\fa_1 \prec_1 \fa_2 \prec_1 \ldots \prec_1 \fa_{p-1} \prec_1 \fa_p$ where $\fa_1 = d_1^\cA$ and $\fa_p = e_1^\cA$.
			\end{description}
			
		\item Inductive case $\ell > 1$.
			\begin{description}
				\item \textit{Ad \ref{enum:IndexSetsProperties:Two:Ia}.}
					Due to the assumption $\cA, [j \Mapsto \fa, \tj \Mapsto \tfa] \models \eq{\ell}_{j,\tj}$ with $\ell > 1$, the construction of $\eq{\ell}_{j,\tj}$ entails
						\begin{align*}
							 \cA, [j \Mapsto \fa, \tj \Mapsto \tfa] \models\;\;  
							 	&\forall i.\, L(\ul{\ell\!-\!1}, i) \;\;\rightarrow\;\; \exists \ti.\; L(\ul{\ell\!-\!1}, \ti) \wedge \eq{\ell\!-\!1}_{i,\ti} \\
								&\hspace{23ex} \wedge \Bigl( \bigl( J(\ul{\ell}, j, i, 0) \leftrightarrow J(\ul{\ell}, \tj, \ti, 0) \bigr) \\
								&\hspace{26ex} \wedge \bigl( J(\ul{\ell}, j, i, 1) \leftrightarrow J(\ul{\ell}, \tj, \ti, 1) \bigr) \Bigr) ~.
						\end{align*}
					By inductive application of \ref{enum:IndexSetsProperties:Two:Ia}, $\cA, [i \Mapsto \fb, \ti \Mapsto \tfb] \models \eq{\ell-1}_{i,\ti}$ entails $\fb \sim_{\ell-1} \tfb$. 
					Inductive application of \ref{enum:IndexSetsProperties:Two:III} implies that $\fb \in \cS_{\ell-1, f}$ for some $f \in \cF_{\ell-1}$.
					By Lemma~\ref{lemma:EquivalenceEntailsSameSlice} together with  $\fb \sim_{\ell-1} \tfb$, we conclude $\tfb \in \cS_{\ell-1, f}$.
					Now, inductive application of \ref{enum:IndexSetsProperties:Two:IV} leads to $\fb = \tfb$.
					This means we in fact have 
						\begin{align*}
							 \cA, [j \Mapsto \fa, \tj \Mapsto \tfa] \models\;\; 
							 	&\forall i.\; L(\ul{\ell\!-\!1}, i) \rightarrow \Bigl( \bigl( J(\ul{\ell}, j, i, 0) \leftrightarrow J(\ul{\ell}, \tj, \ti, 0) \bigr) \\
								&\hspace{17ex} \wedge \bigl( J(\ul{\ell}, j, i, 1) \leftrightarrow J(\ul{\ell}, \tj, \ti, 1) \bigr) \Bigr) ~.
						\end{align*}
					In other words, $\fa \sim_{\ell} \tfa$.

				\item \textit{Ad \ref{enum:IndexSetsProperties:Two:Ib}.}
					By definition of $\sim_\ell$, $\fa \sim_\ell \tfa$ entails that for every $\fb \in \cI_{\ell-1}$ we have
					\begin{itemize}
						\item $\cA, [j \Mapsto \fa, i \Mapsto \fb] \models J(\ul{1}, j, i, 0)$ if and only if $\cA, [\tj \Mapsto \tfa, i \Mapsto \fb] \models J(\ul{1}, \tj, i, 0)$ and
						\item $\cA, [j \Mapsto \fa, i \Mapsto \fb] \models J(\ul{1}, j, i, 1)$ if and only if $\cA, [\tj \Mapsto \tfa, i \Mapsto \fb] \models J(\ul{1}, \tj, i, 1)$.
					\end{itemize}
					By \ref{enum:IndexSetsProperties:Two:Ia} and the fact that $\sim_{\ell-1}$ is an equivalence relation and thus $\fb \sim_{\ell-1} \fb$, we conclude $\cA, [i \Mapsto \fb, \ti \Mapsto \fb] \models \eq{\ell-1}_{i,\ti}$ for every $\fb \in \cI_{\ell-1}$.
					Consequently, 
						$\cA, [j \Mapsto \fa, \tj \Mapsto \tfa] \models \eq{\ell}_{j,\tj}$.					

				\item \textit{Ad \ref{enum:IndexSetsProperties:Two:II}.}		
					By inductive application of \ref{enum:IndexSetsProperties:Two:V}, we know that there is a unique chain $\fb_1 \prec_{\ell-1} \ldots \prec_{\ell-1} \fb_p$ comprising all elements in $\cI_{\ell-1}$. Moreover, we observe $\cA, [i \Mapsto \fb_1] \models \MinIdx(\ul{\ell-1}, i)$ and $\cA, [i \Mapsto \fb_p] \models \MaxIdx(\ul{\ell-1}, i)$.
					Hence, by Lemma~\ref{lemma:ArrangeFunctions}, we can arrange all mappings in $\cF_{\ell-1}$ into a sequence $f_1 \sqsubset_{\ell-1} \ldots \sqsubset_{\ell-1} f_{p'}$ where $p' = 2^{p-1}+1$.
					 
					Clearly, $f_1$ maps every element $\fb\in \cI_{\ell-1}$ to $f_1(\fb) = 0$, and $f_{p'}$ maps every element $\fb_k$ with $k < p$ to $f_{p'}(\fb_k) = 0$ and $\fb_p$ to $f_{p'}(\fb_p) = 1$.
					By $\cA \models \psi_{3} \wedge \psi_{10}$, we know that $\cA, [j \Mapsto d_\ell^\cA, i \Mapsto \fb] \models J(\ul{\ell}, j, i, 0)$ for every $\fb \in \cI_{\ell-1}$. Hence, $d_\ell^\cA \in S_{\ell, f_1}$.
					
					We next show that for every $k$, $1 \leq k < p'$, if $S_{\ell, f_k}$ is nonempty, then $S_{\ell, f_{k+1}}$ is nonempty.
					Let $\fa$ be an element of $S_{\ell, f_k}$.
					Because of $k < p'$, we know that $\cA, [j \Mapsto \fa, i \Mapsto \fb_p] \models J(\ul{\ell}, j, i, 0)$.
					By virtue of \ref{enum:IndexSetsProperties:Two:Ia} and due to $\cA \models \psi_{16}$ we conclude that there are elements $\tfa, \tfa' \in \cI_\ell$ such that $\fa \sim_\ell \tfa$ and $\tfa \prec_\ell \tfa'$. 
					Moreover, Lemma~\ref{lemma:EquivalenceEntailsSameSlice} leads to $\tfa \in S_{\ell, f_k}$.
					By Lemma~\ref{lemma:SuccEntailsProgressInSlice}, this results in $\tfa' \in \cS_{\ell, f_{k+1}}$.					
															
				\item \textit{Ad \ref{enum:IndexSetsProperties:Two:III}.}
					As a consequence of $\cA \models \psi_{13}$
					there is a unique partial mapping $g_\fa : \cI_{\ell-1} \to \{0,1\}$ for every $\fa \in \cI_\ell$ such that for every $\fb \in \cI_{\ell-1}$ we have $\cA, [j \Mapsto \fa, i \Mapsto \fb] \models J(\ul{\ell}, j, i, g_\fa(\fb))$ if and only if $g_\fa$ is defined for $\fb$. Because of $\cA \models \chi_2$, we know that $g_\fa$ must be total.
					
					For every $\fa \in \cI_\ell$ where $g_\fa(e_{\ell-1}^\cA) = 0$ we have $g_\fa \in \cF_{\ell-1}$ and thus also $\fa \in \cS_{\ell, g_\fa}$.
					
					Because of $\cA \models \psi_{5} \wedge \psi_{11} \wedge \psi_{12}$ we know that $\cA \models \MaxIdx(\ul{\ell}, e_\ell)$ and that $e_\ell^\cA$ is the only element in $\cI_\ell$ for which $\cA, [j \Mapsto e_\ell^\cA, i \Mapsto e_{\ell-1}^\cA] \models J(\ul{\ell}, j, i, 1)$.
					
					It remains to show that for every $\fb \in \cI_{\ell-1}$ with $\fb \neq e_{\ell-1}^\cA$ we have $\cA, [j \Mapsto e_\ell^\cA, i \Mapsto \fb] \models J(\ul{\ell}, j, i, 0)$.
					But this is a consequence of \ref{enum:IndexSetsProperties:Two:II} and the fact that the total function $f_*$ mapping all elements $\fb$ in $\cI_{\ell-1}$ but $e_{\ell-1}^\cA$ to $f_*(\fb) = 0$ belongs to $\cF_{\ell-1}$, and thus $\cS_{\ell, f_*}$ is nonempty. In particular, $\cS_{\ell, f_*} = \{e_\ell^\cA\}$.
					
					Consequently, $\cI_\ell$ cannot contain any elements that do not lie in $\bigcup_{f \in \cF_{\ell-1}} \cS_{\ell, f}$.
					
				\item \textit{Ad \ref{enum:IndexSetsProperties:Two:IV}.}
					Consider any set $\cS_{\ell,f}$. By virtue of \ref{enum:IndexSetsProperties:Two:II}, $\cS_{\ell,f}$ contains at least one element.
					Suppose we are given two elements $\fa, \fa'$ in $\cS_{\ell,f}$.
					By virtue of Lemma~\ref{lemma:SameSliceEntailsEquivalence}, this means $\fa \sim_\ell \fa'$.
					
					Because of $\cA \models \chi_3$, there are two elements $\tfa, \tfa' \in \cI_\ell$ for which \ref{enum:IndexSetsProperties:Two:Ia} entails $\fa \sim_\ell \tfa$ and $\fa' \sim_\ell \tfa'$.
					By symmetry and transitivity of $\sim_\ell$, we have $\tfa \sim_\ell \tfa'$.
					Hence, we observe 
						\begin{align*}
							\cA, &[\tj \Mapsto \tfa, \tj' \Mapsto \tfa'] \models \forall \ti.\; L(\ul{\ell-1}, \ti) \rightarrow \bigl( J(\ul{\ell}, \tj, \ti, 0) \leftrightarrow J(\ul{\ell}, \tj', \ti, 0) \bigr) ~.
						\end{align*}	
					Consequently, $\cA \models \chi_3$ leads to $\fa = \fa'$.
					
				\item \textit{Ad \ref{enum:IndexSetsProperties:Two:V}.}
					As we have seen in the proof of Lemma~\ref{lemma:ArrangeFunctions}, we can define a bijective mapping $\mu$ that maps the functions in $\cF_{\ell-1}$ to bit strings of length $\mu$ that either have a $0$ as most significant bit or correspond to $0 \ldots 0 1$ (where the most significant bit is the rightmost one).
					
					Let $p' := 2^{|\cI_{\ell-1}|-1}+1$.
					By virtue of Lemma~\ref{lemma:ArrangeFunctions}, we can uniquely construct a chain
						\[ f_1 \sqsubset_{\ell-1} f_2 \sqsubset_{\ell-1} \ldots \sqsubset_{\ell-1} f_{p'} \]
					comprising all functions in $\cF_{\ell-1}$. 
					
					Properties \ref{enum:IndexSetsProperties:Two:III} and \ref{enum:IndexSetsProperties:Two:IV} together yield that $\cI_\ell = \{ \fa_1, \ldots, \fa_{p'} \}$ where $\fa_k \in \cS_{\ell,f_k}$ for every $k = 1, \ldots, p'$.
					Lemma~\ref{lemma:SuccEntailsProgressInSlice} says that for any $\fa_k, \fa_{k'}$ with $\fa_k \prec_\ell \fa_{k'}$ we observe $f_k \sqsubset_{\ell-1} f_{k'}$.
					By definition of $\sqsubset_{\ell-1}$ and the fact that all the $f_k, f_{k'}$ are distinct, it clearly holds that $f_k \sqsubset_{\ell-1} f_{k'}$ implies $k' = k+1$ and $1\leq k < k' \leq p'$.
					Hence, $\fa_k \prec_\ell \fa_{k'}$ can only hold if $k' = k+1$ and $1\leq k < k' \leq p'$.

					Consider any element $\fa \in \cI_\ell$ for which $\cA, [j \Mapsto \fa, i \Mapsto e_{\ell-1}^\cA] \models J(\ul{\ell}, j, i, 0)$.
					By $\cA \models \psi_{16}$, we know that there are elements $\tfa, \tfa' \in \cI_\ell$ such that $\tfa \prec_\ell \tfa'$ and $\cA, [j \Mapsto \fa, \tj \Mapsto \tfa] \models \eq{\ell}_{j,\tj}$. 
					By \ref{enum:IndexSetsProperties:Two:Ia}, the latter translates to $\fa \sim_\ell \tfa$. 
					
					Let $g, \widetilde{g}$ be functions such that $\fa \in \cS_{\ell, g}$ and $\tfa \in \cS_{\ell, \widetilde{g}}$.
					Such functions exist by virtue of \ref{enum:IndexSetsProperties:Two:III}.
					However, $\fa \sim_\ell \tfa$ entails $g = \widetilde{g}$, by Lemma~\ref{lemma:EquivalenceEntailsSameSlice}.
					Moreover, \ref{enum:IndexSetsProperties:Two:IV} leads to $\fa = \tfa$.
					Consequently, we have $\fa \prec_\ell \tfa'$.					
					By \ref{enum:IndexSetsProperties:Two:IV}, this means that all but one element in $\cI_\ell$ must have a successor in $\cI_\ell$ and all but one element in $\cI_\ell$ are successor in $\cI_\ell$.
					Hence, we obtain the chain $\fa_1 \prec_\ell \fa_2 \prec_\ell \ldots \prec_\ell \fa_{p'-1} \prec_\ell \fa_{p'}$ where $\fa_1 = d_\ell^\cA$ and $\fa_{p'} = e_\ell^\cA$.
					\qedhere
			\end{description}									
	\end{description}
\end{proof}

The following is a corollary of the previous lemma, and so is Lemma~\ref{lemma:IndexSetsProperties}.
\begin{corollary}\label{corollary:IndexSetsProperties}
	For every $\ell = 1, \ldots, \kappa$ we have $|\cI_{\ell}| = p$ where $p := 2^{|\cI_{\ell-1}|-1}+1 = \twoup{\ell}{\mu-1}+1$.
	Moreover, there is a unique chain $\fa_1 \prec_\ell \ldots \prec_\ell \fa_p$ comprising all elements in $\cI_\ell$, and $\cA, [j \Mapsto \fa_1] \models \MinIdx(\ul{\ell}, j)$ and $\cA, [j' \Mapsto \fa_p] \models \MaxIdx(\ul{\ell}, j')$.
\end{corollary}

\section{Details regarding Section~\ref{section:FormalizingTheTiling}}

The following sentences encode a given domino problem $\fD := \<\cD, \cH, \cV\>$ plus an initial condition $\bD$, which is a finite word over $\cD$. 
We try to make as many sentences as possible Horn.
\begin{align*}	
	\eta_{1} \!:=\;	
		& \forall x y x' y'.\; H(x, y, x', y') \;\;\rightarrow\;\; L(\ul{\kappa}, x) \wedge L(\ul{\kappa}, y) \wedge L(\ul{\kappa}, x') \wedge L(\ul{\kappa}, y') \wedge y \approx y' \\
	\eta_{2} \!:=\;	
		& \forall x y x' y' i.\; H(x, y, x', y')  \wedge \MaxIdx(\ul{\kappa-1}, i) \wedge J(\ul{\kappa}, x, i, 0) \;\;\rightarrow\;\; \Succ(\ul{\kappa}, x, x') \\
	\eta_{3} \!:=\;	
		& \forall x y i.\; L(\ul{\kappa}, x)\wedge L(\ul{\kappa}, y) \wedge \MaxIdx(\ul{\kappa-1}, i) \wedge J(\ul{\kappa}, x, i, 0) \\
		&\hspace{8ex} \rightarrow\;\; \exists \tx\, \ty\, \tx'.\; \eq{\kappa}_{x, \tx} \wedge \eq{\kappa}_{y, \ty} \wedge \Bigl( \bigwedge_{D \in \cD} \ul{D}(x,y) \leftrightarrow \ul{D}(\tx, \ty) \Bigr) \wedge H(\tx, \ty, \tx', \ty) \\
	\eta_{4} \!:=\;
		& \forall x y x'.\; \MaxIdx(\ul{\kappa}, x) \wedge \MinIdx(\ul{\kappa}, x') \;\;\rightarrow\;\; H(x, y, x', y) \\
	\eta_{5} \!:=\;
		& \forall x y x' y'.\; H(x, y, x', y') \wedge \MaxIdx(\ul{\kappa}, x) \;\;\rightarrow\;\; \MinIdx(\ul{\kappa}, x') \\
	\eta_{6} \!:=\;
		& \forall x y x' y'.\; H(x, y, x', y') \wedge \MinIdx(\ul{\kappa}, x') \;\;\rightarrow\;\; \MaxIdx(\ul{\kappa}, x) \\
	\eta_{7} \!:=\;	
		& \forall x y x' y'.\; V(x, y, x', y') \;\;\rightarrow\;\; L(\ul{\kappa}, x) \wedge L(\ul{\kappa}, y) \wedge L(\ul{\kappa}, x') \wedge L(\ul{\kappa}, y') \wedge x\approx x' \\
	\eta_{8} \!:=\;	
		& \forall x y x' y'.\; V(x, y, x', y') \wedge \MaxIdx(\ul{\kappa-1}, i) \wedge J(\ul{\kappa}, y, i, 0) \;\;\rightarrow\;\; \Succ(\ul{\kappa}, y, y') \\
	\eta_{9} \!:=\;	
		& \forall x y i.\; L(\ul{\kappa}, x)\wedge L(\ul{\kappa}, y) \wedge \MaxIdx(\ul{\kappa-1}, i) \wedge J(\ul{\kappa}, y, i, 0) \\
		&\hspace{8ex} \rightarrow\;\; \exists \tx\, \ty\, \ty'.\; \eq{\kappa}_{x, \tx} \wedge \eq{\kappa}_{y, \ty} \wedge \Bigl( \bigwedge_{D \in \cD} \bigl( \ul{D}(x,y) \leftrightarrow \ul{D}(\tx, \ty) \bigr) \Bigr) \wedge V(\tx, \ty, \tx, \ty') \\
	\eta_{10} \!:=\;
		& \forall x y y'.\; \MaxIdx(\ul{\kappa}, y) \wedge \MinIdx(\ul{\kappa}, y') \;\;\rightarrow\;\; V(x, y, x, y') \\
	\eta_{11} \!:=\;
		& \forall x x' y y'.\; V(x, y, x', y') \wedge \MaxIdx(\ul{\kappa}, y) \;\;\rightarrow\;\; \MinIdx(\ul{\kappa}, y')  \\
	\eta_{12} \!:=\;
		& \forall x x' y y'.\; V(x, y, x', y') \wedge \MinIdx(\ul{\kappa}, y') \;\;\rightarrow\;\; \MaxIdx(\ul{\kappa}, y) \\
	\eta_{13} \!:=
		& \bigwedge_{D \in \cD} \forall x y.\; \ul{D}(x,y) \;\;\rightarrow\;\; L(\ul{\kappa}, x) \wedge L(\ul{\kappa}, y) \\
	\eta_{14} \!:=
		& \bigwedge_{D \in \cD} \bigwedge_{D' \in \cD \setminus \{D\}} \forall x y.\; \ul{D}(x,y) \;\;\rightarrow\;\; \neg \ul{D}'(x,y) 
\end{align*}		
\begin{align*}
	\eta_{15} \!:=\;
		& \forall x x' y.\; H(x, y, x', y) \;\;\rightarrow\;\; \bigvee_{\<D, D'\> \in \cH} \ul{D}(x,y) \wedge \ul{D}'(x',y) \\
	\eta_{16} \!:=\;
		& \forall x y y'.\; V(x, y, x, y') \;\;\rightarrow\;\; \bigvee_{\<D, D'\> \in \cV} \ul{D}(x,y) \wedge \ul{D}'(x,y') \\
	\eta_{17} \!:=\;
		& \forall z.\; \MinIdx(\ul{\kappa}, z) \;\;\rightarrow\;\; f_1 \approx z \wedge \bigwedge_{i=1}^{n-1} H(f_i, z, f_{i+1}, z) \\
	\eta_{18} \!:=\;
		& \forall z.\; \MinIdx(\ul{\kappa}, z) \;\;\rightarrow\;\; \bigwedge_{i = 1}^{n} \ul{D}_i(f_i, z)
\end{align*}
The sentences $\eta_{17}$ and $\eta_{18}$ express the initial condition $\bD$ where the lower left domino tiles are predefined to be the sequence $\bD = D_1 \ldots D_n$.

Regarding the length of the sentences $\eta_1, \ldots, \eta_{18}$, we observe the following:
	\begin{itemize}
		\item $\len(\eta_1), \len(\eta_2), \len(\eta_4), \ldots, \len(\eta_8), \len(\eta_{10}), \ldots, \len(\eta_{12}) \in \cO(\log \kappa)$
		\item $\len(\eta_3), \len(\eta_9) \in \cO\bigl( \kappa \log \kappa + \mu (\log \kappa + \log \mu) + |\cD| \log |\cD| \bigr)$
		\item $\len(\eta_{13}) \in \cO\bigl( |\cD| (\log |\cD| + \log \kappa) \bigr)$
		\item $\len(\eta_{14}) \in \cO\bigl( |\cD|^2 \log |\cD| \bigr)$
		\item $\len(\eta_{15}) \in \cO\bigl( |\cH| \log |\cD| \bigr) = \cO\bigl( |\cD|^2 \log |\cD| \bigr)$
		\item $\len(\eta_{16}) \in \cO\bigl( |\cH| \log |\cD| \bigr) = \cO\bigl( |\cD|^2 \log |\cD| \bigr)$
		\item $\len(\eta_{17}) \in \cO\bigl( \log \kappa + n \log n \bigr)$
		\item $\len(\eta_{18}) \in \cO\bigl( \log \kappa + n (\log n + \log |\cD|) \bigr)$
	\end{itemize}	
In total, the length of $\eta_1 \wedge \ldots \wedge \eta_{18}$ lies in $\cO \bigl( \widehat{n} \log \widehat{n} \bigr)$, where $\widehat{n} := \max \{ \kappa, \mu, n, |\cD|^2 \}$.

For the remainder of this subsection we consider a model $\cA$ of the sentence $\psi_{1} \wedge \ldots \wedge \psi_{16}\wedge \chi_1 \wedge \chi_2 \wedge \chi_3 \wedge \eta_1 \wedge \ldots \wedge \eta_{18}$.
Moreover, we take over the notation from Definition~\ref{definition:IndexSetsAndRelatedNotation}.

\begin{definition}
	We define the following relations:
	\begin{itemize}
		\item ${\prec^{H}} \subseteq \cI_\kappa^2 \times \cI_\kappa^2$ such that $\<\fa, \fb\> \prec^H \<\fa', \fb'\>$ holds if and only if 
			 $\cA, [x \Mapsto \fa, y \Mapsto \fb, x' \Mapsto \fa', y' \Mapsto \fb'] \models H(x, y, x', y')$.
		\item ${\prec^{V}} \subseteq \cI_\kappa^2 \times \cI_\kappa^2$ such that $\<\fa, \fb\> \prec^V \<\fa', \fb'\>$ holds if and only if  
			$\cA, [x \Mapsto \fa, y \Mapsto \fb, x' \Mapsto \fa', y' \Mapsto \fb'] \models V(x, y, x', y')$.
	\end{itemize}
\end{definition}

\begin{lemma}\label{lemma:TorusRelationsProperties}
	For all pairs $\<\fa, \fb\>, \<\fa', \fb'\> \in \cI_\kappa^2$ we observe the following properties.
	\begin{enumerate}[label=(\roman{*}), ref=(\roman{*})]
		\item\label{enum:TorusRelationsProperties:I} $\<\fa, \fb\> \prec^H \<\fa', \fb'\>$ entails that $\fb = \fb'$ and that either $\fa \prec_\kappa \fa'$ or $\fa = e_{\kappa}^\cA$ and $\fa' = d_{\kappa}^\cA$.
		\item\label{enum:TorusRelationsProperties:II} $\fa = e_{\kappa}^\cA$ and $\fa' = d_{\kappa}^\cA$ entails $\<\fa, \fb\> \prec^H \<\fa', \fb\>$ for every $\fb \in \cI_\kappa$.
		\item\label{enum:TorusRelationsProperties:III} $\fa \prec_\kappa \fa'$ implies $\<\fa, \fb\> \prec^H \<\fa', \fb\>$ for every $\fb \in \cI_\kappa$.
		
		\item\label{enum:TorusRelationsProperties:IV} $\<\fa, \fb\> \prec^V \<\fa', \fb'\>$ entails that $\fa = \fa'$ and that either $\fb \prec_\kappa \fb'$ or $\fb = e_{\kappa}^\cA$ and $\fb' = d_{\kappa}^\cA$.
		\item\label{enum:TorusRelationsProperties:V} $\fb = e_{\kappa}^\cA$ and $\fb' = d_{\kappa}^\cA$ entails $\<\fa, \fb\> \prec^V \<\fa, \fb'\>$ for every $\fa \in \cI_\kappa$.
		\item\label{enum:TorusRelationsProperties:VI} $\fb \prec_\kappa \fb'$ implies $\<\fa, \fb\> \prec^V \<\fa, \fb'\>$ for every $\fa \in \cI_\kappa$.
	\end{enumerate}
\end{lemma}
\begin{proof}
	Property~\ref{enum:TorusRelationsProperties:I} follows by $\cA \models \eta_1 \wedge \eta_2 \wedge \eta_5 \wedge \eta_6$.
	Property~\ref{enum:TorusRelationsProperties:II} follows by $\cA \models \eta_4$.
	
	In order to show Property~\ref{enum:TorusRelationsProperties:III}, we have to argue a bit more.
	First of all, we conclude $\fa \neq e_\kappa^\cA$, by $\cA \models \psi_2 \wedge \psi_4$.
	Because of $\cA \models \eta_3$ and Lemma~\ref{lemma:IndexSetsProperties:Two}\ref{enum:IndexSetsProperties:Two:Ia}, for all $\fa, \fb$ with $\fa \neq e_\kappa^\cA$ there must exist $\tfa, \tfb, \tfa' \in \cI_\kappa$ such that $\fa \sim_\kappa \tfa$, $\fb \sim_\kappa \tfb$, and $\<\tfa, \tfb\> \prec^H \<\tfa', \tfb\>$.
	Due to Lemma~\ref{lemma:EquivalenceEntailsSameSlice} in combination with Lemma~\ref{lemma:IndexSetsProperties:Two}\ref{enum:IndexSetsProperties:Two:IV}, we get $\fa = \tfa$ and $\fb = \tfb$.
	Thus, we have $\<\fa, \fb\> \prec^H \<\tfa', \fb\>$.
	Moreover, $\fa \neq e_\kappa^\cA$ together with \ref{enum:TorusRelationsProperties:I} leads to $\fa \prec_\kappa \tfa'$.
	Since we assumed $\fa \prec_\kappa \fa'$, Lemma~\ref{lemma:IndexSetsProperties:Two}\ref{enum:IndexSetsProperties:Two:V} says that $\fa'$ is the only element satisfying $\fa \prec_\kappa \fa'$, i.e.\ $\tfa' = \fa'$.
	Consequently, it in fact holds $\<\fa, \fb\> \prec^H \<\fa', \fb\>$.
	
	Properties~\ref{enum:TorusRelationsProperties:IV} to \ref{enum:TorusRelationsProperties:VI} can be proved analogously to the first three properties using $\cA \models \eta_7 \wedge \ldots \wedge \eta_{12}$.
\end{proof}

\begin{lemma}\label{lemma:TorusBijection}
	Let $r := \twoup{\kappa}{\mu-1}+1$.		
	There is a bijective mapping $\rho: \Int_r^2 \to \cI_\kappa^2$ such that
	$\rho(0,0) = \< d_{\kappa}^\cA, d_{\kappa}^\cA \>$
	and
	for every pair $\<s,t\> \in \Int_r^2$ we have
		\begin{itemize}		
			\item $\rho(s,t) \prec^H \rho(s+1,t)$ and
			\item $\rho(s,t) \prec^V \rho(s,t+1)$.
		\end{itemize}	
	where $+$ stands for addition modulo $r$.
\end{lemma}
\begin{proof}
	By Corollary~\ref{corollary:IndexSetsProperties} we know that there is a unique chain $\fa_1 \prec _\kappa \ldots \prec_\kappa \fa_r$ comprising all elements in $\cI_\kappa$.
	Notice that $\fa_k, \fa_{k'}$ with $k \neq k'$ are distinct.
	We define $\rho$ so that $\rho(s,t) := \< \fa_{s+1}, \fa_{t+1} \>$ for all $s, t \in \Int_r = \{0, \ldots r-1\}$.
	
	Obviously, $\rho$ is bijective.
	Since $\fa_1$ is the only element in $\cI_\kappa$ for which there is no $\fb$ in the above chain with $\fb \prec_\kappa \fa_1$, $\cA \models \psi_2 \wedge \psi_3$ enforces $\fa_1 = d_{\kappa}^\cA$.
	Hence, $\rho(0,0) = \<d_{\kappa}^\cA, d_{\kappa}^\cA\>$.
	
	Since $\fa_r$ is the only element in $\cI_\kappa$ for which there is no $\fb'$ in the above chain with $\fa_r \prec_\kappa \fb'$, $\cA \models \psi_4 \wedge \psi_5$ enforces $\fa_r = e_{\kappa}^\cA$.
	Hence, Lemma~\ref{lemma:TorusRelationsProperties}\ref{enum:TorusRelationsProperties:II} entails $\rho(r-1, t) \prec^H \rho(0,t)$ for every $t \in \Int_r$.
	Moreover, the existence of the above chain together with Lemma~\ref{lemma:TorusRelationsProperties}\ref{enum:TorusRelationsProperties:III} leads to $\rho(s, t) \prec^H \rho(s+1,t)$ for every $s \in \Int_r \setminus \{r-1\}$ and every $t \in \Int_r$.
	Consequently, we observe $\rho(s,t) \prec^H \rho(s+1,t)$---modulo $r$---for every pair $\<s,t\> \in \Int_r^2$.
		
	By similar arguments, we may infer $\rho(s,t) \prec^V \rho(s,t+1)$ for every pair $\<s,t\> \in \Int_r^2$.
\end{proof}

\begin{lemma}\label{lemma:CompletenessAndUniquenessOfInducedTiling}
	Assume that $\cD$, $\cH$, and $\cV$ are nonempty.
	For all pairs $\<\fa, \fb\> \in \cI_\kappa$ we have $\cA, [x \Mapsto \fa, y \Mapsto \fb] \models \ul{D}(x,y)$ for exactly one $D \in \cD$.
\end{lemma}
\begin{proof}	
	Due to $\cA \models \eta_{15} \wedge \eta_{16}$, we observe the following properties for all pairs $\<\fa, \fb\>, \<\fa', \fb'\> \in \cI_\kappa^2$:
	\begin{itemize}
		\item $\<\fa,\fb\>\! \prec^H \!\!\<\fa',\fb'\>$ implies that there are $D, D' \in \cD$ such that $\<D, D'\> \in \cH$ and $\cA, [x \Mapsto \fa, y \Mapsto \fb] \models \ul{D}(x,y)$ and $\cA, [x \Mapsto \fa', y \Mapsto \fb'] \models \ul{D}'(x,y)$.
		
		\item $\<\fa,\fb\>\! \prec^V \!\!\<\fa',\fb'\>$ implies that there are $D, D' \in \cD$ such that $\<D, D'\> \in \cV$ and $\cA, [x \Mapsto \fa, y \Mapsto \fb] \models \ul{D}(x,y)$ and $\cA, [x \Mapsto \fa', y \Mapsto \fb'] \models \ul{D}'(x,y)$.
	\end{itemize}
	By virtue of Lemma~\ref{lemma:TorusBijection}, we know that there is a bijection $\rho$ such that for every pair $\<\fa, \fb\>$ in the image of $\rho$ there is another pair $\<\fa', \fb'\>$ such that $\<\fa, \fb\> \prec^H \<\fa', \fb'\>$ or $\<\fa, \fb\> \prec^V \<\fa', \fb'\>$.
	Since the image of $\rho$ is $\cI_\kappa^2$, this means that there is at least one $D \in \cD$ for every pair $\<\fa, \fb\> \in \cI_\kappa^2$ such that $\cA, [x \Mapsto \fa, y \Mapsto \fb] \models \ul{D}(x,y)$.

	Finally, because of $\cA \models \eta_{14}$ we know that there is at most one $D \in \cD$ for every pair $\<\fa, \fb\> \in \cI_\kappa^2$ such that $\cA, [x \Mapsto \fa, y \Mapsto \fb] \models \ul{D}(x,y)$.
\end{proof}

\setcounter{auxTheorem}{\value{theorem}}
\setcounter{theorem}{24}

\begin{lemma}
	Let $r := \twoup{\kappa}{\mu-1}+1$ and assume that $\cD$, $\cH$, and $\cV$ are nonempty.
	$\cA$ induces a tiling $\tau$ of $\Int_r^2$ with initial condition $\bD := D_1, \ldots, D_n$.
\end{lemma}
\begin{proof}
	Let $\rho$ be a bijection according to Lemma~\ref{lemma:TorusBijection}.
	We define the mapping $\tau$ such that $\tau(s,t) := D$ if and only if $\rho(s,t) = \<\fa, \fb\>$ and $\cA, [x \Mapsto \fa, y \Mapsto \fb] \models \ul{D}(x,y)$.
	By Lemma~\ref{lemma:CompletenessAndUniquenessOfInducedTiling}, we know that $\tau$ is well defined. 

	By $\cA \models \eta_{17} \wedge \eta_{18}$, we know that $\cA, [x \Mapsto \fa, y \Mapsto \fb] \models \ul{D}_i(x, y)$ for $\<\fa, \fb\> = \rho(i,0)$ and $i = 0, \ldots, n-1$.
	Hence, $\tau$ satisfies the initial condition.
	
	By definition of $\rho$ and because of $\cA \models \eta_{15} \wedge \eta_{16}$, we observe the following:
	\begin{itemize}
		\item For every pair $\<s, t\> \in \Int_r^2$ there are pairs $\<D, D'\> \in \cH$, $\<\fa, \fb\> = \rho(s,t)$, and $\<\fa', \fb'\> = \rho(s+1,t)$ such that 
			\begin{itemize}
				\item $\cA, [x \Mapsto \fa, y \Mapsto \fb] \models \ul{D}(x,y)$ and 
				\item $\cA, [x \Mapsto \fa', y \Mapsto \fb'] \models \ul{D}'(x,y)$.
			\end{itemize}	

		\item For every pair $\<s, t\> \in \Int_r^2$ there are pairs $\<D, D'\> \in \cV$, $\<\fa, \fb\> = \rho(s,t)$, and $\<\fa', \fb'\> = \rho(s,t+1)$ such that 
			\begin{itemize}
				\item $\cA, [x \Mapsto \fa, y \Mapsto \fb] \models \ul{D}(x,y)$ and 
				\item $\cA, [x \Mapsto \fa', y \Mapsto \fb'] \models \ul{D}'(x,y)$.
			\end{itemize}	
	\end{itemize}
	Consequently, the mapping $\tau$ constitutes a proper tiling of $\Int_r^2$.
\end{proof}
\setcounter{theorem}{\value{auxTheorem}}

\section{Details regarding Section~\ref{section:ReplaceEqualityInLowerBoundProof}}
The following sentences comprise the axioms of reflexivity, symmetry, transitivity, and several axioms of congruence ($\psi'_1, \ldots, \psi'_{12}, \eta'_{1}, \ldots, \eta'_{10}$) for the binary predicate $E$ that is introduced to replace equality.
\begin{align*}
	\psi'_1 :=\;
		& \forall j.\; E(j,j) \\
	\psi'_2 :=\;
		& \forall j j'.\; E(j,j') \;\rightarrow\; E(j',j) \\
	\psi'_3 :=\;
		& \forall j j' j''.\; E(j,j') \wedge E(j',j'') \;\rightarrow\; E(j,j'') \\
	\psi'_4 :=\;
		& \forall j j'. \bigwedge_{\ell = 0}^\kappa E(j,j) \wedge L(\ul{\ell}, j) \;\rightarrow\; L(\ul{\ell},j') \\
	\psi'_5 :=\;
		& \forall j j'. \bigwedge_{\ell = 0}^\kappa E(j,j) \wedge \MinIdx(\ul{\ell}, j) \;\;\rightarrow\;\; \MinIdx(\ul{\ell},j') \\
	\psi'_6 :=\;
		& \forall j j'. \bigwedge_{\ell = 0}^\kappa E(j,j) \wedge \MaxIdx(\ul{\ell}, j) \;\;\rightarrow\;\; \MaxIdx(\ul{\ell},j') \\
	\psi'_7 :=\;
		& \forall j j' i. \bigwedge_{\ell = 0}^\kappa E(j,j) \wedge \Succ(\ul{\ell}, j, i) \;\;\rightarrow\;\; \Succ(\ul{\ell},j', i) \\
	\psi'_8 :=\;
		& \forall j i i'. \bigwedge_{\ell = 0}^\kappa E(j,j) \wedge \Succ(\ul{\ell}, j, i) \;\;\rightarrow\;\; \Succ(\ul{\ell},j, i') \\
	\psi'_9 :=\;
		& \forall j j' i. \bigwedge_{\ell = 0}^\kappa \bigl( E(j,j) \wedge J(\ul{\ell}, j, i, 0) \;\rightarrow\; J(\ul{\ell},j', i, 0) \bigr) \\[-1ex]
			&\hspace{10ex} \wedge\;\; \bigl( E(j,j) \wedge J(\ul{\ell}, j, i, 1) \;\rightarrow\; J(\ul{\ell},j', i, 1) \bigr) \\
	\psi'_{10} :=\;
		& \forall j i i'. \bigwedge_{\ell = 0}^\kappa \bigl( E(j,j) \wedge J(\ul{\ell}, j, i, 0) \;\rightarrow\; J(\ul{\ell},j, i', 0) \bigr) \\[-1ex]
			&\hspace{10ex} \wedge\;\; \bigl( E(j,j) \wedge J(\ul{\ell}, j, i, 1) \;\rightarrow\; J(\ul{\ell},j, i', 1) \bigr) \\
	\psi'_{11} :=\;
		& \forall j j' i. \bigwedge_{\ell = 0}^\kappa \bigl( E(j,j) \wedge J^*(\ul{\ell}, j, i, 0) \;\rightarrow\; J^*(\ul{\ell},j', i, 0) \bigr) \\[-1ex]
			&\hspace{10ex} \wedge\;\; \bigl( E(j,j) \wedge J^*(\ul{\ell}, j, i, 1) \;\rightarrow\; J^*(\ul{\ell},j', i, 1) \bigr) \\
	\psi'_{12} :=\;
		& \forall j i i'. \bigwedge_{\ell = 0}^\kappa \bigl( E(j,j) \wedge J^*(\ul{\ell}, j, i, 0) \;\rightarrow\; J^*(\ul{\ell},j, i', 0) \bigr) \\[-1ex]
			&\hspace{10ex} \wedge\;\; \bigl( E(j,j) \wedge J^*(\ul{\ell}, j, i, 1) \;\rightarrow\; J^*(\ul{\ell},j, i', 1) \bigr) 
\end{align*}
\begin{align*}		
	\eta'_1 :=\;
		& \forall x y u v x'.\, E(x,x') \wedge H(x, y, u, v) \;\;\rightarrow\;\; H(x', y, u, v) \\
	\eta'_2 :=\;
		& \forall x y u v y'.\, E(y,y') \wedge H(x, y, u, v) \;\;\rightarrow\;\; H(x, y', u, v) \\
	\eta'_3 :=\;
		& \forall x y u v u'.\, E(u,u') \wedge H(x, y, u, v) \;\;\rightarrow\;\; H(x, y, u', v) \\
	\eta'_4 :=\;
		& \forall x y u v v'.\, E(v,v') \wedge H(x, y, u, v) \;\;\rightarrow\;\; H(x, y, u, v') \\
	\eta'_5 :=\;
		& \forall x y u v x'.\,  E(x,x') \wedge V(x, y, u, v) \;\;\rightarrow\;\; V(x', y, u, v) \\
	\eta'_6 :=\;
		& \forall x y u v y'.\, E(y,y') \wedge V(x, y, u, v) \;\;\rightarrow\;\; V(x, y', u, v) \\
	\eta'_7 :=\;
		& \forall x y u v u'.\, E(u,u') \wedge V(x, y, u, v) \;\;\rightarrow\;\; V(x, y, u', v) \\
	\eta'_8 :=\;
		& \forall x y u v v'.\, E(v,v') \wedge V(x, y, u, v) \;\;\rightarrow\;\; V(x, y, u, v') \\
	\eta'_9 :=\;
		& \forall x y x'. \bigwedge_{D \in \cD} E(x,x') \wedge \ul{D}(x, y) \;\;\rightarrow\;\; \ul{D}(x', y) \\
	\eta'_{10} :=\;
		& \forall x y y'. \bigwedge_{D \in \cD} E(y,y') \wedge \ul{D}(x, y) \;\;\rightarrow\;\; \ul{D}(x, y')
\end{align*}
Regarding the length of the above sentences, we observe the following:
	\begin{itemize}
		\item $\len(\psi'_1), \len(\psi'_2), \len(\psi'_3) \in \cO(1)$,
		\item $\len(\eta'_1), \ldots, \len(\eta'_8) \in \cO(1)$,
		\item $\len(\psi'_4), \ldots, \len(\psi'_{12}) \in \cO(\kappa \log \kappa)$,
		\item $\len(\eta'_9), \len(\eta'_{10}) \in \cO\bigl( |\cD| \log |\cD| \bigr)$.
	\end{itemize}

\bigskip
\setcounter{auxTheorem}{\value{theorem}}
\setcounter{theorem}{25}

\begin{lemma}
	Let $\varphi := \psi_1 \wedge \ldots \psi_{16} \wedge \chi_1 \wedge \chi_2 \wedge \chi_3 \wedge \eta_1 \wedge \ldots \wedge \eta_{18}$.
	Let $\hphi$ be the result of replacing every atom $s \approx t$ in $\varphi$ with $E(s,t)$.
	Moreover, let $\psi' := \psi'_1 \wedge \ldots \wedge \psi'_{12}$ and $\eta' := \eta'_1 \wedge \ldots \wedge \eta'_{10}$.
	
	\begin{enumerate}[label=(\roman{*}), ref=(\roman{*})]
		\item\label{enum:replaceEquality:I} Every model $\cA \models \varphi$ can be extended to a model $\cB \models \hphi \wedge \psi' \wedge \eta'$ over the same domain.
		\item\label{enum:replaceEquality:II} From every model $\cB \models \hphi \wedge \psi' \wedge \eta'$ we can construct a model $\cA \models \varphi$.
	\end{enumerate}
\end{lemma}
\begin{proof}~
	\begin{description}
		\item Ad \ref{enum:replaceEquality:I}.
			Let $\cA$ be a model of $\varphi$. 
			We define $\cB$ exactly like $\cA$ except for the interpretation of $E$, which we define to be $E^\cB := \{\<\fa, \fa\> \mid \fa \in \fU_\cA\}$.
			Clearly, we have $\cB \models \varphi$ and also $\cB \models \hphi$.
			Moreover, it is easy to see that $\cB \models \psi' \wedge \eta'$.
			
		\item Ad \ref{enum:replaceEquality:II}.
			Let $\cB$ be a model of $\hphi \wedge \psi' \wedge \eta'$.
			For all domain elements $\fa, \fb \in \fU_\fB$ we write $\fa \sim_E \fb$ if and only if $\cB, [j \Mapsto \fa, j' \Mapsto \fb] \models E(j,j')$.
			Due to $\cB \models \psi'_1 \wedge \psi'_2 \wedge \psi'_3$, we know that $\sim_E$ constitutes an equivalence relation on $\fU_\cB$.
			Let $\fU_\cB /_{\sim_E}$ be the set of all equivalence classes induced by $\sim_E$.			

			\begin{description}
				\item \underline{Claim I:} 
					Consider two variable assignments $\beta, \beta'$ such that for every variable $j$ we have $\beta(j) \sim_E \beta'(j)$.
					For every atom $A$ occurring in $\hphi$ we have $\cB, \beta \models A$ if and only if $\cB, \beta' \models A$.
					
				\item \underline{Proof:}
					We distinguish two cases:
					Let $A = P(\ldots)$, where $P$ is one of the predicate symbols $L$, $\MinIdx$, $\MaxIdx$, $\Succ$, $J$, $J^*$, $H$, $V$, or $\ul{D}$ with $D \in \cD$.
					Then the claim follows due to $\cB \models \psi' \wedge \eta'$ and the definition of $\sim_E$.
				
					Let $A = E(j,i)$. 
					Assume $\cB,\beta \models E(j,i)$.
					Hence, we have $\beta'(j) \sim_E \beta(j) \sim_E \beta(i) \sim_E \beta'(i)$.
					By transitivity of $\sim_E$ it follows that $\beta'(j) \sim_E \beta'(i)$, and thus we have $\cB, \beta' \models E(j,i)$.
					
					Assume $\cB,\beta' \models E(j,i)$.
					Hence, we have $\beta(j) \sim_E \beta'(j) \sim_E \beta'(i) \sim_E \beta(i)$.
					By transitivity of $\sim_E$ it follows that $\beta(j) \sim_E \beta(i)$, and thus we have $\cB, \beta \models E(j,i)$.
				\strut\hfill$\Diamond$	
			\end{description}
									
			We define the structure $\cA$ as follows:
				\begin{itemize}
					\item We define $\fU_{\cA}$ such that it contains exactly one representative from every equivalence class in $\fU_\cB/_{\sim_E}$.
					\item For every constant symbol $c$ occurring in $\hphi$ we set $c^{\cA} := \fc$, where $\fc$ is the representative in $\fU_{\cA}$ such that it represents the $\sim_E$-class of $c^\cB$, i.e.\ $[\fc]_{\sim_E} = [c^\cB]_{\sim_E}$.
					\item For every predicate symbol $P$ occurring in $\hphi$ we set $P^{\cA} := P^{\cB} \cap \fU_{\cA}^{m}$, where $m$ is the arity of $P$.
				\end{itemize}
		
			\begin{description}
				\item\underline{Claim II:} 
					$\cA \models \hphi$.
				
				\item\underline{Proof:}
					By construction of $\fU_\cA$, we can find for every variable assignment $\beta$ mapping variables to elements in $\fU_\cB$ another variable assignment $\beta'$ mapping every variable to an element in $\fU_\cA$ such that $\beta(j) \sim_E \beta'(j)$ holds for every variable $j$.
					Consequently, using Claim I, we can show that $\cA \models \hphi$ holds by induction on the structure of $\hphi$.
				\strut\hfill$\Diamond$			
			\end{description}	
			
			From the definition of $\cA$ it follows that $E^\cA = \{\<\fa, \fa\> \mid \fa \in \fU_\cA\}$.
			Hence, for all $\fa, \fb \in \fU_\cA$ we have $\cA, [j \Mapsto \fa, j' \Mapsto \fb] \models E(j, j')$ if and only if $\cA, [j \Mapsto \fa, j' \Mapsto \fb] \models j \approx j'$.
			This observation together with Claim II entails $\cA \models \varphi$.				
			\qedhere
	\end{description}
\end{proof}
\setcounter{theorem}{\value{auxTheorem}}

\end{document}